\documentclass[longauth]{aa} 
\usepackage[utf8]{inputenc}
\usepackage{graphicx}
\usepackage{amsmath}
\usepackage{txfonts}
\usepackage{fourier}
\usepackage{url}
\usepackage{longtable}
\usepackage{tikz}
\usepackage{chemformula}
\usepackage[breaklinks, colorlinks, citecolor=blue, linkcolor=blue]{hyperref}
\usepackage[normalem]{ulem}
\usepackage{fancyhdr}
\usepackage{placeins}
\usepackage{natbib}
\begin{document}

\newcommand{\teff}{T$_{\rm eff}$}
\newcommand{\logg}{$\log${(g)}}
\newcommand{\met}{$[$Fe/H$]$}
\newcommand{\wftb}{WASP-43b}
\newcommand{\wft}{WASP-43}
\newcommand{\cheops}{CHEOPS}
\newcommand{\tess}{TESS}
\newcommand{\uvis}{HST~WFC3/UVIS}
\newcommand{\aas}{AAS}
\newcommand{\vsini}{\ensuremath{v \sin i_\star}\xspace}
\newcommand{\vmic}{$V_{\rm mic}$}
\newcommand{\vmac}{$V_{\rm mac}$}
\newcommand{\kms}{km\,s$^{-1}$}
\newcommand{\ag}{A$_{\rm g}$}

\graphicspath{{figures/}}
\let\orgautoref\autoref
\makeatletter
\renewcommand*\aa@pageof{, page \thepage{} of \pageref*{LastPage}}
\def\instrefs#1{{\def\scsep{\def\scsep{,}}\@for\w:=#1\do{\scsep\ref{inst:\w}}}}
\makeatother

   \title{The phase curve and the geometric albedo of \wftb\ measured with \cheops, \tess\ and \uvis\thanks{The \cheops\ program ID is CH\_PR100016. The \cheops\ photometry discussed in this paper is available in electronic form at the CDS via anonymous ftp to cdsarc.u-strasbg.fr (130.79.128.5) or via \url{http://cdsweb.u-strasbg.fr/cgi-bin/qcat?J/A+A/}}}


   \author{
G. Scandariato\inst{1} $^{\href{https://orcid.org/0000-0003-2029-0626}{\includegraphics[scale=0.5]{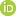}}}$,
V. Singh\inst{1} $^{\href{https://orcid.org/0000-0002-7485-6309}{\includegraphics[scale=0.5]{figures/orcid.jpg}}}$,
D. Kitzmann\inst{2} $^{\href{https://orcid.org/0000-0003-4269-3311}{\includegraphics[scale=0.5]{figures/orcid.jpg}}}$,
M. Lendl\inst{3} $^{\href{https://orcid.org/0000-0001-9699-1459}{\includegraphics[scale=0.5]{figures/orcid.jpg}}}$,
A. Brandeker\inst{4} $^{\href{https://orcid.org/0000-0002-7201-7536}{\includegraphics[scale=0.5]{figures/orcid.jpg}}}$,
G. Bruno\inst{1} $^{\href{https://orcid.org/0000-0002-3288-0802}{\includegraphics[scale=0.5]{figures/orcid.jpg}}}$,
A. Bekkelien\inst{3}, 
W. Benz\inst{5,6} $^{\href{https://orcid.org/0000-0001-7896-6479}{\includegraphics[scale=0.5]{figures/orcid.jpg}}}$,
P. Gutermann\inst{7,8}, 
P. F. L. Maxted\inst{9} $^{\href{https://orcid.org/0000-0003-3794-1317}{\includegraphics[scale=0.5]{figures/orcid.jpg}}}$,
A. Bonfanti\inst{10} $^{\href{https://orcid.org/0000-0002-1916-5935}{\includegraphics[scale=0.5]{figures/orcid.jpg}}}$,
S. Charnoz\inst{11} $^{\href{https://orcid.org/0000-0002-7442-491X}{\includegraphics[scale=0.5]{figures/orcid.jpg}}}$,
M. Fridlund\inst{12,13} $^{\href{https://orcid.org/0000-0002-0855-8426}{\includegraphics[scale=0.5]{figures/orcid.jpg}}}$,
K. Heng\inst{6,14} $^{\href{https://orcid.org/0000-0003-1907-5910}{\includegraphics[scale=0.5]{figures/orcid.jpg}}}$,
S. Hoyer\inst{7} $^{\href{https://orcid.org/0000-0003-3477-2466}{\includegraphics[scale=0.5]{figures/orcid.jpg}}}$,
I. Pagano\inst{1} $^{\href{https://orcid.org/0000-0001-9573-4928}{\includegraphics[scale=0.5]{figures/orcid.jpg}}}$,
C. M. Persson\inst{13}, 
S. Salmon\inst{3} $^{\href{https://orcid.org/0000-0002-1714-3513}{\includegraphics[scale=0.5]{figures/orcid.jpg}}}$,
V. Van Grootel\inst{15} $^{\href{https://orcid.org/0000-0003-2144-4316}{\includegraphics[scale=0.5]{figures/orcid.jpg}}}$,
T. G. Wilson\inst{16} $^{\href{https://orcid.org/0000-0001-8749-1962}{\includegraphics[scale=0.5]{figures/orcid.jpg}}}$,
J. Asquier\inst{17}, 
M. Bergomi\inst{18}, 
L. Gambicorti\inst{19}, 
J. Hasiba\inst{20}, 
Y. Alibert\inst{5} $^{\href{https://orcid.org/0000-0002-4644-8818}{\includegraphics[scale=0.5]{figures/orcid.jpg}}}$,
R. Alonso\inst{21,22} $^{\href{https://orcid.org/0000-0001-8462-8126}{\includegraphics[scale=0.5]{figures/orcid.jpg}}}$,
G. Anglada\inst{23,24} $^{\href{https://orcid.org/0000-0002-3645-5977}{\includegraphics[scale=0.5]{figures/orcid.jpg}}}$,
T. Bárczy\inst{25} $^{\href{https://orcid.org/0000-0002-7822-4413}{\includegraphics[scale=0.5]{figures/orcid.jpg}}}$,
D. Barrado y Navascues\inst{26} $^{\href{https://orcid.org/0000-0002-5971-9242}{\includegraphics[scale=0.5]{figures/orcid.jpg}}}$,
S. C. C. Barros\inst{27,28} $^{\href{https://orcid.org/0000-0003-2434-3625}{\includegraphics[scale=0.5]{figures/orcid.jpg}}}$,
W. Baumjohann\inst{10} $^{\href{https://orcid.org/0000-0001-6271-0110}{\includegraphics[scale=0.5]{figures/orcid.jpg}}}$,
M. Beck\inst{3} $^{\href{https://orcid.org/0000-0003-3926-0275}{\includegraphics[scale=0.5]{figures/orcid.jpg}}}$,
T. Beck\inst{5}, 
N. Billot\inst{3} $^{\href{https://orcid.org/0000-0003-3429-3836}{\includegraphics[scale=0.5]{figures/orcid.jpg}}}$,
X. Bonfils\inst{29} $^{\href{https://orcid.org/0000-0001-9003-8894}{\includegraphics[scale=0.5]{figures/orcid.jpg}}}$,
C. Broeg\inst{5,6} $^{\href{https://orcid.org/0000-0001-5132-2614}{\includegraphics[scale=0.5]{figures/orcid.jpg}}}$,
J. Cabrera\inst{30} $^{\href{https://orcid.org/0000-0001-6653-5487}{\includegraphics[scale=0.5]{figures/orcid.jpg}}}$,
A. Collier Cameron\inst{16} $^{\href{https://orcid.org/0000-0002-8863-7828}{\includegraphics[scale=0.5]{figures/orcid.jpg}}}$,
Sz. Csizmadia\inst{30} $^{\href{https://orcid.org/0000-0001-6803-9698}{\includegraphics[scale=0.5]{figures/orcid.jpg}}}$,
M. B. Davies\inst{31} $^{\href{https://orcid.org/0000-0001-6080-1190}{\includegraphics[scale=0.5]{figures/orcid.jpg}}}$,
M. Deleuil\inst{7} $^{\href{https://orcid.org/0000-0001-6036-0225}{\includegraphics[scale=0.5]{figures/orcid.jpg}}}$,
A. Deline\inst{3}, 
L. Delrez\inst{32,33} $^{\href{https://orcid.org/0000-0001-6108-4808}{\includegraphics[scale=0.5]{figures/orcid.jpg}}}$,
O. Demangeon\inst{27,28} $^{\href{https://orcid.org/0000-0001-7918-0355}{\includegraphics[scale=0.5]{figures/orcid.jpg}}}$,
B.-O. Demory\inst{6} $^{\href{https://orcid.org/0000-0002-9355-5165}{\includegraphics[scale=0.5]{figures/orcid.jpg}}}$,
A. Erikson\inst{30}, 
A. Fortier\inst{5,6} $^{\href{https://orcid.org/0000-0001-8450-3374}{\includegraphics[scale=0.5]{figures/orcid.jpg}}}$,
L. Fossati\inst{10} $^{\href{https://orcid.org/0000-0003-4426-9530}{\includegraphics[scale=0.5]{figures/orcid.jpg}}}$,
D. Gandolfi\inst{34} $^{\href{https://orcid.org/0000-0001-8627-9628}{\includegraphics[scale=0.5]{figures/orcid.jpg}}}$,
M. Gillon\inst{32} $^{\href{https://orcid.org/0000-0003-1462-7739}{\includegraphics[scale=0.5]{figures/orcid.jpg}}}$,
M. Güdel\inst{35}, 
K. G. Isaak\inst{36} $^{\href{https://orcid.org/0000-0001-8585-1717}{\includegraphics[scale=0.5]{figures/orcid.jpg}}}$,
L. L. Kiss\inst{37,38}, 
J. Laskar\inst{39} $^{\href{https://orcid.org/0000-0003-2634-789X}{\includegraphics[scale=0.5]{figures/orcid.jpg}}}$,
A. Lecavelier des Etangs\inst{40} $^{\href{https://orcid.org/0000-0002-5637-5253}{\includegraphics[scale=0.5]{figures/orcid.jpg}}}$,
C. Lovis\inst{3} $^{\href{https://orcid.org/0000-0001-7120-5837}{\includegraphics[scale=0.5]{figures/orcid.jpg}}}$,
D. Magrin\inst{18} $^{\href{https://orcid.org/0000-0003-0312-313X}{\includegraphics[scale=0.5]{figures/orcid.jpg}}}$,
V. Nascimbeni\inst{18} $^{\href{https://orcid.org/0000-0001-9770-1214}{\includegraphics[scale=0.5]{figures/orcid.jpg}}}$,
G. Olofsson\inst{4} $^{\href{https://orcid.org/0000-0003-3747-7120}{\includegraphics[scale=0.5]{figures/orcid.jpg}}}$,
R. Ottensamer\inst{41}, 
E. Pallé\inst{21} $^{\href{https://orcid.org/0000-0003-0987-1593}{\includegraphics[scale=0.5]{figures/orcid.jpg}}}$,
H. Parviainen\inst{21} $^{\href{https://orcid.org/0000-0001-5519-1391}{\includegraphics[scale=0.5]{figures/orcid.jpg}}}$,
G. Peter\inst{42} $^{\href{https://orcid.org/0000-0001-6101-2513}{\includegraphics[scale=0.5]{figures/orcid.jpg}}}$,
G. Piotto\inst{18,43} $^{\href{https://orcid.org/0000-0002-9937-6387}{\includegraphics[scale=0.5]{figures/orcid.jpg}}}$,
D. Pollacco\inst{14}, 
D. Queloz\inst{44,45} $^{\href{https://orcid.org/0000-0002-3012-0316}{\includegraphics[scale=0.5]{figures/orcid.jpg}}}$,
R. Ragazzoni\inst{18,43} $^{\href{https://orcid.org/0000-0002-7697-5555}{\includegraphics[scale=0.5]{figures/orcid.jpg}}}$,
N. Rando\inst{17}, 
H. Rauer\inst{30,46,47} $^{\href{https://orcid.org/0000-0002-6510-1828}{\includegraphics[scale=0.5]{figures/orcid.jpg}}}$,
I. Ribas\inst{23,24} $^{\href{https://orcid.org/0000-0002-6689-0312}{\includegraphics[scale=0.5]{figures/orcid.jpg}}}$,
N. C. Santos\inst{27,28} $^{\href{https://orcid.org/0000-0003-4422-2919}{\includegraphics[scale=0.5]{figures/orcid.jpg}}}$,
D. Ségransan\inst{3} $^{\href{https://orcid.org/0000-0003-2355-8034}{\includegraphics[scale=0.5]{figures/orcid.jpg}}}$,
L. M. Serrano\inst{34} $^{\href{https://orcid.org/0000-0001-9211-3691}{\includegraphics[scale=0.5]{figures/orcid.jpg}}}$,
A. E. Simon\inst{5} $^{\href{https://orcid.org/0000-0001-9773-2600}{\includegraphics[scale=0.5]{figures/orcid.jpg}}}$,
A. M. S. Smith\inst{30} $^{\href{https://orcid.org/0000-0002-2386-4341}{\includegraphics[scale=0.5]{figures/orcid.jpg}}}$,
S. G. Sousa\inst{27} $^{\href{https://orcid.org/0000-0001-9047-2965}{\includegraphics[scale=0.5]{figures/orcid.jpg}}}$,
M. Steller\inst{10} $^{\href{https://orcid.org/0000-0003-2459-6155}{\includegraphics[scale=0.5]{figures/orcid.jpg}}}$,
Gy. M. Szabó\inst{48,49}, 
N. Thomas\inst{5}, 
S. Udry\inst{3} $^{\href{https://orcid.org/0000-0001-7576-6236}{\includegraphics[scale=0.5]{figures/orcid.jpg}}}$,
B. Ulmer\inst{42}, 
N. Walton\inst{50}
   }

   \institute{
\label{inst:1} INAF, Osservatorio Astrofisico di Catania, Via S. Sofia 78, 95123 Catania, Italy \and
\label{inst:2} Center for Space and Habitability, University of Bern, Gesellschaftsstrasse 6, Bern, 3012, Switzerland \and
\label{inst:3} Observatoire Astronomique de l'Université de Genève, Chemin Pegasi 51, Versoix, Switzerland \and
\label{inst:4} Department of Astronomy, Stockholm University, AlbaNova University Center, 10691 Stockholm, Sweden \and
\label{inst:5} Physikalisches Institut, University of Bern, Gesellsschaftstrasse 6, 3012 Bern, Switzerland \and
\label{inst:6} Center for Space and Habitability, Gesellsschaftstrasse 6, 3012 Bern, Switzerland \and
\label{inst:7} Aix Marseille Univ, CNRS, CNES, LAM, 38 rue Frédéric Joliot-Curie, 13388 Marseille, France \and
\label{inst:8} Division Technique INSU, CS20330, 83507 La Seyne sur Mer cedex, France \and
\label{inst:9} Astrophysics Group, Keele University, Staffordshire, ST5 5BG, United Kingdom \and
\label{inst:10} Space Research Institute, Austrian Academy of Sciences, Schmiedlstrasse 6, A-8042 Graz, Austria \and
\label{inst:11} Université de Paris, Institut de physique du globe de Paris, CNRS, F-75005 Paris, France \and
\label{inst:12} Leiden Observatory, University of Leiden, PO Box 9513, 2300 RA Leiden, The Netherlands \and
\label{inst:13} Department of Space, Earth and Environment, Chalmers University of Technology, Onsala Space Observatory, 439 92 Onsala, Sweden \and
\label{inst:14} Department of Physics, University of Warwick, Gibbet Hill Road, Coventry CV4 7AL, United Kingdom \and
\label{inst:15} Space sciences, Technologies and Astrophysics Research (STAR) Institute, Universit{\'e} de Liège, Allée du 6 Août 19C, 4000 Liège, Belgium \and
\label{inst:16} Centre for Exoplanet Science, SUPA School of Physics and Astronomy, University of St Andrews, North Haugh, St Andrews KY16 9SS, UK \and
\label{inst:17} ESTEC, European Space Agency, 2201AZ, Noordwijk, NL \and
\label{inst:18} INAF, Osservatorio Astronomico di Padova, Vicolo dell'Osservatorio 5, 35122 Padova, Italy \and
\label{inst:19} ATG Europe B.V. on behalf of ESA - European Space Agency ESA - ESTEC / TEC - MMO, Keplerlaan 1 - P.O. Box 299, NL-2200 AG Noordwijk, The Netherlands \and
\label{inst:20} Space Research Institute, Austrian Academy of Sciences, Schmiedlstrasse 6, 8042 Graz, Austria \and
\label{inst:21} Instituto de Astrofisica de Canarias, 38200 La Laguna, Tenerife, Spain \and
\label{inst:22} Departamento de Astrofisica, Universidad de La Laguna, 38206 La Laguna, Tenerife, Spain \and
\label{inst:23} Institut de Ciencies de l'Espai (ICE, CSIC), Campus UAB, Can Magrans s/n, 08193 Bellaterra, Spain \and
\label{inst:24} Institut d'Estudis Espacials de Catalunya (IEEC), 08034 Barcelona, Spain \and
\label{inst:25} Admatis, 5. Kandó Kálmán Street, 3534 Miskolc, Hungary \and
\label{inst:26} Depto. de Astrofisica, Centro de Astrobiologia (CSIC-INTA), ESAC campus, 28692 Villanueva de la Cañada (Madrid), Spain \and
\label{inst:27} Instituto de Astrofisica e Ciencias do Espaco, Universidade do Porto, CAUP, Rua das Estrelas, 4150-762 Porto, Portugal \and
\label{inst:28} Departamento de Fisica e Astronomia, Faculdade de Ciencias, Universidade do Porto, Rua do Campo Alegre, 4169-007 Porto, Portugal \and
\label{inst:29} Université Grenoble Alpes, CNRS, IPAG, 38000 Grenoble, France \and
\label{inst:30} Institute of Planetary Research, German Aerospace Center (DLR), Rutherfordstrasse 2, 12489 Berlin, Germany \and
\label{inst:31} Centre for Mathematical Sciences, Lund University, Box 118, 221 00 Lund, Sweden \and
\label{inst:32} Astrobiology Research Unit, Université de Liège, Allée du 6 Août 19C, B-4000 Liège, Belgium \and
\label{inst:33} Space sciences, Technologies and Astrophysics Research (STAR) Institute, Université de Liège, Allée du 6 Août 19C, 4000 Liège, Belgium \and
\label{inst:34} Dipartimento di Fisica, Universita degli Studi di Torino, via Pietro Giuria 1, I-10125, Torino, Italy \and
\label{inst:35} University of Vienna, Department of Astrophysics, Türkenschanzstrasse 17, 1180 Vienna, Austria \and
\label{inst:36} Science and Operations Department - Science Division (SCI-SC), Directorate of Science, European Space Agency (ESA), European Space Research and Technology Centre (ESTEC),
Keplerlaan 1, 2201-AZ Noordwijk, The Netherlands \and
\label{inst:37} Konkoly Observatory, Research Centre for Astronomy and Earth Sciences, 1121 Budapest, Konkoly Thege Mikl\'os \'ut 15-17, Hungary \and
\label{inst:38} ELTE E\"otv\"os Lor\'and University, Institute of Physics, P\'azm\'any P\'eter s\'et\'any 1/A, 1117 Budapest, Hungary \and
\label{inst:39} IMCCE, UMR8028 CNRS, Observatoire de Paris, PSL Univ., Sorbonne Univ., 77 av. Denfert-Rochereau, 75014 Paris, France \and
\label{inst:40} Institut d'astrophysique de Paris, UMR7095 CNRS, Université Pierre \& Marie Curie, 98bis blvd. Arago, 75014 Paris, France \and
\label{inst:41} Department of Astrophysics, University of Vienna, Tuerkenschanzstrasse 17, 1180 Vienna, Austria \and
\label{inst:42} Institute of Optical Sensor Systems, German Aerospace Center (DLR), Rutherfordstrasse 2, 12489 Berlin, Germany \and
\label{inst:43} Dipartimento di Fisica e Astronomia "Galileo Galilei", Universita degli Studi di Padova, Vicolo dell'Osservatorio 3, 35122 Padova, Italy \and
\label{inst:44} ETH Zurich, Department of Physics, Wolfgang-Pauli-Strasse 2, CH-8093 Zurich, Switzerland \and
\label{inst:45} Cavendish Laboratory, JJ Thomson Avenue, Cambridge CB3 0HE, UK \and
\label{inst:46} Zentrum f{\"u}r Astronomie und Astrophysik, Technische Universit{\"a}t Berlin, Hardenbergstr. 36, D-10623 Berlin, Germany \and
\label{inst:47} Institut f{\"u}r Geologische Wissenschaften, Freie Universit{\"a}t Berlin, 12249 Berlin, Germany \and
\label{inst:48} ELTE Eötvös Loránd University, Gothard Astrophysical Observatory, 9700 Szombathely, Szent Imre h. u. 112, Hungary \and
\label{inst:49} MTA-ELTE Exoplanet Research Group, 9700 Szombathely, Szent Imre h. u. 112, Hungary \and
\label{inst:50} Institute of Astronomy, University of Cambridge, Madingley Road, Cambridge, CB3 0HA, United Kingdom
}


 
  \abstract{Observations of the phase curves and secondary eclipses of extrasolar planets provide a window on the composition and thermal structure of the planetary atmospheres. For example, the photometric observations of secondary eclipses lead to the measurement of the planetary geometric albedo $A_g$, which is an indicator of the presence of clouds in the atmosphere.}
  {In this work we aim to measure the $A_g$ in the optical domain of \wftb, a moderately irradiated giant planet with an equilibrium temperature of $\sim$1400~K.}
  {To this purpose, we analyze the secondary eclipse light curves collected by \cheops, together with \tess\ observations of the system and the publicly available photometry obtained with \uvis. We also analyze the archival infrared observations of the eclipses and retrieve the thermal emission spectrum of the planet. By extrapolating the thermal spectrum to the optical bands, we correct the optical eclipses for thermal emission and derive the optical $A_g$.}
  {The fit of the optical data leads to a marginal detection of the phase curve signal, characterized by an amplitude of $160\pm60$~ppm and 80$^{+60}_{-50}$~ppm in the \cheops\ and \tess\ passband respectively, with an eastward phase shift of $\sim50^\circ$ (1.5$\sigma$ detection). The analysis of the infrared data suggests a non-inverted thermal profile and solar-like metallicity. The combination of optical and infrared analysis allows us to derive an upper limit for the optical albedo of \ag$<0.087$ with a confidence of 99.9\%.
  }
  {Our analysis of the atmosphere of \wftb\ places this planet in the sample of irradiated hot Jupiters, with monotonic temperature-pressure profile and no indication of condensation of reflective clouds on the planetary dayside.}


\titlerunning{Phase curve and \ag\ of \wftb}
\authorrunning{Scandariato et al.}

   \maketitle
%

\section{Introduction}
Phase curve and secondary eclipse observations are among the main avenues to characterize extra-solar planet atmospheres, as they provide a window on their composition and thermal structure. Thanks to these observations, we can probe planetary brightness temperatures at different wavelengths, and constrain molecular abundances at different pressure levels \citep[e.g.][]{Alonso2009,Kreidberg2014,Stevenson2014,Foote2022}. Additionally, secondary eclipse depth measurements are sensitive to temperature gradients with atmospheric pressure (i.e. altitude), to the presence of molecules which absorb ultraviolet-to-visible stellar radiation, as well as to temperature inversion that manifests through emission lines \citep[e.g.][and references therein]{Mansfield2018,Baxter2020,Garhart2020}.

One of the main parameters obtained by eclipse depth measurements is the planetary albedo, which describes the body's surface or atmosphere reflectivity \citep{Seager2010}. This latter, in turn, is an indicator of the presence of reflective clouds, currently a poorly constrained component in our understanding of exoplanet atmospheres and an important source of limitation in our measurement of molecular mixing ratios \citep[e.g.][and references therein]{Sing2016,Pinhas2019}.

Unlike most ultra-short period (P $\leq$1\,day) Hot Jupiters, \wftb\ is only moderately irradiated to an equilibrium temperature of $\sim 1400$~K. It thus resides in a temperature range where cloud condensation can occur on the planetary dayside, setting the object apart from ultra-hot Jupiters with extremely hot and therefore likely clear daysides \citep[e.g.][]{Helling2021}. In an effort to understand its atmospheric physical-chemical environment, this planet has been targeted by Spitzer \citep[e.g.][]{Stevenson2017}, HST \citep[e.g.][]{Stevenson2014,Fraine2021} and ground-based telescopes \citep{Weaver2019}. In particular, 3D Global Circulation Models with $5\times$ solar metallicity without clouds match the HST WFC3/G141 infrared observations of the planetary dayside \citep{Kataria2015}. However, later 3D atmospheric models including cloud condensation processes by \citet{Helling2020} and \citet{Venot2020} predict the presence of several species of clouds (most importantly silicate and metal-oxide components) on the dayside of \wftb. If present at observable altitudes, their reflectance would contribute to a significantly enhanced geometric albedo \citep[see e.g.][]{Marley1999,Sudarsky2000,Parmentier2016}.

The CHaracterizing ExOplanet Satellite (\cheops) is a 30-cm photometric space telescope dedicated to the characterization of known transiting exoplanets through precise optical-light photometry \citep{Benz2021}. One of the main themes of its Science Program is the characterization of exoplanet atmospheres \citep{Benz2021}. This consists of observations of secondary eclipses of hot Jupiters \citep[e.g.][]{Lendl2020, Hooton2021} across a wide range of temperatures and planetary surface gravities, full or partial phase curves of the most compelling targets \citep[e.g.][]{Deline2022}, and detailed observations of ultra-hot super-Earths \citep[e.g.][]{Morris2021}. 

In this paper we study the phase curve and measure the geometric albedo of \wftb. To this purpose, we analyze \cheops\ observations of the optical secondary eclipses of \wftb, jointly with data obtained by the Transiting Exoplanet Survey Satellite (\tess\, \citealt{Ricker2014}) and with a revision of the publicly available \uvis\ observations of the planetary eclipse. We also homogeneously analyze the archival near-IR eclipse observations of the system to estimate the thermal emission in the optical bands. This analysis is necessary to disentangle the reflective contribution to the observed eclipse depth from the thermal emission component. All the datasets are presented in Sect.~\ref{sec:observations} together with the data reduction. In Sect.~\ref{sec:spectroscopy} we provide the spectroscopic characterization of the host star. We describe the modeling of the light curves (LCs) in Sect.~\ref{sec:modeling}, while in Sect.~\ref{sec:discussion} we discuss the results of our analysis. Finally in Sect.~\ref{sec:conclusion} we draw our conclusions.

\section{Observations and data reduction}\label{sec:observations}

\subsection{\textit{CHEOPS} observations}\label{sec:CHEOPSobs}

The \cheops\ satellite is dedicated to the observations of exoplanetary systems. It is equipped with an f/8 Ritchey-Chr\'etien on-axis telescope having an effective diameter of $\sim$30 cm, which projects the field of view on a single frame-transfer back-side illuminated charge-coupled device (CCD) detector \citep{Benz2021}.

\cheops\ observed the \wft\ system during eleven secondary eclipses of the planet \wftb\ in order to measure the eclipse depth, a measurable directly linked to the brightness of the planet. To this purpose, each visit is $\sim$3.5--6~hr long, sampling a time interval around the eclipse that is $\sim$3--4 times longer than the expected transit/eclipse duration of $\sim$1.2~hr \citep{Hellier2011,Esposito2017}. 

The observations were carried out as part of the Guaranteed Time Observation (GTO) program, and are summarized in Table~\ref{tab:obs}. The first two of them were observed in April 2020 and are characterized by large interruptions due to Earth occultations and crossings of the South Atlantic Anomaly (SAA). This explains their low efficiency (<70\%), that is the fraction of allocated time which is spent on source. Moreover, the coverage of the eclipses turned out to be extremely low, particularly for the second visit. For these reasons we decided to exclude them from our analysis. The remaining nine visits were performed between February and April 2021, with a better efficiency (>70\%).

\begin{table*}
\caption{Logbook of the \cheops\ observations of \wft. The filekey is the unique identifier associated with each dataset processed by the \cheops\ Data Reduction Pipeline (DRP).}             
\label{tab:obs}      
\centering          
\begin{tabular}{c c c c c c c}     
\hline\hline       
Filekey & Visit & Start time & End time & Exposure time\tablefootmark{a} & N. frames & Efficiency\\
& ID & \multicolumn{2}{c}{(UTC)} & (s) &  &  (\%)\\
\hline                    
PR100016\_TG007801\_V0200  & - & 2020-04-24 04:04:30  &  2020-04-24 07:37:35  &  60.0  &  137  &  64.3 \\
PR100016\_TG007802\_V0200  & - & 2020-04-27 10:40:30  &  2020-04-27 14:13:36  &  60.0  &  134  &  62.9 \\
PR100016\_TG012201\_V0200  & V1 & 2021-02-24 01:10:11  &  2021-02-24 04:43:17  &  60.0  &  170  &  79.8 \\
PR100016\_TG012202\_V0200  & V2 & 2021-02-24 21:08:11  &  2021-02-25 00:41:17  &  60.0  &  154  &  72.3 \\
PR100016\_TG012203\_V0200  & V3 & 2021-03-02 13:19:11  &  2021-03-02 16:52:17  &  60.0  &  193  &  90.6 \\
PR100016\_TG012204\_V0200  & V4 & 2021-03-04 23:34:11  &  2021-03-05 03:07:17  &  60.0  &  210  &  98.5 \\
PR100016\_TG012701\_V0200  & V5 &  2021-03-09 00:16:11  &  2021-03-09 06:56:22  &  60.0  &  395  &  98.7 \\
PR100016\_TG012702\_V0200  & V6 & 2021-03-11 10:19:10  &  2021-03-11 16:36:20  &  60.0  &  363  &  96.2 \\
PR100016\_TG012703\_V0200  & V7 & 2021-03-22 20:08:11  &  2021-03-23 03:33:23  &  60.0  &  408  &  91.6 \\
PR100016\_TG012704\_V0200  & V8 & 2021-03-25 06:40:11  &  2021-03-25 13:22:22  &  60.0  &  319  &  79.3 \\
PR100016\_TG012705\_V0200  & V9 & 2021-04-10 12:33:10  &  2021-04-10 18:36:21  &  60.0  &  266  &  73.2 \\

\hline                  
\end{tabular}
\tablefoot{
        \tablefoottext{a}{{The cadence is 0.025 s longer than the exposure time due to the transfer time needed from the image section to the storage section of the CCD.}}
}
\end{table*}

The data are reduced using version 13 of the \cheops\ Data Reduction Pipeline \citep[DRP,][]{Hoyer2020}. This pipeline performs the standard calibration steps (bias, gain, non-linearity, dark current and flat fielding), corrects for environmental effects (cosmic rays, smearing trails from nearby stars, and background), and then extracts aperture photometry using three fixed aperture sizes, along with a fourth aperture which is automatically selected by the algorithm to optimize the photometric extraction.

The DRP also estimates the contaminating flux inside the aperture from nearby sources by simulating the \cheops\ field of view based on the GAIA DR2 star catalog \citep{Gaia2021} and a template of the extended \cheops\ PSF. The computation of the contaminant flux is performed assuming that target and background stars have constant flux. This is not the case of \wft, as it is an active star which exhibits clear signatures of activity during our observational campaigns. For this reason we adopted a different approach. We extracted the LC of each visit using the PSF photometry package PIPE\footnote{\url{https://github.com/alphapsa/PIPE}} \citep{Brandeker2022,Morris2021,Szabo2021}, specifically designed for \cheops. {Using the star catalogue produced by the DRP, PIPE models and removes background stars from the subimages before extracting photometry from the target by fitting a PSF. Since the PSF fitting is weighted by the signal and noise of each pixel, this extraction is much less sensitive than aperture photometry to background star contamination.} In the specific case of \wft, PIPE reduced by a factor of $\sim2$ the correlated noise due to the instrumental systematics compared to the optimal aperture selected by the DRP. Tests on the extracted light curves indicated that the {improvement is indeed mainly due to PIPE being less sensitive to contaminating flux}. 

The frames acquired close to the Earth occultation display anomalously high value of the background flux, due to stray-light from Earth itself. The corresponding background-subtracted photometry is noisier than the rest of the LC. {Moreover, the light curve of the background flux changes, both in average value and in shape, from visit to visit, due to the changing angular separation between the target and Earth. We thus adopted a dynamic approach to clean the LC of each visit by clipping all the frames with high background:} we empirically set a background threshold which corresponds to twice the minimum background measurement returned by the pipeline. If this threshold rejected more than 10\% of the data, then the rejection threshold was raised to the 90\% quantile of the background measurements. This selection criterion thus removed at most 10\% of the data, and proved to be effective in the rejection of noisy data. Finally, we rejected the remaining outliers in the LCs by smoothing the data with a Savitzky-Golay filter, computing the residuals with respect to the smoothed LC and sigma-clipping the data at the 5$\rm\sigma$level. This last rejection criterion excluded a handful of data points in each LC.

Finally, we normalized the LCs by the median value of the photometry. The final \cheops\ LCs are shown in Figs.~\ref{fig:cheopsVisitFits13}--\ref{fig:cheopsVisitFits79} and publicly available at CDS.

\subsection{\textit{TESS} observations}\label{sec:TESSobs}

\tess\ \citep{Ricker2014} observed the \wft\ system in sector 9 (from February,28 to March,26 2019, orbits 25 and 26) and in sector 35 (from February,9 to March,06 2021, orbits 77 and 78). Using the package \texttt{lightkurve} \citep{lightkurve2018}, we retrieved the 2 min cadence Simple Aperture Photometry (SAP) and Pre-search Data Conditioning Single Aperture Photometry (PDCSAP), which is corrected for instrumental systematics and for contamination from some nearby stars \citep{Smith2012,Stumpe2012,Stumpe2014}. We rejected all the data that are flagged by the pipeline. During sector 35 the telescope experienced a technical issue with the thermal stability and the pointing, which introduced trends and additional noise in the LC\footnote{\url{https://archive.stsci.edu/missions/tess/doc/tess_drn/tess_sector_35_drn51_v02.pdf}}. To avoid these data, we also clipped out the photometry collected between BJD 2,459,255 and 2,459,256 and between BJD 2,459,266.2 and 2,459,272 (at the beginning and at the end of orbit 77).

For each orbit, we performed a preliminary normalization of the photometry by using the median value. The normalization will be refined in Sect.~\ref{sec:TESSfit}.

\subsection{\textit{UVIS} observations}\label{sec:UVISobs}

\citet{Fraine2021} dedicated four HST orbits to observe a secondary eclipse of the \wftb\ system on 2019 July 3 using WFC3/UVIS with the F350LP filter in scanning mode. We retrieved the reduced photometry published by the authors and we kept the forward and reverse scan separated, {so to detrend the two series independently}. We did not apply any extra manipulation to the data.

\subsection{Spectroscopic observations}\label{sec:spectra}

{We searched public archives for high resolution spectroscopic data for \wft. We ended up using three exposures observed with the Ultraviolet and Visual Echelle Spectrograph (UVES) which are public in the ESO archive and were taken under the program 090.C-0146. The instrumental configuration was Red Arm $@$580nm with a slit of 0.3", which provided a coverage 472--683~nm with a resolution of $R=107200$. The exposure time was 691~s for each spectrum. The individual SNR is $\sim$40, while the SNR of the combined spectrum is $\sim$70.}

\section{Characterization of the host star} \label{sec:spectroscopy}

To determine the stellar parameters, we used the astroARIADNE package\footnote{\url{https://github.com/jvines/astroARIADNE}} \citep{Acton+2020}, a recent code that automatically retrieves photometric data (when existing) from catalogs such as ALL-WISE, APASS, Pan-STARRS1, SDSS, 2MASS and Tycho-2. With distances from Gaia, as well as available maps of dust distribution, the photometric data is fitted to different stellar models using Bayesian Model Averaging in order to find the best model of the stellar parameters. With this method we obtained \teff=$4240^{+30}_{-20}$~K, \logg=4.71$\pm$0.09~cgs, [Fe/H]=$0.15^{+0.10}_{-0.04}$~dex, d=87.2$\pm$0.5~pc and $A_{\rm V}=0.05^{+0.04}_{-0.02}$, where the errors are internal to the statistical calculations.

\begin{figure}
    \centering
    \includegraphics[width=\linewidth]{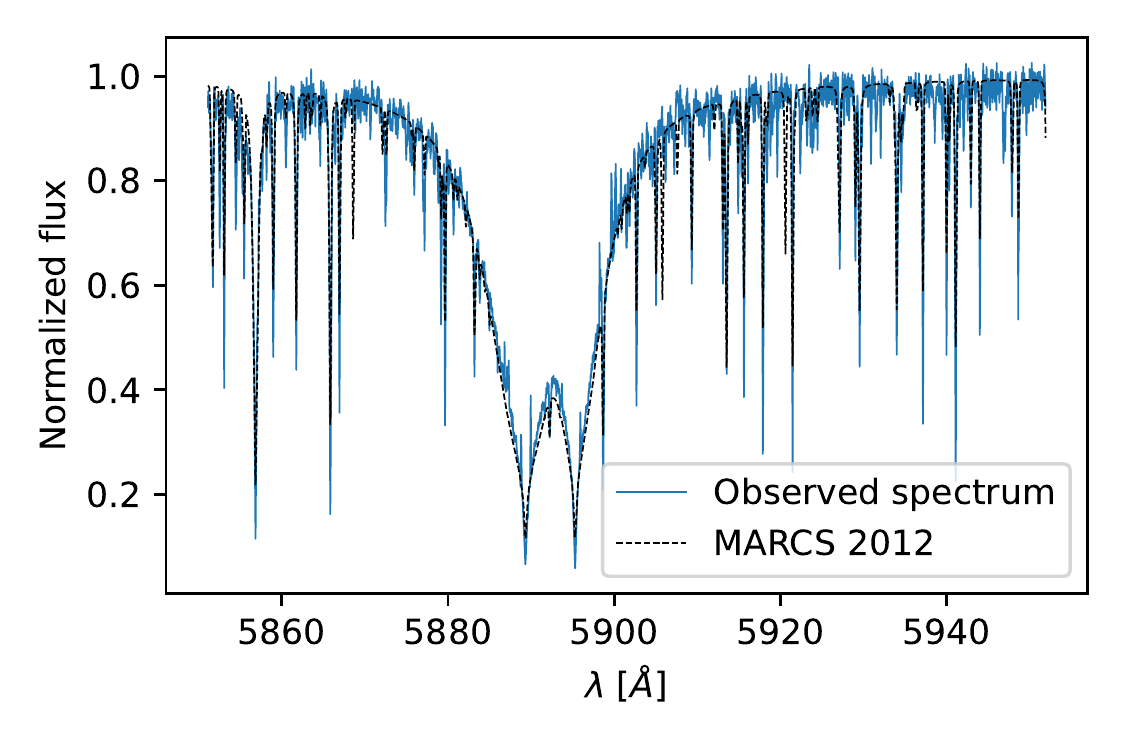}\\
    \caption{Comparison between the spectrum used for the stellar characterization around the \ion{Na}{i} D$_{\rm 1,2}$ doublet and its bestfit model.}\label{fig:smeSpectrum}
\end{figure}

We also used the IDL package Spectroscopy Made Easy (SME) to synthesize models of the observed spectrum. SME tests several atmospheric models \citep{Valenti1996, Piskunov2017}, choosing eventually the one best agreeing with the observations. The code utilizes atomic and molecular line lists from VALD \citep{Piskunov1995}. In the case of \wft\ we found that the model best agreeing with the observed spectrum are the MARCS 2012 models \citep{Gustafsson2008} (Fig.~\ref{fig:smeSpectrum}). While keeping the turbulent velocities \vmac\ and \vmic\ fixed at values from \citet{gray08}, we determined the \vsini=3$\pm$1 km s$^{-1}$, while the other derived spectroscopic parameters have larger uncertainties than the previously determined ones, which are thus preferred.

To compute the radius of \wft, we used a Markov-Chain Monte Carlo (MCMC) modified infrared flux method (IRFM; \citealt{Blackwell1977,Schanche2020}). By building spectral energy distributions (SEDs) from stellar atmospheric models and stellar parameters derived via our spectral analysis, we compared the determined synthetic fluxes to observed broadband photometry to calculate the apparent bolometric flux, and hence the stellar angular diameter and effective temperature. For \wft, we retrieved data taken from the most recent data releases for the following bandpasses; {\it Gaia} G, G$_{\rm BP}$, and G$_{\rm RP}$, 2MASS J, H, and K, and {\it WISE} W1 and W2 \citep{Skrutskie2006,Wright2010,Gaia2021} and used stellar atmospheric models from the \textsc{atlas} Catalogues \citep{Castelli2003}. We converted our stellar angular diameter to the stellar radius using the offset-corrected {\it Gaia} EDR3 parallax \citep{Lindegren2021} and obtained $R_\star=0.698\pm0.014\, R_{\odot}$.

{To compute the stellar mass $M_\star$ and age $t_\star$ we used two different stellar evolutionary models adopting the stellar effective temperature $T_{\mathrm{eff}}$, metallicity [Fe/H], and radius $R_\star$ as the basic input set. In detail, we applied the isochrone placement technique \citep{Bonfanti2015,Bonfanti2016} to pre-computed grids of PARSEC v1.2S\footnote{\textit{PA}dova and T\textit{R}ieste \textit{S}tellar \textit{E}volutionary \textit{C}ode: \url{http://stev.oapd.inaf.it/cgi-bin/cmd}} \citep{Marigo2017} isochrones and tracks to retrieve a first pair of mass and age estimates $(M_{\star,1}; t_{\star,1})$. During interpolation, the basic input set was complemented with our internal estimate of \vsini\ and with the stellar rotation period from \citet{Bonomo2017} in order to improve the fitting process convergence as described in \citet{Bonfanti2016}. We obtained $M_{\star,1}=0.701\pm0.030\,M_{\odot}$ and $t_{\star,1}=6.0\pm3.9$ Gyr.

We determined a second pair of mass and age estimates $(M_{\star,2}; t_{\star,2})$ using the CLES \citep[Code Liègeois d'Évolution Stellaire,][]{Scuflaire2008} code, which directly fits the basic input set within the Li\`ege evolutionary models to retrieve the best-fit outcomes according to the Levenberg-Marquadt minimization scheme \citep{Salmon2021}. This on-the-fly computation yielded to $M_{\star,2}=0.728\pm0.042\,M_{\odot}$ and $t_{\star,2}=10.1\pm8.0$ Gyr.

As thoroughly described in \citet{Bonfanti2021}, we finally merged the two respective pairs of age and mass values after carefully checking their mutual consistency through a $\chi^2$-based criterion and we obtained $M_\star=0.712_{-0.036}^{+0.041}\,M_{\odot}$ and $t_\star=7.4_{-5.3}^{+6.4}$ Gyr.

The characterization of \wft\ is summarized in Table~\ref{tab:parameters}. The estimated parameters are in general agreement within uncertainties with previous characterizations \citep{Bonomo2017,Esposito2017,Stassun2019}.

}

\begin{table*}
\caption{Stellar and system parameters.}             
\label{tab:parameters}      
\centering          
\begin{tabular}{l c c c r}     
\hline\hline       
Parameter & Symbol & Units & Value & Ref.\\
\hline
Spectral Type & & & K7V & \citet{Hellier2011}\\
Effective temperature & $\rm T_{eff}$ & K & $4240^{+30}_{-20}$ & this work\\
Surface gravity & $\log g$ & $\log{\mathrm{cm/s^2}}$ & $4.71\pm0.09$ & this work\\
Metallicity & [Fe/H] & --- & $0.15^{+0.10}_{-0.04}$ & this work\\
Distance & $d$ & pc & $87.2\pm0.5$ & this work\\
Interstellar extinction & $A_{\rm V}$ & --- & 0.05$^{+0.04}_{-0.02}$ & this work\\
Projected rotational velocity & \vsini & km/s & $3\pm1$ & this work\\
Stellar rotation period & $P_{\mathrm{rot}}$ & d & $15.6\pm0.4$ & \citet{Hellier2011} \\
Stellar radius & $\rm R_\star$ & $\rm R_\sun$ & $0.698\pm0.014$ & this work\\
Stellar mass & $\rm M_\star$ & $\rm M_\sun$ & $0.712_{-0.036}^{+0.041}$ & this work\\
Stellar age & $t_\star$ & Gyr & $7.4_{-5.3}^{+6.4}$ & this work \\
Radial velocity semi-amplitude & $\rm K_{RV}$ & m/s & $551.5\pm4.7$ & \citet{Bonomo2017}\\
\hline                  
\end{tabular}
\end{table*}

\section{Light curve fitting}\label{sec:modeling}

To fit the LCs we adopted the same model $F(\phi)$ as \citet{Esteves2013} (see also references therein), that is the sum of the transit model $F_{\rm tr}(\phi)$, the eclipse model $F_{\rm ecl}(\phi)$, the planet's phase curve $F_{p}(\phi)$, the Doppler boosting of the stellar flux $F_{\rm d}(\phi)$ and the stellar ellipsoidal variations $F_{\rm e}(\phi)$:
\begin{equation}
F(\phi)=F_{\rm tr}(\phi)+F_{\rm ecl}(\phi)+F_{\rm p}(\phi)+F_{\rm d}(\phi)+F_{\rm e}(\phi),\label{eq:planet}
\end{equation}
where $\phi$ is the planet's orbital phase. We do not go into the mathematical details of these terms \citep[we refer the reader to][and references therein]{Esteves2013, Singh2022}, but we give only a brief description and the list of parameters that define each of them.

To model the transits in the LC, we used the quadratic limb darkening (LD) law indicated by \citet{Mandel2002} with the reparametrization of the LD coefficients suggested by \citet{Kipping2013}. We also assumed that the orbit of \wftb\ is circular following \citet{Hellier2011,Gillon2012,Bonomo2017}. Under this hypothesis, the model $F_{\rm tr}(\phi)$ thus depends on the time of transit $\rm T_0$, the orbital frequency $\nu_{\rm orb}$ (the inverse of the orbital period $P_{\rm orb}$), the stellar density $\rho_\star$, the ratio between the planetary and stellar radii $R_{\rm p}/R_\star$, the impact parameter $b$ and the reparametrized quadratic LD coefficients $q_1$ and $q_2$. Also the eclipse model $F_{\rm ecl}(\phi)$ is formalized following \citet{Mandel2002}, but assuming that the planetary dayside hiding behind the stellar disc is uniformly bright.

{The phase curve $F_{\rm p}(\phi)$ quantifies the amount of light that the planet emits towards the observer, either by reflection of the incoming stellar light or by thermal emission. In the hypothesis that the planet's surface follows Lambert's reflection law, the phase curve depends on the impact parameter $b$ and the amplitude of the phase curve $A_{\rm p}$ relative to the stellar flux. We also allowed the phase curve to peak at an offset $\Delta\phi$ from the eclipse center ($\phi$=0.5) in order to model any longitudinal asymmetry of the planet's brightness.

In principle, $F_{\rm p}(\phi)$ should also account for the planetary thermal emission, both from the dayside and the nightside. For the planetary dayside, the shape of the thermal emission phase curves depend on how light is emitted by the planetary surface. Regardless of the level of anisotropy in the thermal emission and scattering, the thermal and reflected components have in common that they peak at or near secondary eclipse and smoothly decrease as the planet is closer to transit. For the thermal component, we estimated the dayside temperature of \wftb\ of $\sim1600~K$ using the formalism in \citet{CowanAgol2011}, the orbital parameters provided by \citet{Esposito2017}, the Bond albedo $A_{\rm B}$=0.19 and the recirculation efficiency $\epsilon$=0 estimated by \citet{Stevenson2017}. If we approximate the planet and the star as black bodies, this dayside temperature translates into a planet-to-star contrast $A_{\rm day}=38$~ppm in the \tess\ band and 19~ppm in the \uvis\ and \cheops\ bands}. If we also assume that the dayside of \wftb\ reflects as a Lambertian surface, then the geometric albedo is expected to be \ag$\sim$0.13, which translates into a planet-to-star flux ratio of $A_{\rm refl}\sim$140~ppm, that is 7 times larger than the contrast of the planetary thermal radiation. The thermal phase curve of the planetary dayside thus provides a small contribution to the Lambertian phase curve, and the quality of the data analyzed in this paper does not allow to appreciate such small deviations from the reflection-only scenario. Thus, to reduce the number of free parameters, we fixed $A_{\rm day}=0$ letting the Lambertian phase curve absorb the thermal emission signal. In Sect.~\ref{sec:geometricAlbedo} we will discuss our findings on thermal emission and reflection from the planetary dayside.

As for the nightside, it has already been reported in the literature \citep{Schwartz2015,Stevenson2017,Irwin2020} that the temperature is expected to be $\lesssim$1000~K, which corresponds to a negligible flux contrast of less than 1 ppm compared with the stellar flux. By consequence, in our data modeling we also excluded any nightside emission.

The Doppler boosting $F_{d}(\phi)$ is the modulation of the stellar flux due to the radial velocity of the star with respect to the observer. The observed flux increases when the star is moving towards the observer and decreases when it is moving away. This effect is thus phased with the orbital motion of the planet and depends on the radial velocity semi-amplitude $K_{\rm RV}$ and on the bandpass-integrated average spectral index $\alpha_d$ of the star.

The ellipsoidal variations are periodic modulations in the stellar light due to fluctuations in the shape of the stellar visible hemisphere, distorted by the tidal pull from the planet. $F_{e}(\phi)$ thus depends on the planet-to-star mass ratio $\mu$ (which is a function of $K_{\rm RV}$, $M_\star$ and the orbital parameters of the planet), the linear LD coefficient $u_{\rm LLD}$ and the gravity darkening coefficient $y_{\rm GD}$. As for the phase variations, the model also allows for an angular lag with respect to the planet's orbital phase through the extra parameter $\Theta_{\rm lag}$.

The complete model thus depends on a total of 15 parameters. To reduce the dimensionality of the problem, and to avoid degeneracies which may occur among the model parameters, we locked some of them by means of prior knowledge coming from previous studies of the system. We fixed both $F_{d}(\phi)$ and $F_{e}(\phi)$ in Eq.~\ref{eq:planet} using the estimates of $K_{\rm RV}$ and $M_\star$ reported in Table~\ref{tab:parameters}. For the other input parameters, we proceeded as follows. We used the throughput of \cheops\ and \tess\ publicly available at the SVO Filter Profile Service \citep{Rodrigo2012,Rodrigo2020} and computed the passband-dependent $u_{\rm LLD}$ using the \texttt{LDTk} package \footnote{\url{https://github.com/hpparvi/ldtk}}. We computed $y_{\rm GD}$ in the \tess\ passband by a trilinear interpolation of the table provided by \citet{Claret2017}, assuming the stellar parameters listed in Table~\ref{tab:parameters}. Similarly, we performed the same computation using the table provided by \citet{Claret2021} for the \cheops\ passband, using the gravity darkening exponent predicted by \citet{Claret2000}. For WFC3/UVIS we used similar tables that were provided in a private communication. We also fixed the spectral indices $\alpha_d$ (one for each instrument) by using the BT-Settl spectral model \citep{Allard2012} with $\rm T_{eff}=4200~K$, $\log g=4.5$, $\rm [Fe/H]=0.3$ and no alpha elements enhancement. Finally, assuming that the tidal axis is aligned, we fixed $\Theta_{\rm lag}=0$. The expected amplitudes for the reflection phase curve, the thermal emission from the planetary dayside and nightside, the ellipsoidal variations and the Doppler boosting are summarized in Table~\ref{tab:amplitudes}. 

\begin{table}
\begin{center}
\caption{Expected amplitudes in ppm for the components in the LC model in the \uvis, \cheops\ and \tess\ passbands.}\label{tab:amplitudes}
\begin{tabular}{l|rrr}
\hline
 & WFC3/UVIS & \cheops & \tess\\
\hline
Dayside thermal emission & 19 & 19 & 38 \\
Nightside thermal emission & 0 & 0 & 0 \\
Ellipsoidal variations & 48 & 48 & 43 \\
Doppler boosting & 8 & 8 & 8 \\
\hline
\end{tabular}

\end{center}
\end{table}


Another important point to take care of is the presence in the LCs of instrumental and/or stellar activity signals, which must be corrected in order to improve the quality of the LC modeling. The strategy to adopt depends on the characteristics of the datasets and is discussed separately for each instrument.

\subsection{TESS light curves}\label{sec:TESSfit}

\begin{figure}
    \centering
    \includegraphics[width=\linewidth]{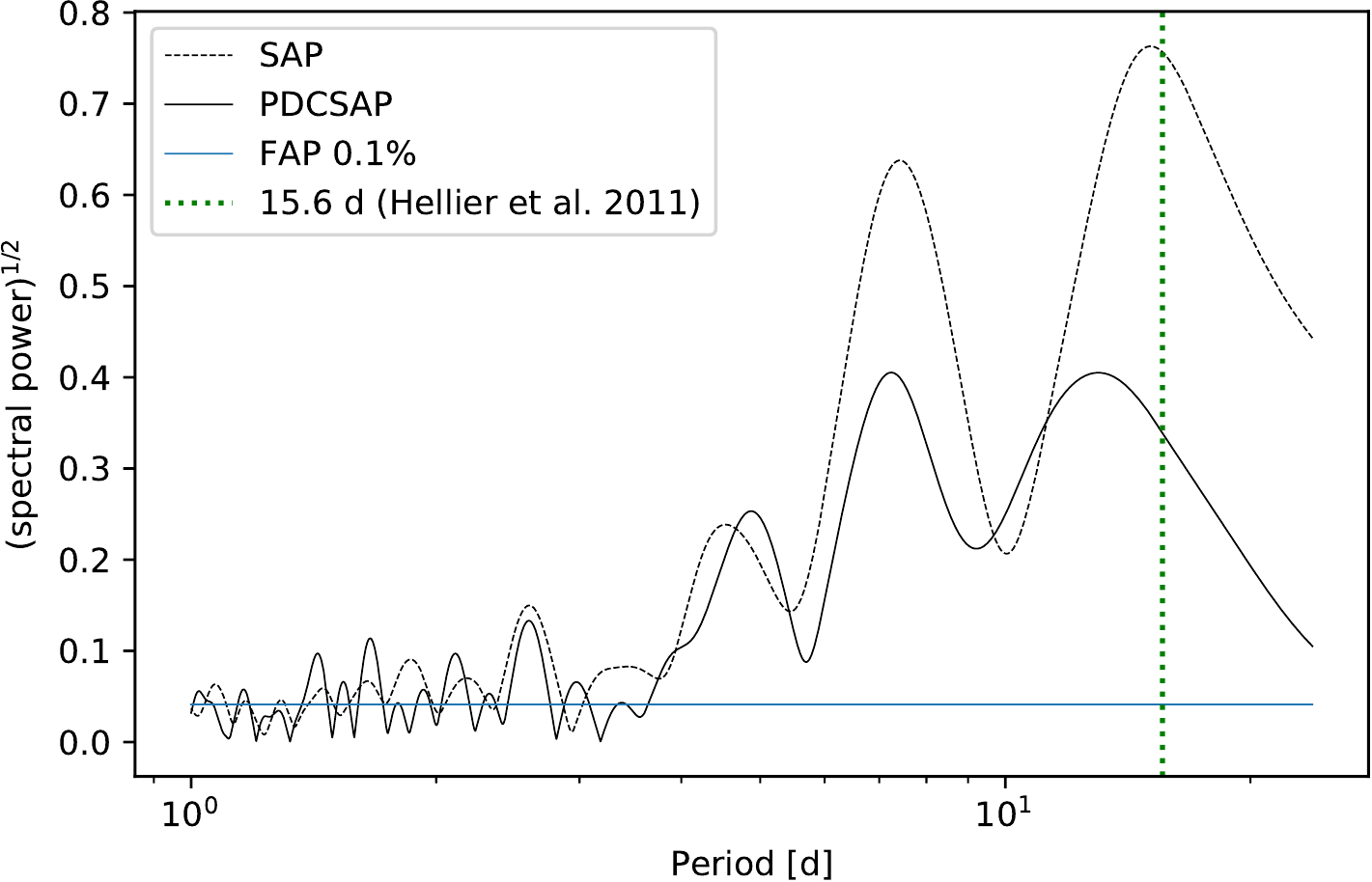}\\
    \includegraphics[width=\linewidth]{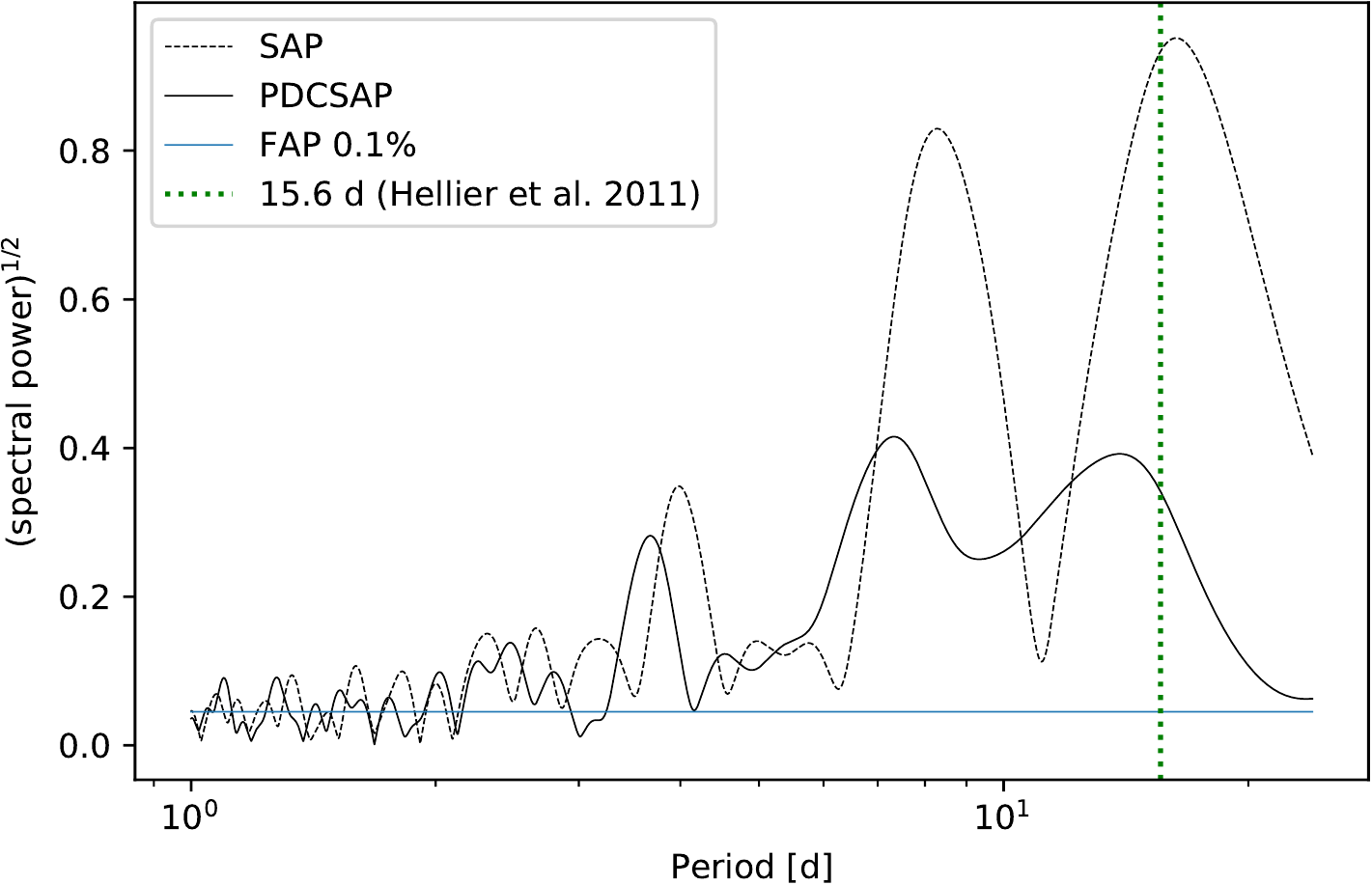}
    \caption{Periodogram of the \tess\ photometry (SAP and PDCSAP) of \wft, sectors 9 and 35 from top to bottom. In each panel, the vertical dashed line marks the 15.6~days rotation period proposed by \citet{Hellier2011}.}
    \label{fig:periodogram}
\end{figure}
\begin{figure*}
    \centering
    \includegraphics[width=.45\linewidth]{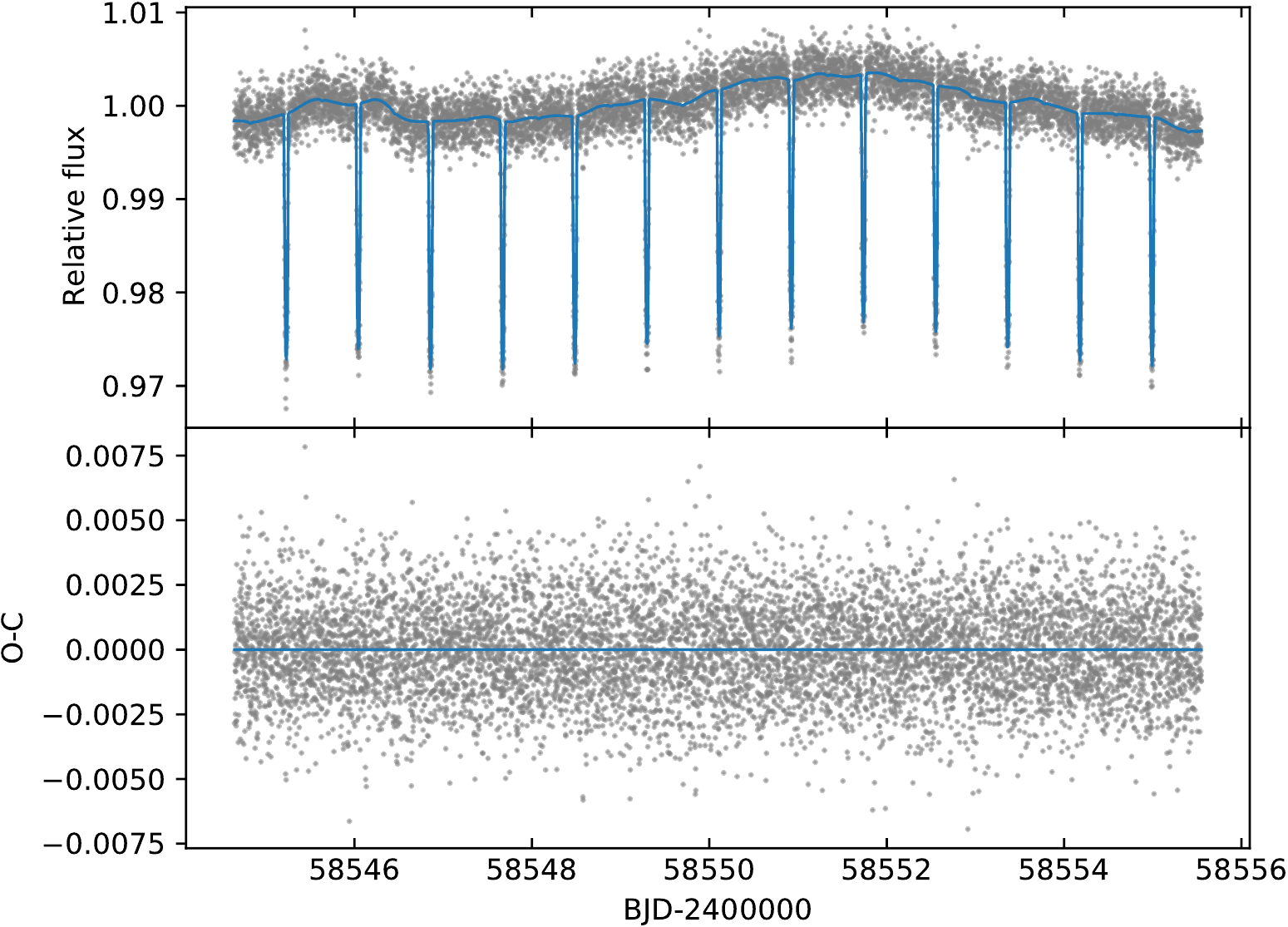}
    \includegraphics[width=.45\linewidth]{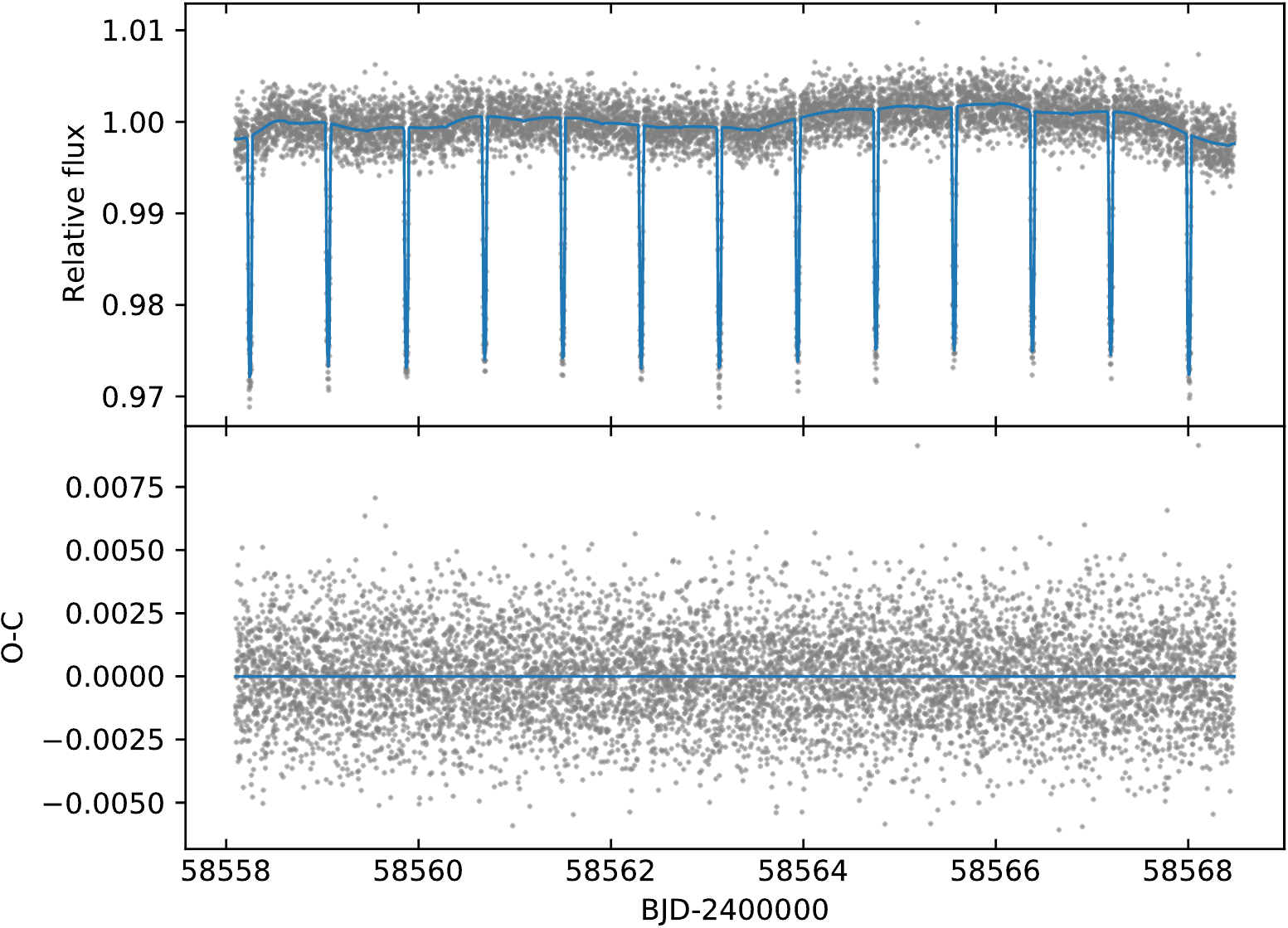}\\
    \includegraphics[width=.45\linewidth]{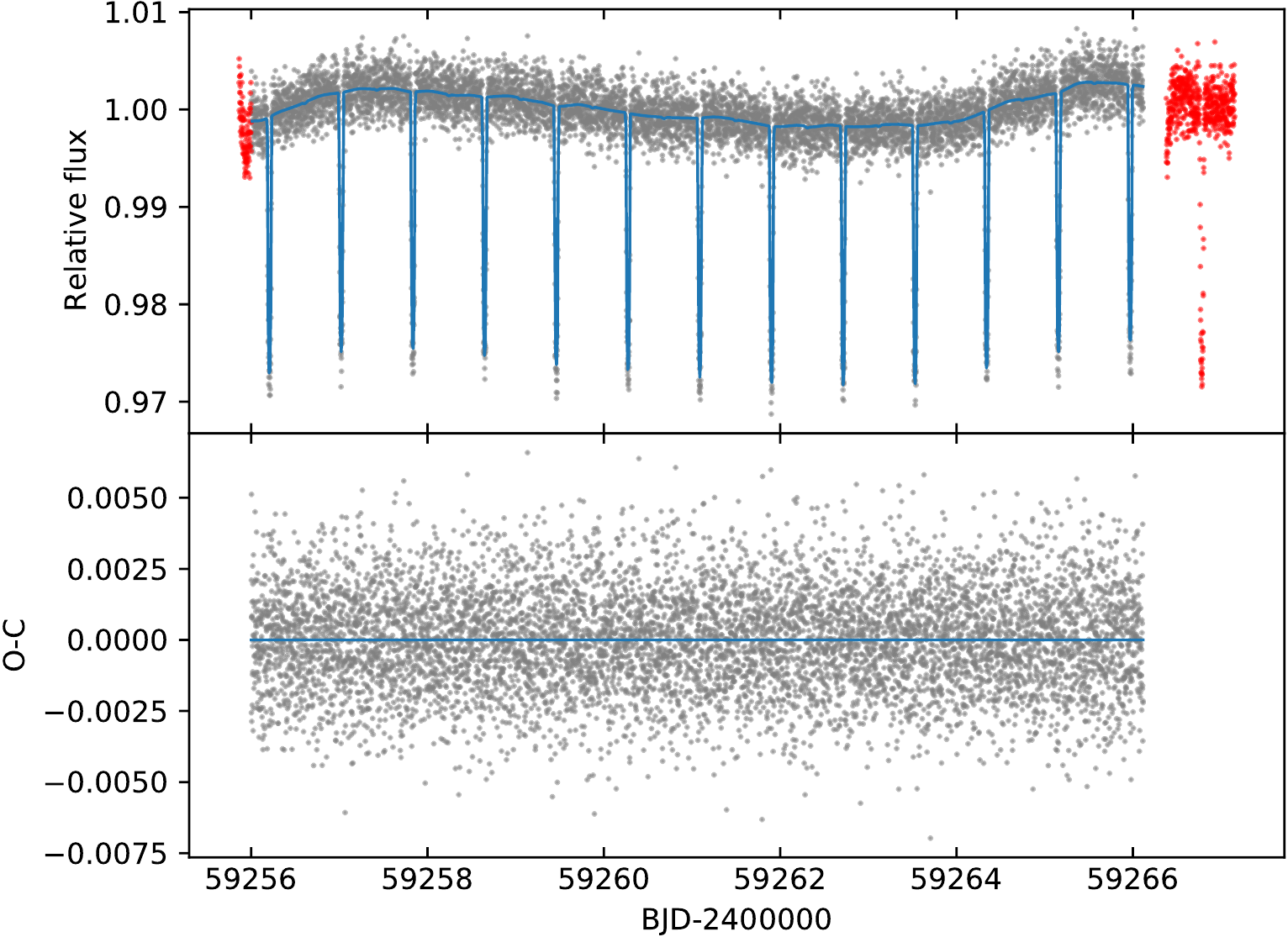}
    \includegraphics[width=.45\linewidth]{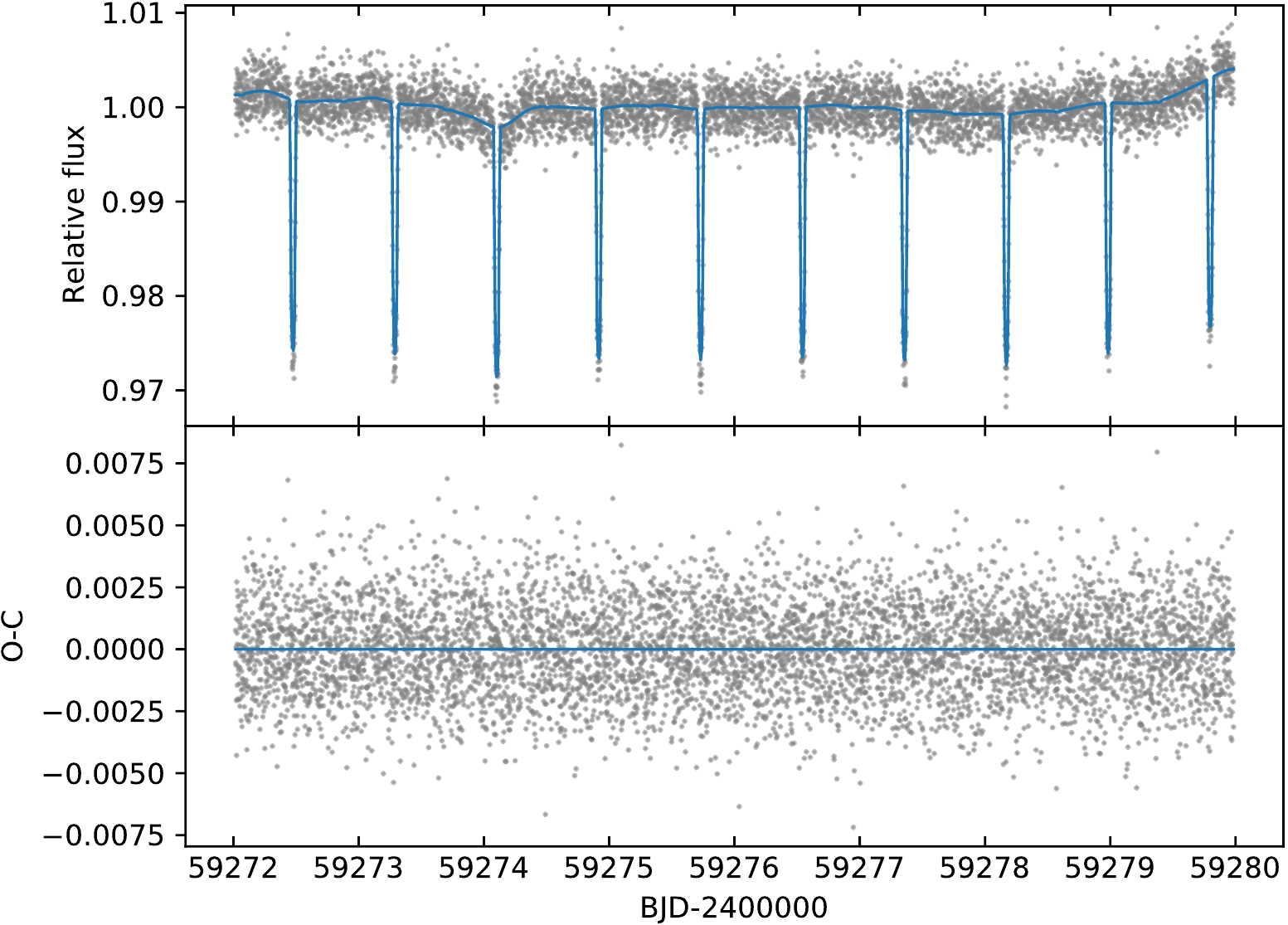}
    \caption{\tess\ LCs for orbits 25 (top left), 26 (top right), 77 (bottom left) and 78 (bottom right). For each panel, the top plot shows also the best fit model as a solid blue line, while the residuals are shown in the bottom plot. In the top plot of the bottom left panel we show in red the data rejected points. See Sect.~\ref{sec:TESSfit} for details.}\label{fig:tessOrbitFits}
\end{figure*}

The \tess\ PDCSAP photometry is shown in Fig.\ref{fig:tessOrbitFits}. It is clear that it is affected by red noise in the form of waves all along the LCs. We identified two possible origins for the trends. The first possibility is that they are artificially introduced by the detrending algorithm in the official reduction pipeline of \tess, as already claimed in previous works \citep[for example][]{Shporer2019,Wong2020,Daylan2021}. As a sanity check, since the correlated noise in the LCs seems to be periodic, for each sector we analyzed the Generalized Lomb Scargle Periodogram \citep{Zechmeister2009} of both the SAP and PDCSAP photometry, after removing the transits. We found that all the periodograms look alike (Fig.\ref{fig:periodogram}): they show a significant peak ({ FAP<0.1\%}) at $\sim12.5$~d and its first harmonic ($\sim6.3$~d, { FAP<0.1\%}) for both SAP and PDCSAP photometry. For sector 9 we also found a peak near 4~d, which corresponds to second harmonic of the 12.5~d period, while for sector 35 there is a peak near 3~d that is the fourth harmonic. We thus postulate that the detrending of the LCs is not the origin of the correlated noise in the data: if it is an artificial signal, then it must be already in the raw photometry, and not related to background or spacecraft jitter as that would have been cleaned up by the PDCSAP. Nonetheless, it is unlikely that the two light curves, obtained two years apart, are affected by instrumental issues leading to similar periodograms.

Therefore, the second possibility is that the red noise is of stellar origin. As a matter of fact, if active regions are present on the stellar surface, then the stellar rotation typically introduces some periodic-like signals in the LCs, which can be identified by means of periodogram analysis. The stellar rotation has already been detected by \citet{Hellier2011}, who claim a rotation period of $15.6\pm0.4$~d with a photometric amplitude of 0.006~mag. Nonetheless, this rotation period does not seem a good match for the peaks in the periodograms of the \tess\ data. One possibility is that, due to differential rotation, we recovered a rotation period slightly shorter than what is indicated by \citet{Hellier2011}. Moreover, if differential rotation and migration of active latitudes are in place, it is unlikely that we found the same rotation period in sectors which are two years apart.

Based on the discussion above, we are left with the ambiguity that the correlated noise in the LCs is of stellar origin or just instrumental effects not correctly removed, or maybe a combination of the two. Regardless of the source of the trends, we modeled them as a Gaussian Process \citep[GP,][]{Rasmussen2006,Gibson2012} with a Mat\'ern~3/2 kernel, that quantifies the covariance between two observations at times $t_{i_1}$ and $t_{i_2}$ as:
\begin{equation}
k(t_{i_1},t_{i_2})=h^2\left(1+\frac{\sqrt{3}|t_{i_1}-t_{i_2}|}{\rho}\right)\exp\left(-\frac{\sqrt{3}|t_{i_1}-t_{i_2}|}{\lambda}\right)+j_o^2\delta_{i_1,i_2},\label{eq:kernelTESS}
\end{equation}
where $h$ is the amplitude of the GP and $\lambda$ is its timescale\footnote{See also \url{https://celerite2.readthedocs.io/en/latest/api/python/}}. In order to take into account any additional white noise, either instrumental or astrophysical, not included in the uncertainties, we added the diagonal elements $j_o^2\delta_{i_1,i_2}$ where $o$ denotes the \tess\ orbit 25, 26, 77 and 78.

To trace any long-term activity signal, we also included in the model a linear function of time $t$ for each \tess\ orbit:
\begin{equation}
l_o(t)=c_{0,o}+c_{1,o}\cdot(t-t_{m,o}),\label{eq:lineartermTESS}
\end{equation}
where the subscript $o$ denotes again the orbit of interest, while $t_{m,o}$ is the mid-time { ($(t_{max}+t_{min})/2$)} of the corresponding orbit $o$. With this representation, it is straightforward that the constant term refines the normalization of the LC, while the linear term traces any slope in the data.

We fit the data by maximizing in the parameter space the log-likelihood function given by:
\begin{equation}
\ln L=-\frac{n}{2} \ln(2\pi)-\frac{1}{2}\ln(\det{\bf K})-\frac{1}{2}{\bf \overline{r}^T}\cdot{\bf K}\cdot{\bf \overline{r}},\label{eq:logl}
\end{equation}
where $\overline{\bf r}$ is the array of length $n$ containing the residuals with respect to the model, while $\bf K$ is the covariance matrix obtained with the kernel in Eq.~\ref{eq:kernelTESS}.

For the log-likelihood maximization we used a home made code written in the \texttt{Python3} language. Our code reads in the prior distributions, one for each free parameter in the model, which define the boundaries of the parameter space within which the maximum likelihood location is searched for (Table~\ref{tab:TESSfit}). Then, using the python package \texttt{PyDE}\footnote{\url{https://github.com/hpparvi/PyDE}}, the code searches for the maximum likelihood location, which is used to initialize the MonteCarlo fit. Finally the algorithm samples the posterior probability distribution of the model parameters in a MCMC framework using the \texttt{emcee} package version 3.0.2 \citep{Foreman2013}. The GP is implemented by using the \texttt{celerite2} package version 0.0.2 \citep{Foreman2017,Foreman2018}. Given the complexity and the demand of resources of the model fitting, we ran the code in the HOTCAT computing infrastructure \citep{Bertocco2020,Taffoni2020}.

\begin{figure*}
    \centering
    \includegraphics[width=.49\linewidth]{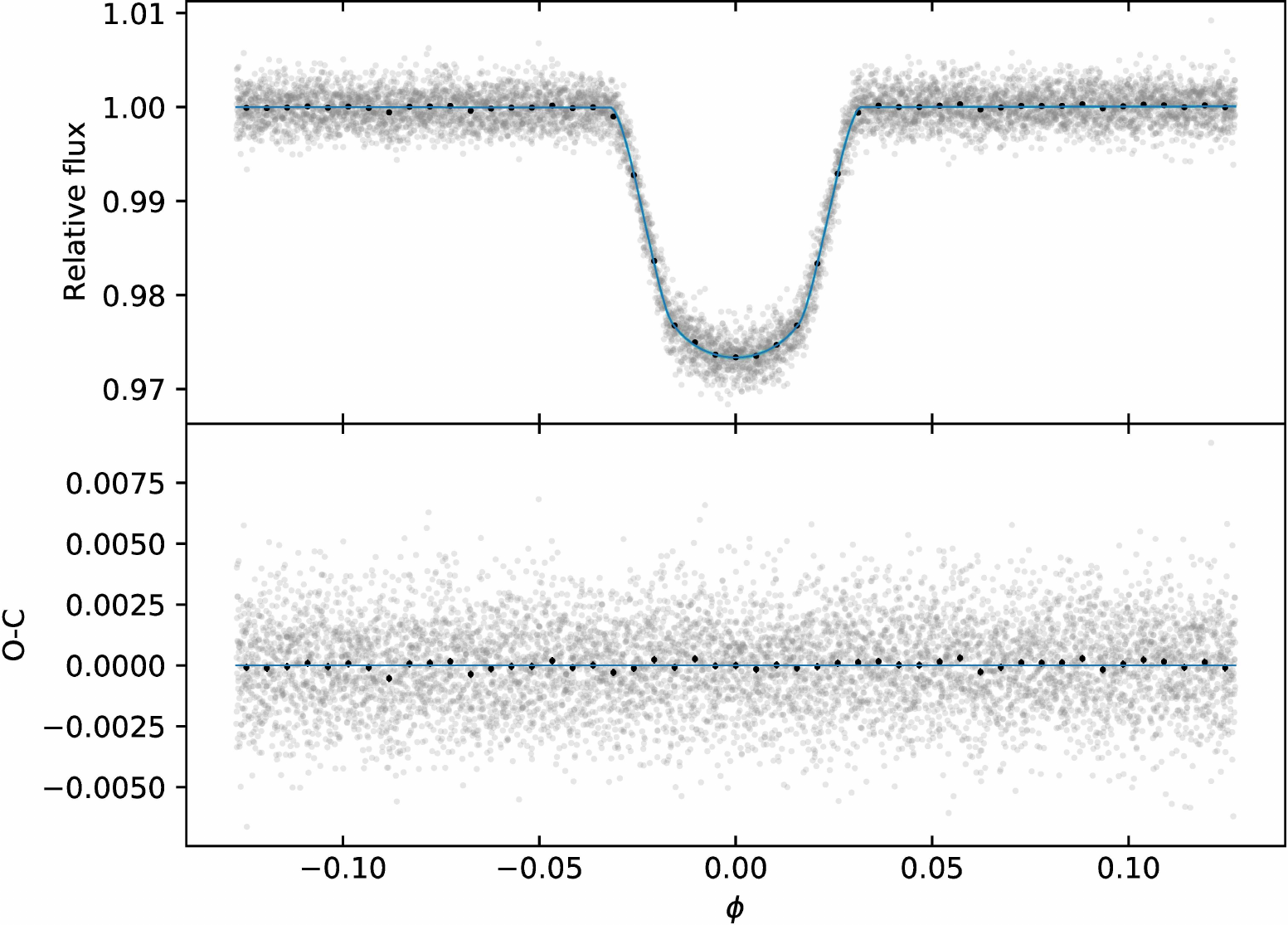}
    \includegraphics[width=.49\linewidth]{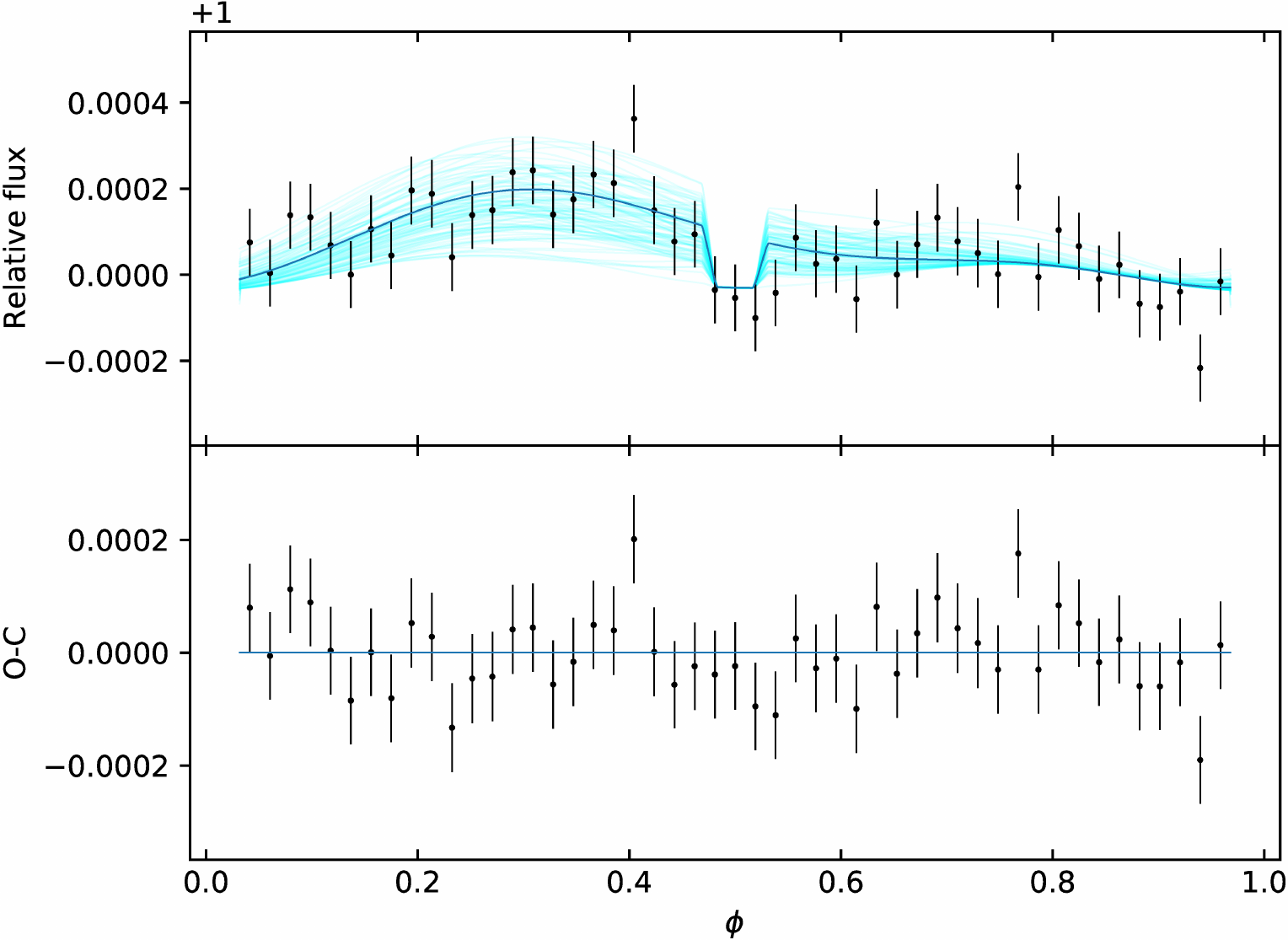}
    \caption{\textit{Left - }Close-up of the phase folded planetary transits observed with \tess, after the removal of the stellar activity (top panel). The solid blue line is the best fit model, while the black dots represent the rebinned photometry. The corresponding residuals are shown in the bottom panel. \textit{Right - } Same as in the left panel, but centered on the eclipse. For visualization purposes, the vertical axis has been zoomed in and only the rebinned photometry is shown. { To give an idea of the uncertainty on the best fit model, the shaded cyan lines in the top panel correspond to the model computed using 100 random steps in the MCMC chain.}} \label{fig:phaseFoldedTess}
\end{figure*}

\begin{figure*}
    \begin{tikzpicture}

    \centering
    \node(a){\includegraphics[width=\linewidth]{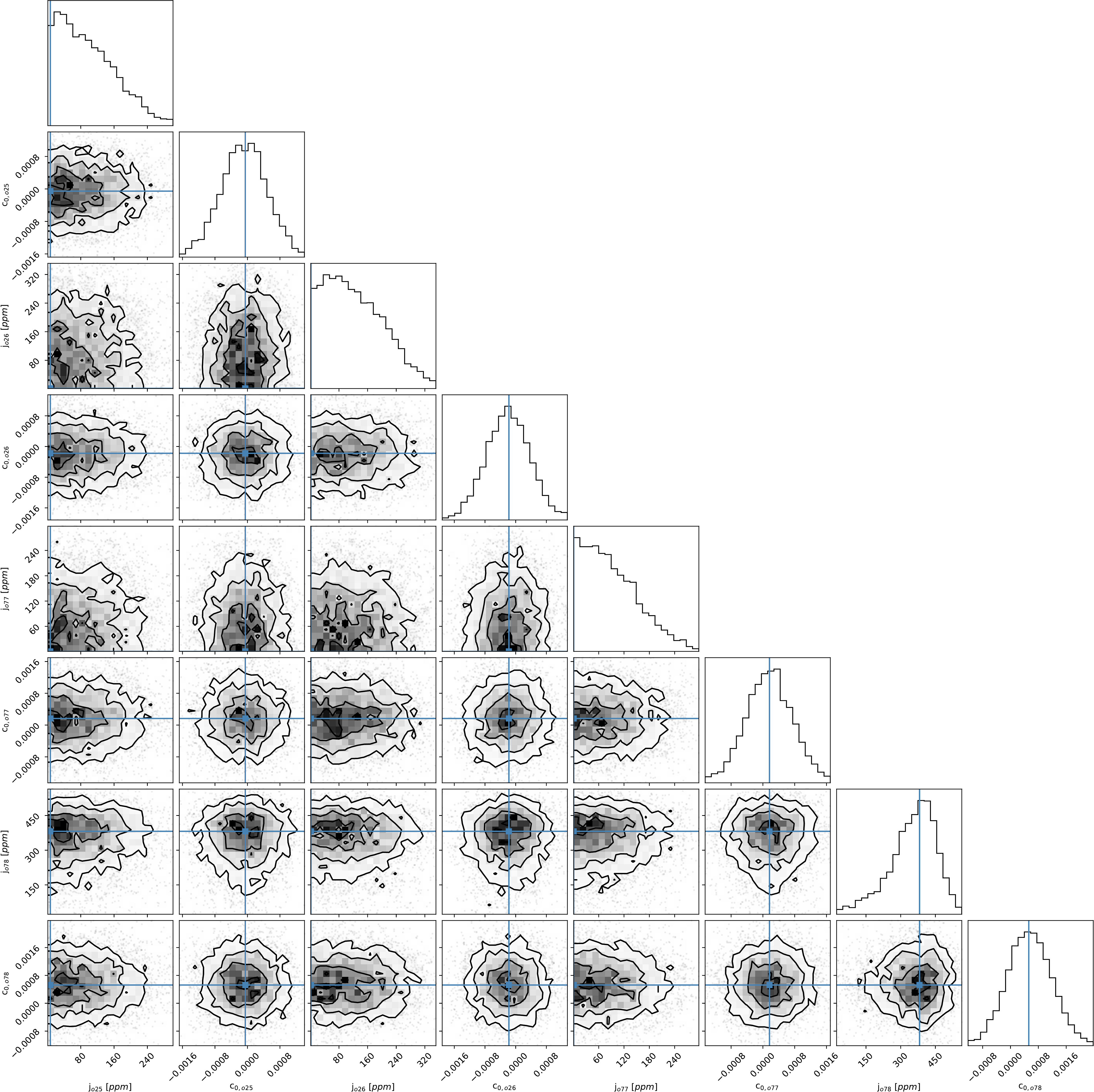}};
    \node at (a.north east)
    [
    anchor=north east,
    xshift=0mm,
    yshift=0mm
    ]
    {
    \includegraphics[width=.4\linewidth]{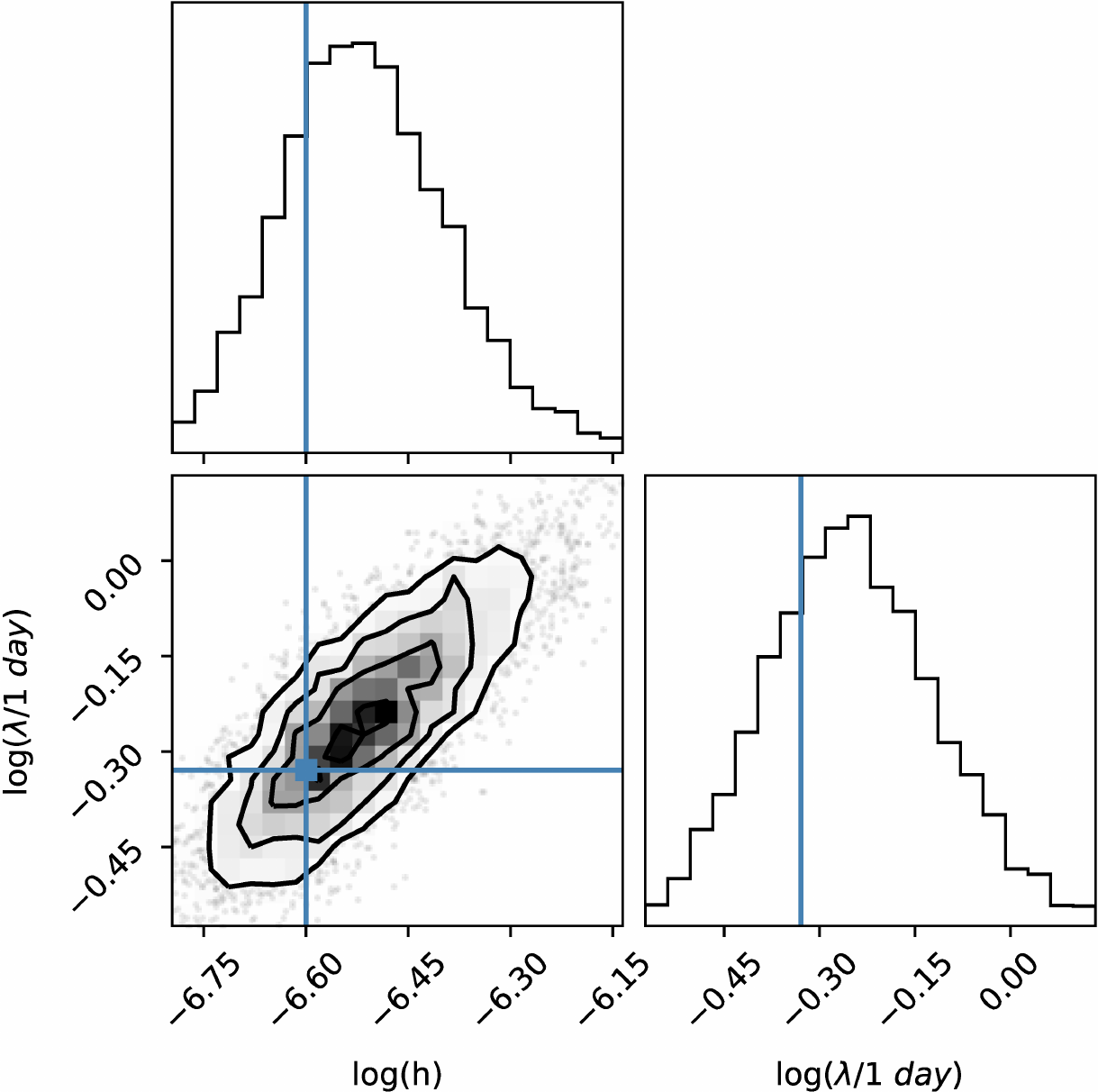}
    };
    \end{tikzpicture}
   \caption{Corner plot of the MCMC chains of the jitter terms and normalization coefficients from the fit of the \tess\ LCs (see Sect.~\ref{sec:TESSfit}). The top right inset shows the corner plot for the GP parameters. In each plot, the solid blue lines mark the MAP values.}\label{fig:TESSfit1}

\end{figure*}

\begin{figure*}
    \begin{tikzpicture}

    \centering
    \node(a){\includegraphics[width=\linewidth]{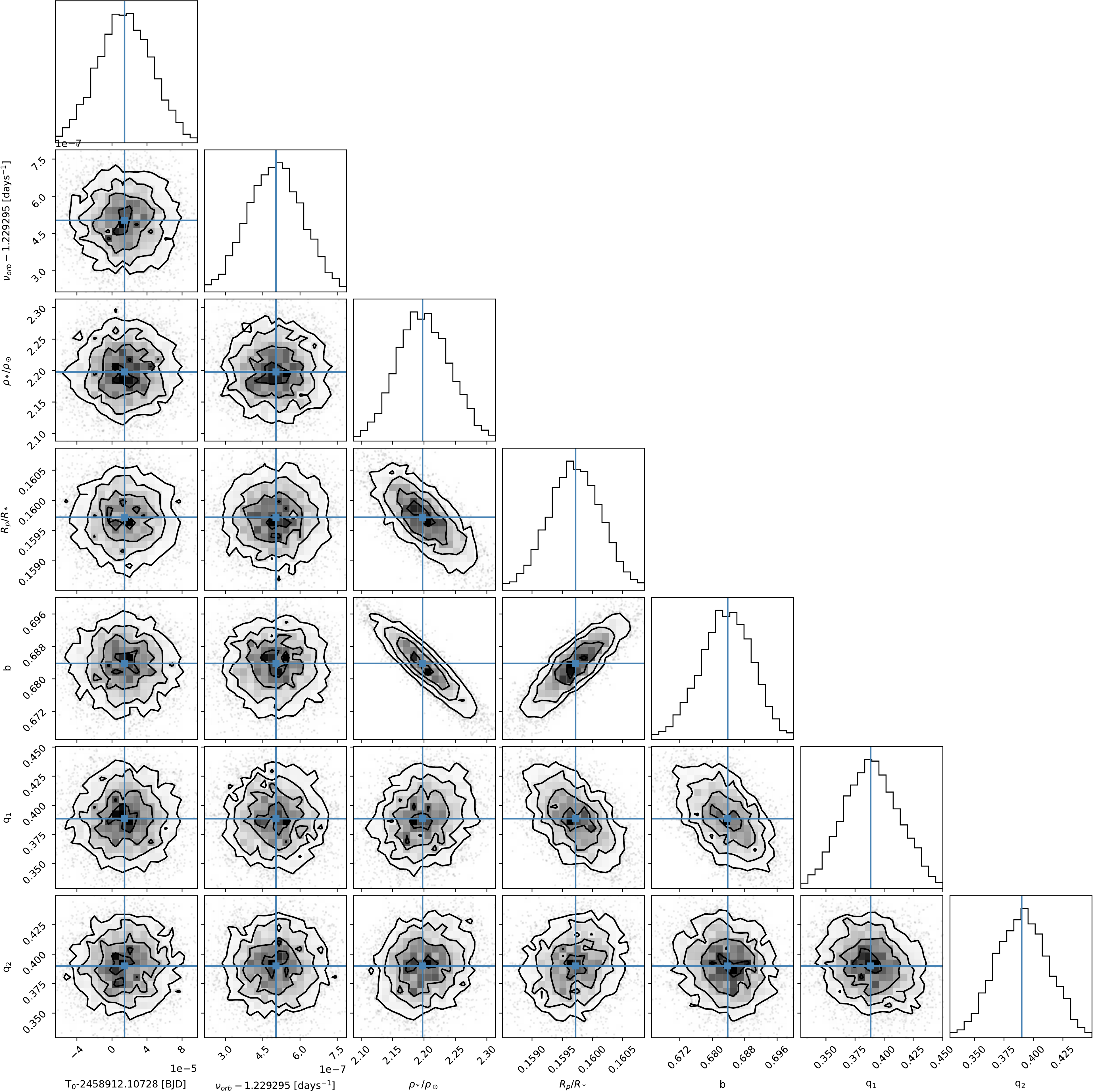}};
    \node at (a.north east)
    [
    anchor=north east,
    xshift=0mm,
    yshift=0mm
    ]
    {
    \includegraphics[width=.4\linewidth]{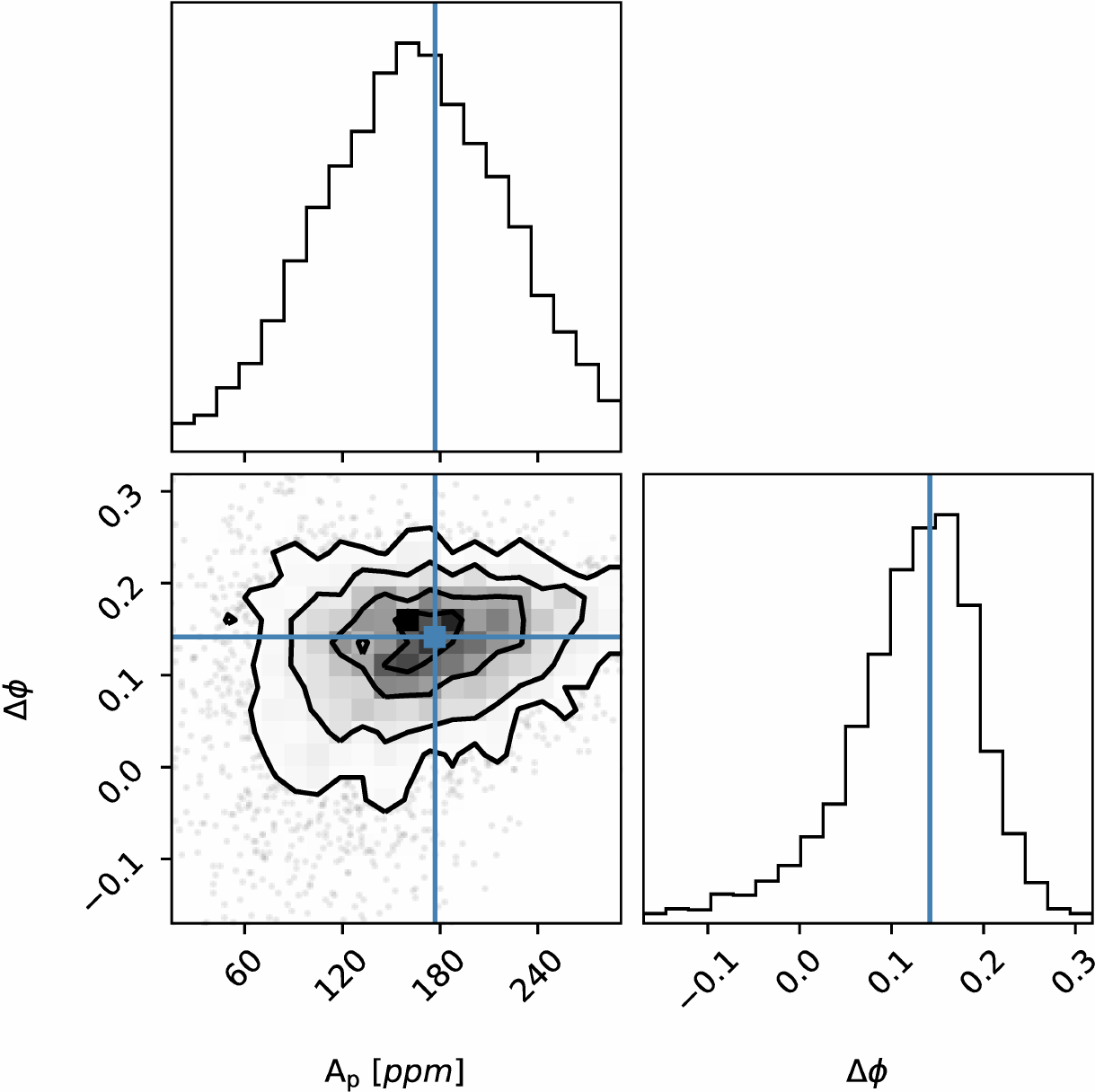}
    };
    \end{tikzpicture}
    \caption{Corner plot of the MCMC chains of planetary parameters from the fit of the \tess\ LCs (see Sect.~\ref{sec:TESSfit}). The top right inset shows the corner plot for the phase curve amplitude and the phase offset. In each plot, the solid blue lines mark the MAP values.}\label{fig:TESSfit2}
\end{figure*}

Our Monte Carlo fit consisted of 100,000 steps, which corresponded to $\sim$300 times the auto-correlation length of the chains, estimated following \citet{Goodman2010}. This suggests that the fit successfully converged\footnote{\url{https://dfm.io/posts/autocorr/}}. We did not find any evidence of a linear long term trend in the data, thus we fixed $c_{1,o}=0$ in Eq.~\ref{eq:lineartermTESS} for each of the four \tess\ orbits. The final list of free parameters, together with the corresponding priors, confidence intervals (C.I.) and Maximum A Posteriori (MAP) values, is reported in Table~\ref{tab:TESSfit}, while the posterior distributions are shown in Figs.~\ref{fig:TESSfit1}--\ref{fig:TESSfit2}. The best fit of the data is shown in Fig.\ref{fig:tessOrbitFits}, while in Fig.~\ref{fig:phaseFoldedTess} we show the data corrected for stellar activity and phase folded to the best fit orbital period. { Our orbital solution is in general agreement with what is present in literature \citep{Hellier2011, Bonomo2017, Esposito2017}.}

{ To test the robustness of our findings, we also tried other simplified models by assuming iteratively a flat out-of-transit LC, a phase curve with no reflection component from the planet, a model with no ellipsoidal variation or doppler boosting and a complete phase curve with no phase offset in the reflection component. We compared all the models using the Akaike Information Criterion \citep[AIC,][]{Burnham2002}. We found that the most likely model is the one described so far, while the second model ranked by the AIC is the one with $\Delta\phi=0$, with a relative likelihood of 14\%. The AIC and 2-$\sigma$ detection on $\Delta\phi$ in Table~\ref{tab:TESSfit} thus make the detection of the phase offset only marginal. All the other models have likelihood $<2\%$ and therefore we omit them from further consideration.}

\begin{table*}
\begin{center}
\caption{Model parameters for the fit of the \tess\ data.}\label{tab:TESSfit}
\begin{tabular}{lllllll}
\hline\hline
Jump parameters & Symbol & Units & MAP & C.I.\tablefootmark{a} & Prior\\
\hline
GP amplitude & $\log h$ & --- & -6.6 & -6.5(1) & $U$(-7,-3) \\
GP timescale & $\log\frac{\lambda}{\rm 1~day}$ & --- & -0.3 & -0.3(1) & $U$(-8,4) \\
Time of transit & $T_0$ & BJD$_{\rm TDB}$-2400000 & 58912.10729 & 58912.10729(3) & $U$(58912.1071,58912.1075) \\
Orbital frequency & $\nu_{\rm orb}$ & days$^{-1}$ & 1.2292955 & 1.2292955(1) & $U$(1.229295,1.229296) \\
Stellar density & $\rho_\star$ & $\rho_\sun$ & 2.20 & 2.20(4)\tablefootmark{b} & $U$(2,2.5) \\
Radii ratio & $R_p/R_\star$ & --- & 0.1597 & 0.1597(4) & $U$(0.154,0.165) \\
Impact parameter & b & --- & 0.684 & 0.684(7) & $U$(0.64,0.73) \\
First LD coef. & q$_{\rm 1}$ & --- & 0.39 & 0.39(2) & $N$(0.454,0.013) \\
Second LD coef. & q$_{\rm 2}$ & --- & 0.39 & 0.39(2) & $N$(0.401,0.024) \\
Phase curve amplitude & A$_{\rm p}$ & ppm & 180 & 160(60) & $U$(0,300) \\
Phase offset & $\Delta\phi$ & --- & 0.14 & $0.13^{+0.06}_{-0.08}$ & $U$(-0.2,.5) \\
Jitter in orbit 25 & j$_{\rm o25}$ & ppm & 0 & $90^{+80}_{-60}$ & $U$(0,2000) \\
Normalization of orbit 25 & c$_{\rm 0,o25}$ & --- & -0.0001 & -0.0001(6) & $U$(-0.02,0.02)\\
Jitter in orbit 26  & j$_{\rm o26}$ & ppm & 0 & $100^{+100}_{-70}$ & $U$(0,2000)\\
Normalization of orbit 26 & c$_{\rm 0,o26}$ & --- & -0.0002 & -0.0002(6) & $U$(-0.02,0.02)\\
Jitter in orbit 77  & j$_{\rm o77}$ & ppm & 0 & $80^{+80}_{-60}$ & $U$(0,2000) \\
Normalization of orbit 77 & c$_{\rm 0,o77}$ & --- & 0.0002 & 0.0002(6) & $U$(-0.02,0.02)\\
Jitter in orbit 78  & j$_{\rm o78}$ & ppm & 400 & $370^{+80}_{-100}$ & $U$(0,2000)\\
Normalization of orbit 78 & c$_{\rm 0,o78}$ & --- & 0.0005 & 0.0005(6) & $U$(-0.02,0.02)\\
\hline
\\
\hline
\hline
Fixed parameters & Symbol & Units & Value & & Notes \\
\hline
Linear trend for orbit 25 & c$_{\rm 1,o25}$ & days$^{-1}$ & 0 &  & \\
Linear trend for orbit 26 & c$_{\rm 1,o26}$ & days$^{-1}$ & 0 &  & \\
Linear trend for orbit 77 & c$_{\rm 1,o77}$ & days$^{-1}$ & 0 &  & \\
Linear trend for orbit 78 & c$_{\rm 1,o78}$ & days$^{-1}$ & 0 &  & \\
Stellar mass & M$_{\star}$ & M$_{\rm\sun}$ & 0.71 & & see Table~\ref{tab:parameters} \\
RV semi-amplitude & K$_{\rm RV}$ & m/s & 551.7 & & see Table~\ref{tab:parameters} \\
Linear LD coef. & u$_{\rm LLD}$ & --- & 0.548 & & computed with \texttt{LDTk} \\
GD coef. & y$_{\rm GD}$ & --- &  0.426 & & from \citet{Claret2017} \\
Tidal lag & $\Theta$ & rad & 0 &  &  \\
\hline
\\
\hline
\hline
Derived parameters & Symbol & Units & MAP & C.I. & Notes\\
\hline
Planetary radius & $R_{\rm p}$ & $R_{\rm J}$ & 1.08 & 1.08(2)\tablefootmark{c} & \\
Orbital period & P$_{\rm orb}$ & day & 0.81347406 & 0.81347406(7) & \\
Transit duration & T$_{\rm 14}$ & hr & 1.242 & 1.242(4) & \\
Scaled semi-major axis & $a/R_\star$ & --- & 4.77 & 4.77(3) & \\
Orbital inclination & $i$ & degrees & 81.8 & 81.8(1) & \\
Eclipse depth & $\delta_{\rm ecl}$ & ppm & 120 & 110(50) & 260 (99.9\% upper limit)\\
\hline
\end{tabular}
\tablefoot{
        \tablefoottext{a}{Uncertainties expressed in parentheses refer to the last digit(s).}
        \tablefoottext{b}{Consistent within uncertainty with the spectroscopically derived stellar mass and radius in Table~\ref{tab:parameters}.}
        \tablefoottext{c}{the uncertainty includes the error on $R_\star$ (Table~\ref{tab:parameters}).}
}
\end{center}
\end{table*}

\subsection{CHEOPS light curves}\label{sec:CHEOPSfit}


We analyzed the \cheops\ LCs using the same Monte Carlo framework described in Sect.~\ref{sec:TESSfit}, taking into account the fact that the visits last only a few hours around the secondary eclipses of \wftb. Thus, the dataset does not allow the reconstruction of the full phase curve, but only the flux drop during the eclipse. Moreover, given the eclipse depth measured in the TESS band (Sect.~\ref{sec:TESSfit}), we did not expect the eclipse signal in the CHEOPS LCs to be strong enough to constrain the ephemeris of the system. For these reasons we fixed the orbital parameters and $\rho_\star$ to their MAP values in Table~\ref{tab:TESSfit}. We also fixed $\Delta\phi=0$: this makes a perfect correspondence between the amplitude of the phase curve $A_{\rm p}$ and the eclipse depth $\delta_{\rm ecl}$.

Since the \cheops\ LCs are only a few hours long, much shorter than the stellar rotation period, the rotation signal cannot be modeled with a periodic function. We thus modeled the activity signal in each visit with a linear term of the form:
\begin{equation}
l_v(t)=c_{0,v}+c_{1,v}\cdot(t-t_{m,v}),\label{eq:lineartermCHEOPS}
\end{equation}
where the subscript $v$ indicates the CHEOPS visit ID from V1 to V9 (see Table~\ref{tab:obs}), while $t_{m,v}$ is the mid-time of the corresponding visit $v$.

As discussed in previous analyses \citep[e.g. ][]{Deline2022,Hooton2021,Wilson2022}, \cheops\ photometry is affected by variable contamination from other stars in the field of view. The variability of this contamination is due to the fact that the Point Spread Function (PSF) of the stars is not circular, and the overlap with the target PSF depends on the roll angle of the telescope \citep{Benz2021}. Both the DRP and PIPE have a module that decontaminates the aperture used for the photometric extraction, but residual signals are present in the LCs due to inaccuracies in the correction. This spurious signal $\Phi$ is phased with the roll angle $\phi_{\rm CH}\in[0,2\pi]$ of the \cheops\ satellite, and changes from visit to visit. To remove this signal, we included in our algorithm a module which fits independently for each visit $v$ the harmonic expansions of a periodic signal up to the first five harmonics, the fundamental harmonic having period $\rm2\pi$:
\begin{equation}
\Phi_v(\phi_{\rm CH})=\Sigma_{i=1}^5\left[a_{i,v}\sin(i\cdot\phi_{\rm CH})+b_{i,v}\cos(i\cdot\phi_{\rm CH})\right]\label{eq:fourier}
\end{equation}

Finally, to take into account any white noise not included in the formal photometric uncertainties, we added to our model a diagonal GP kernel of the form:
\begin{equation}
k(t_{i_1},t_{i_2})=j_v^2\delta_{i_1,i_2},\label{eq:kernelCHEOPS}
\end{equation}
with an independent jitter term $j_v$ for each CHEOPS visit V1 to V9.

\begin{figure}
    \centering
    \includegraphics[width=\linewidth]{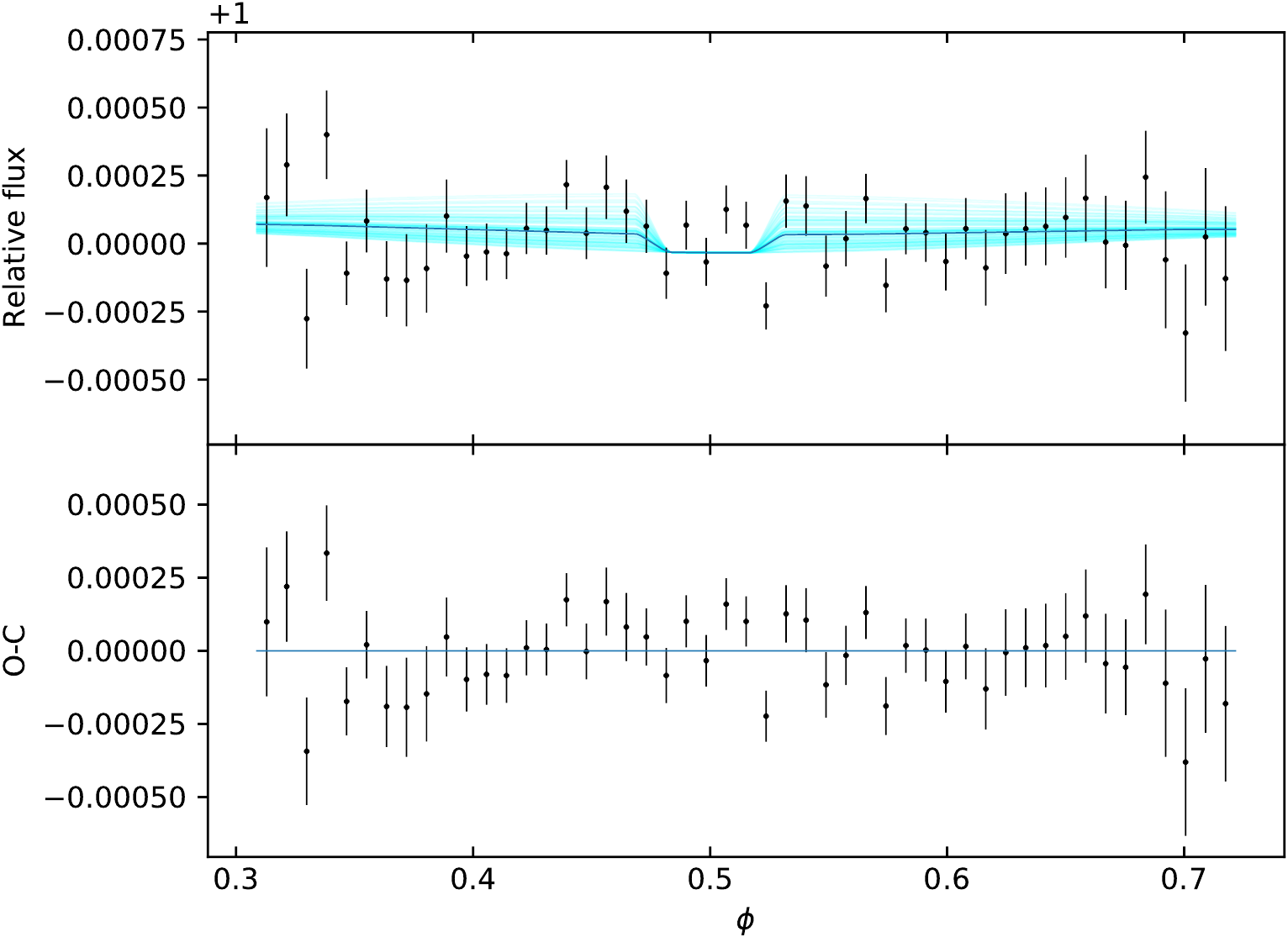}
    \caption{Phase folded \cheops\ photometry of the secondary eclipses of \wftb, corrected for stellar and instrumental correlated noise. The solid line is the best fit model. The corresponding residuals are shown in the bottom panel. For visualization purposes, in both panels the data are binned at regular steps. { The shaded cyan lines in the top panel correspond to the model computed using 100 random steps in the MCMC chain.} }\label{fig:phaseFoldedCheops}
\end{figure}

\begin{figure}
    \centering
    \includegraphics[width=\linewidth]{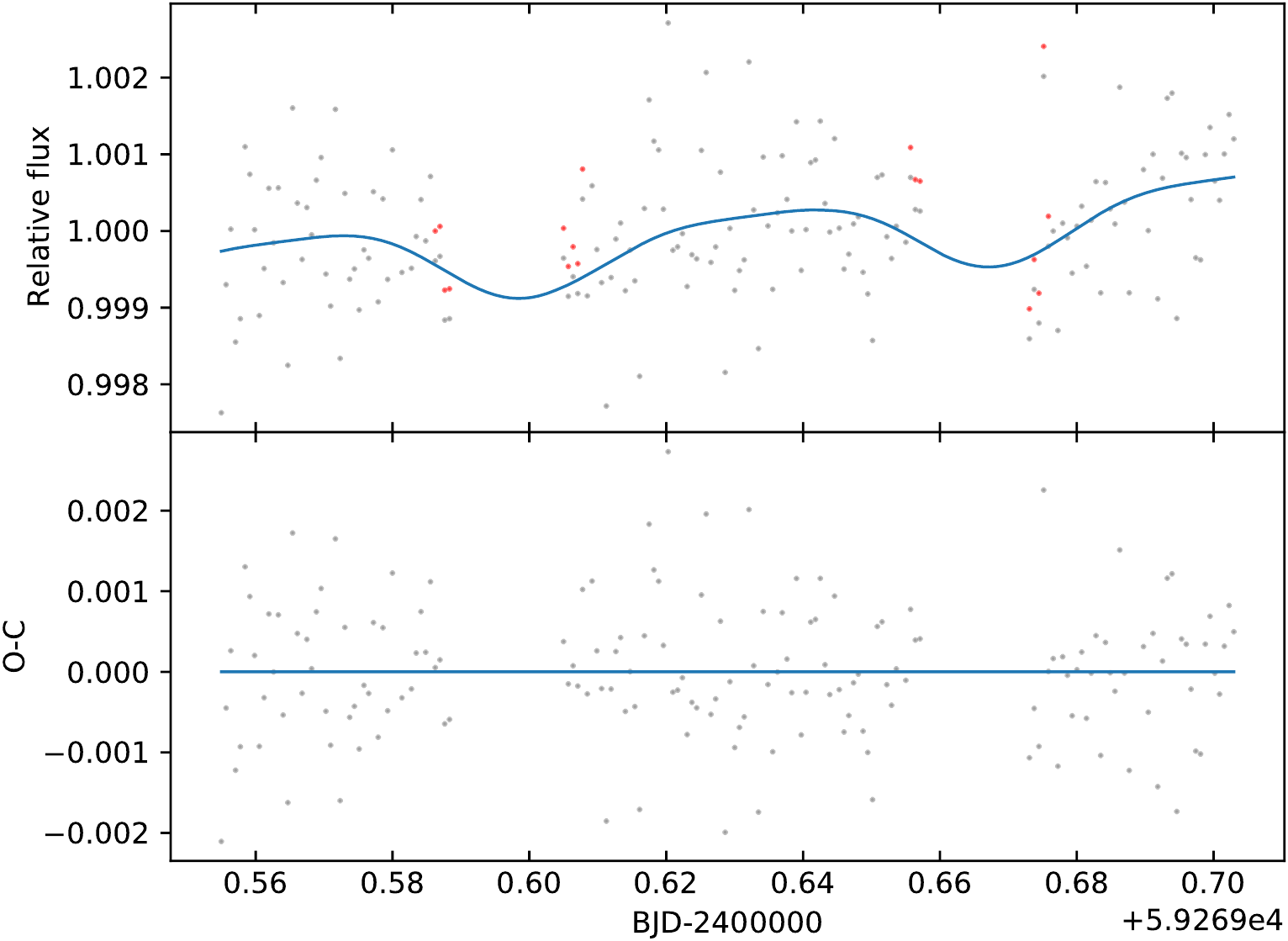}
    \includegraphics[width=\linewidth]{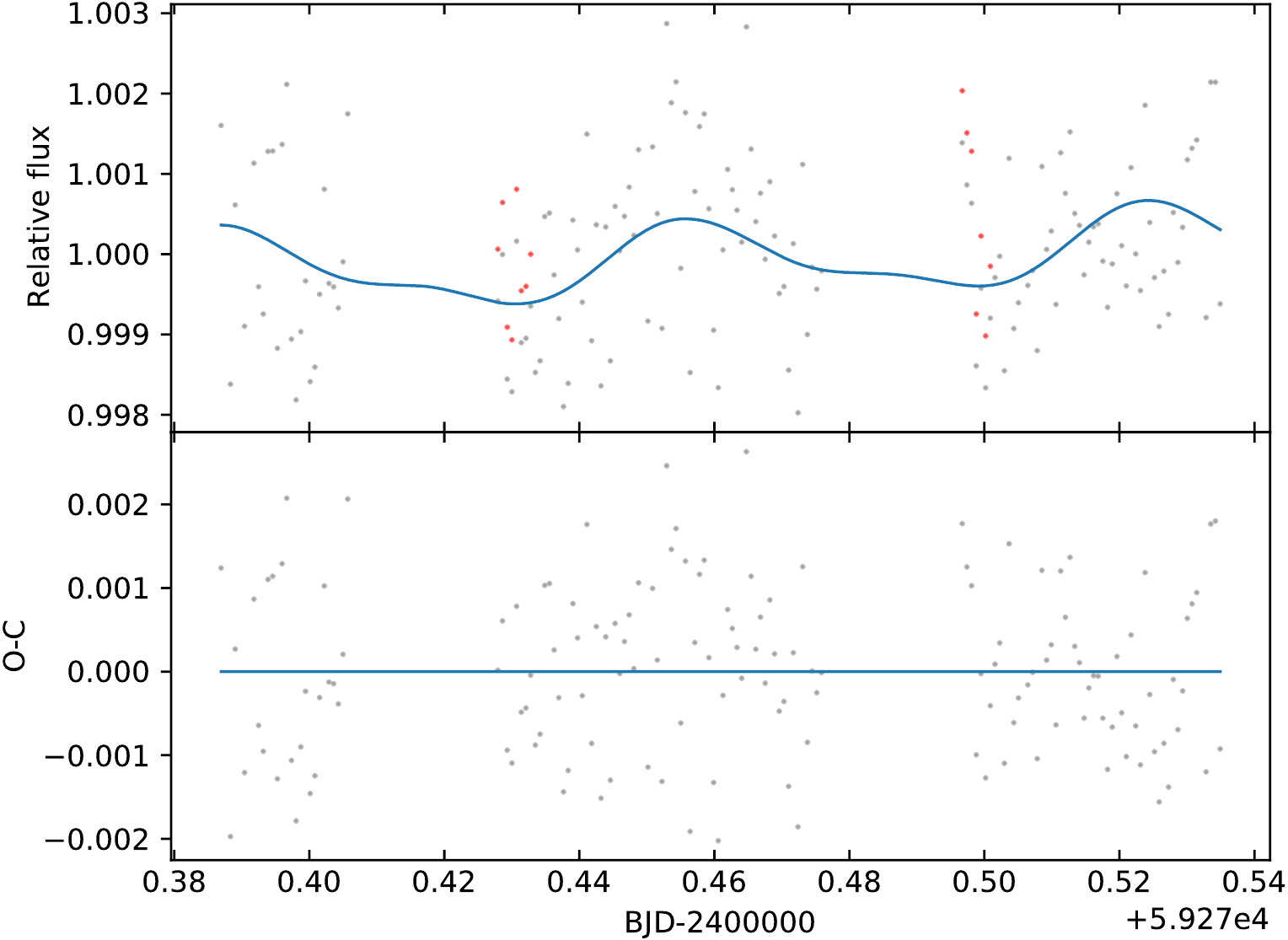}
    \includegraphics[width=\linewidth]{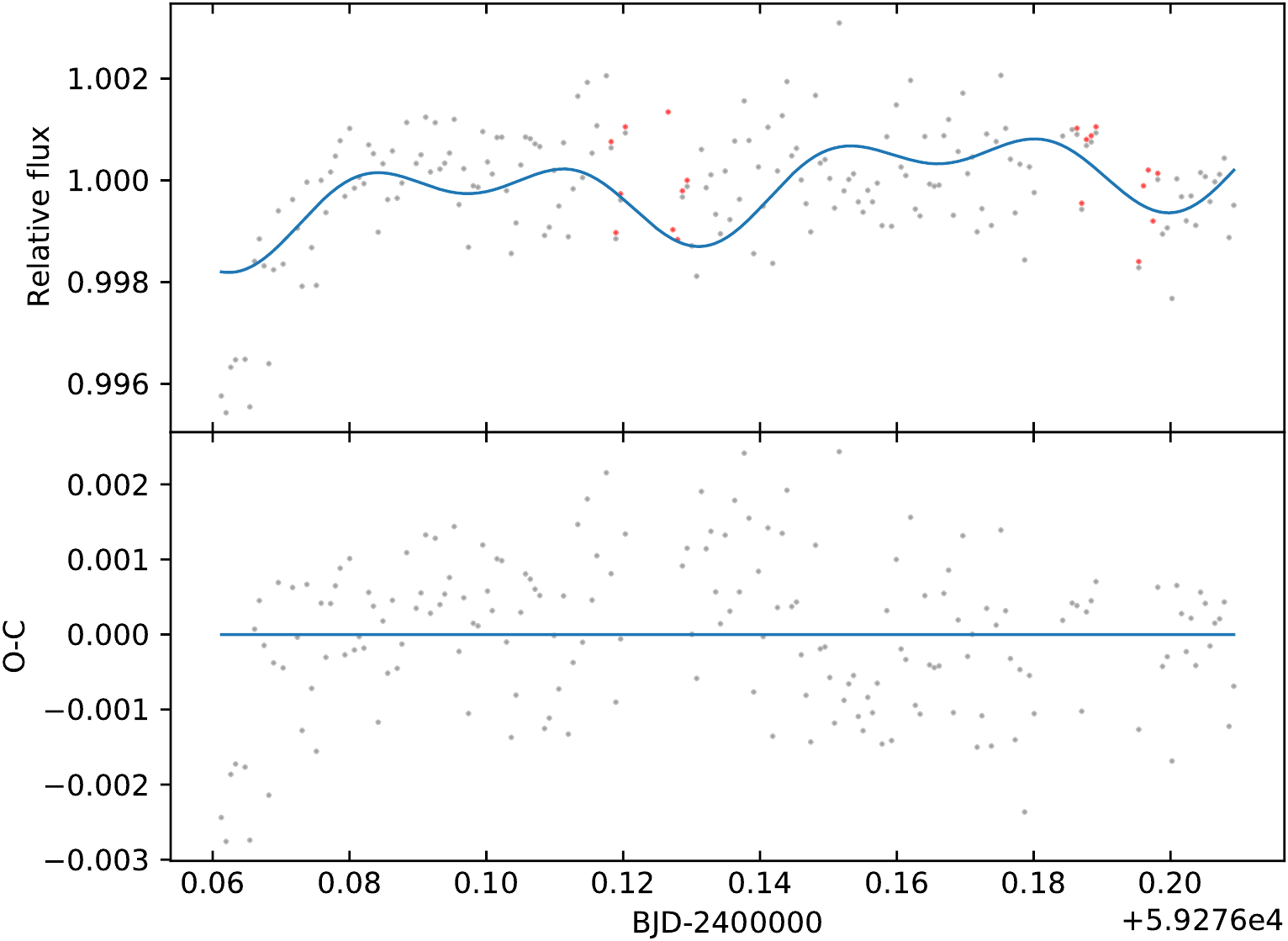}
    \caption{\cheops\ LCs and corresponding best fit for visits V1 to V3 from top to bottom respectively. For each panel, the top plot shows also the best fit model as a solid blue line, while the residuals are shown in the bottom plot. The red data points in the top panel show the rejected data. See Sect.~\ref{sec:CHEOPSfit} for details.}\label{fig:cheopsVisitFits13}
\end{figure}

\begin{figure}
    \centering
    \includegraphics[width=\linewidth]{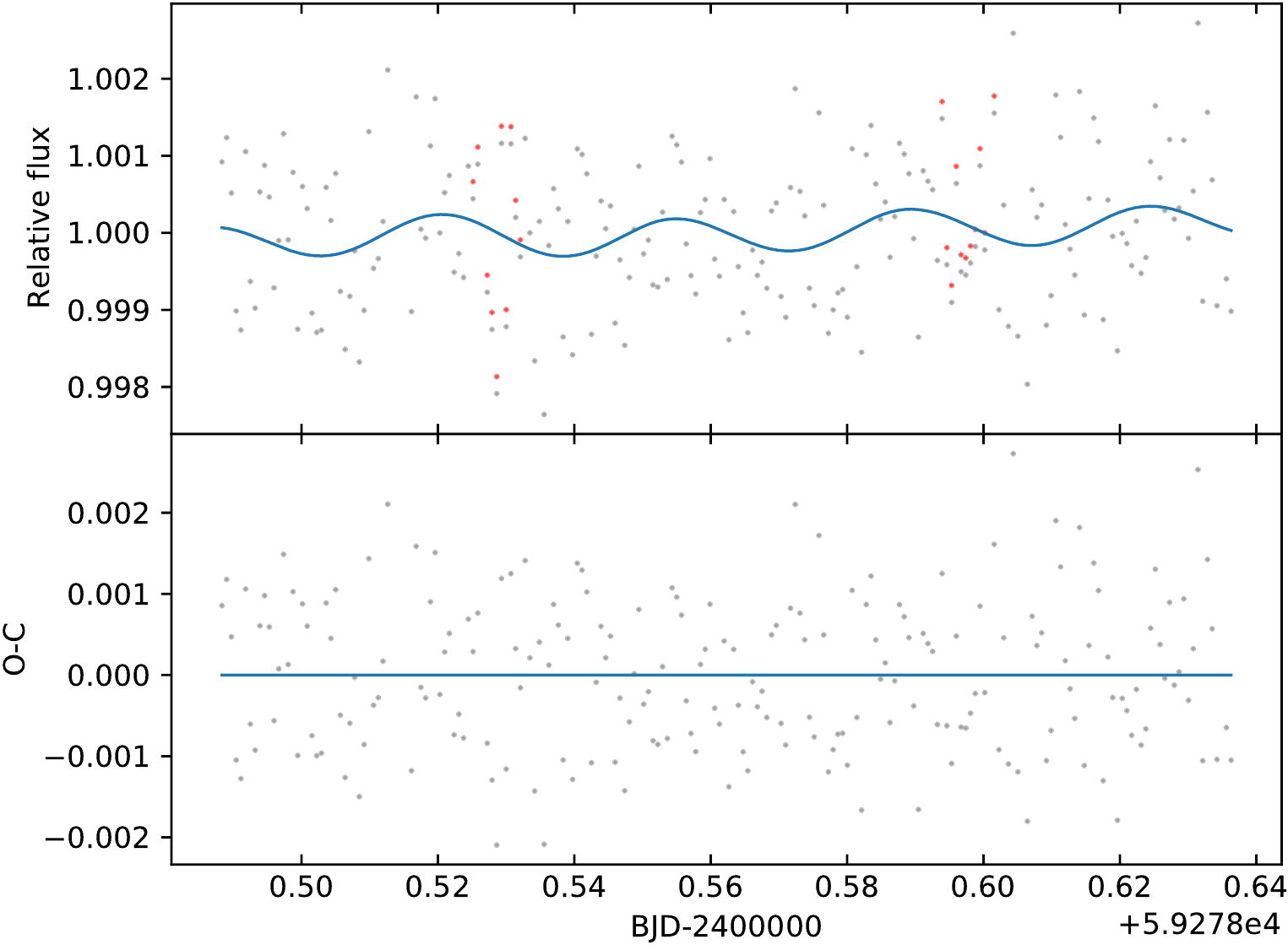}
    \includegraphics[width=\linewidth]{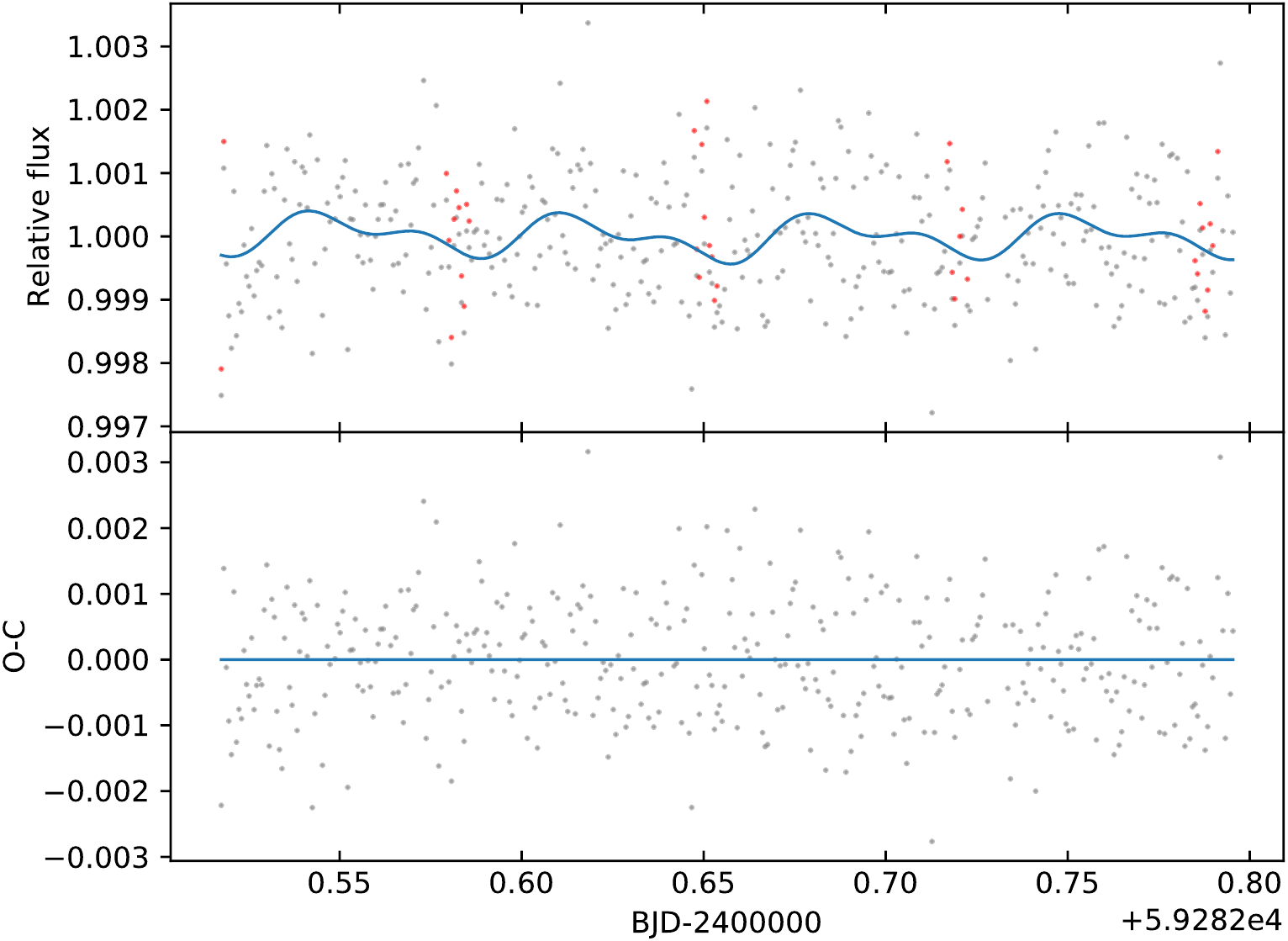}
    \includegraphics[width=\linewidth]{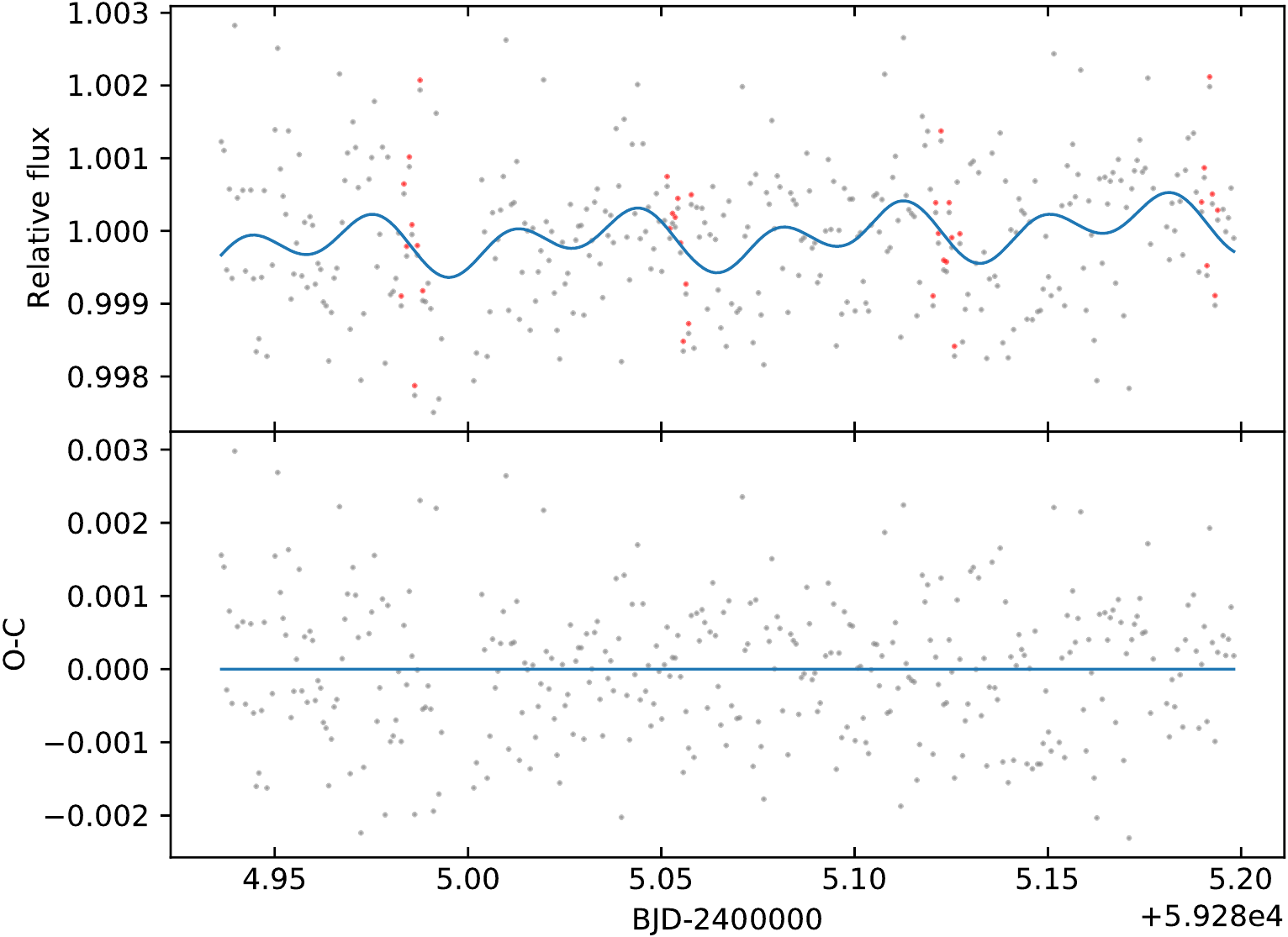}
    \caption{Same as in Fig.~\ref{fig:cheopsVisitFits13} for visits V4 to V6 from top to bottom respectively.}\label{fig:cheopsVisitFits46}
\end{figure}

\begin{figure}
    \centering
    \includegraphics[width=\linewidth]{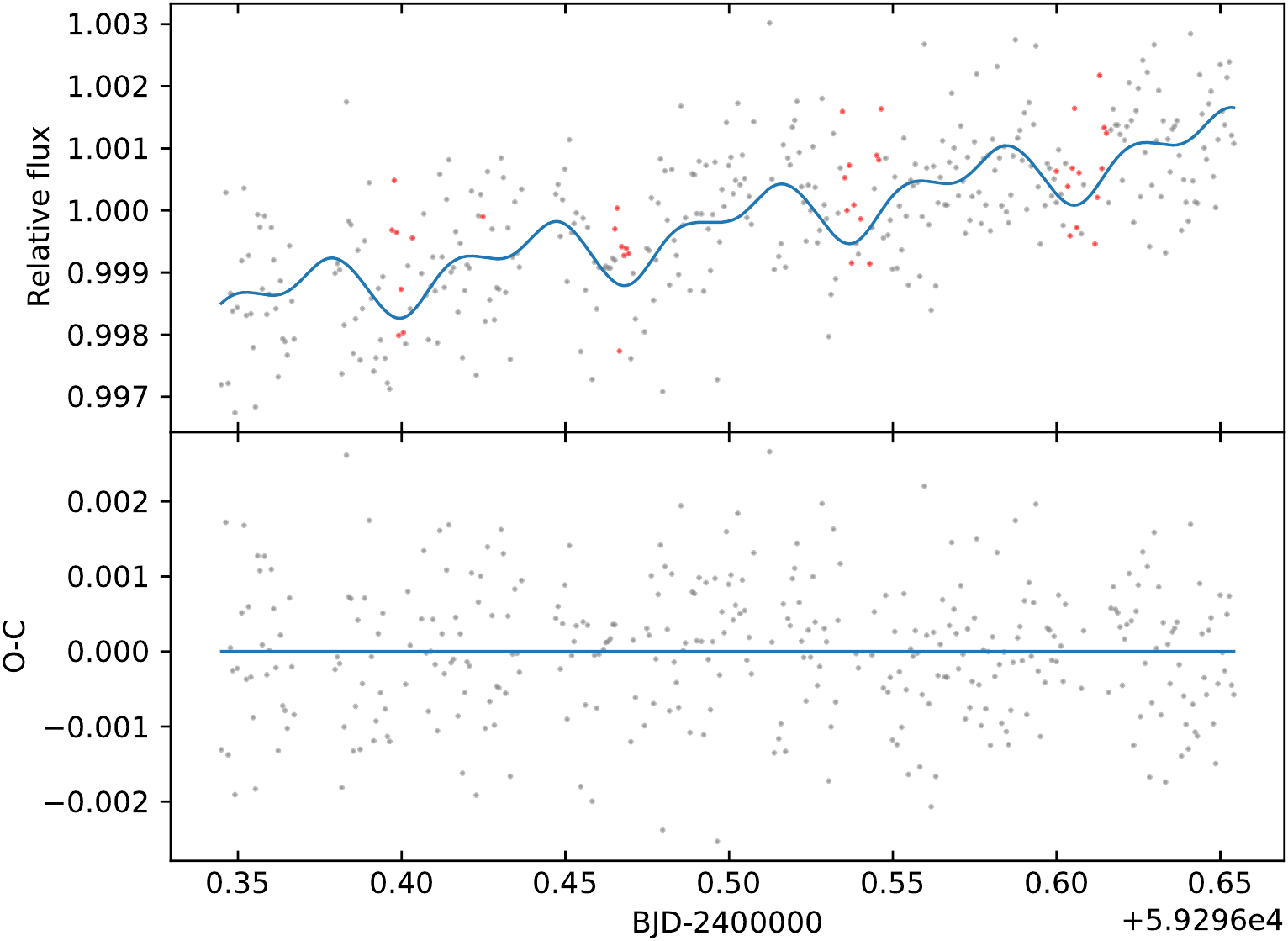}
    \includegraphics[width=\linewidth]{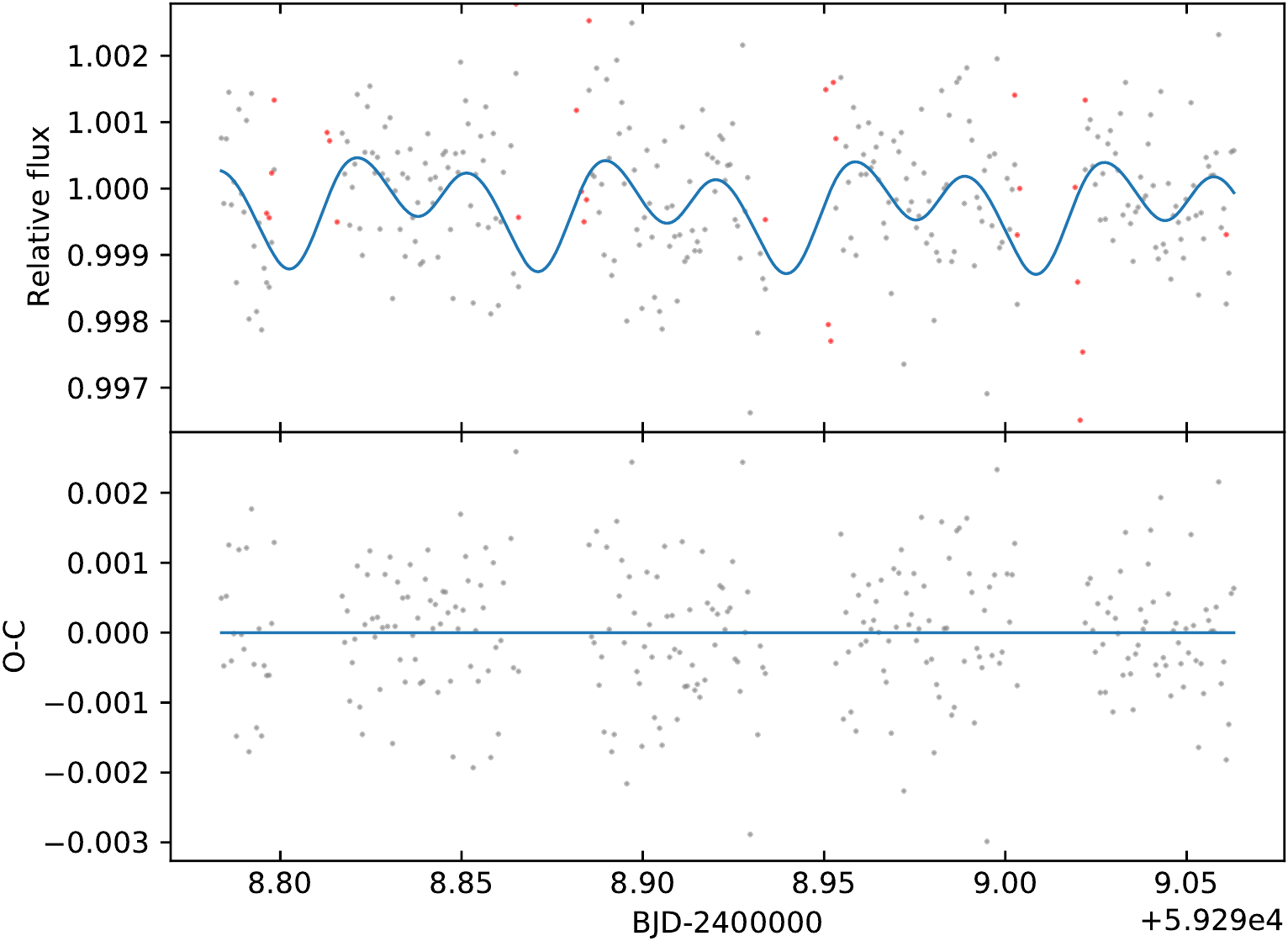}
    \includegraphics[width=\linewidth]{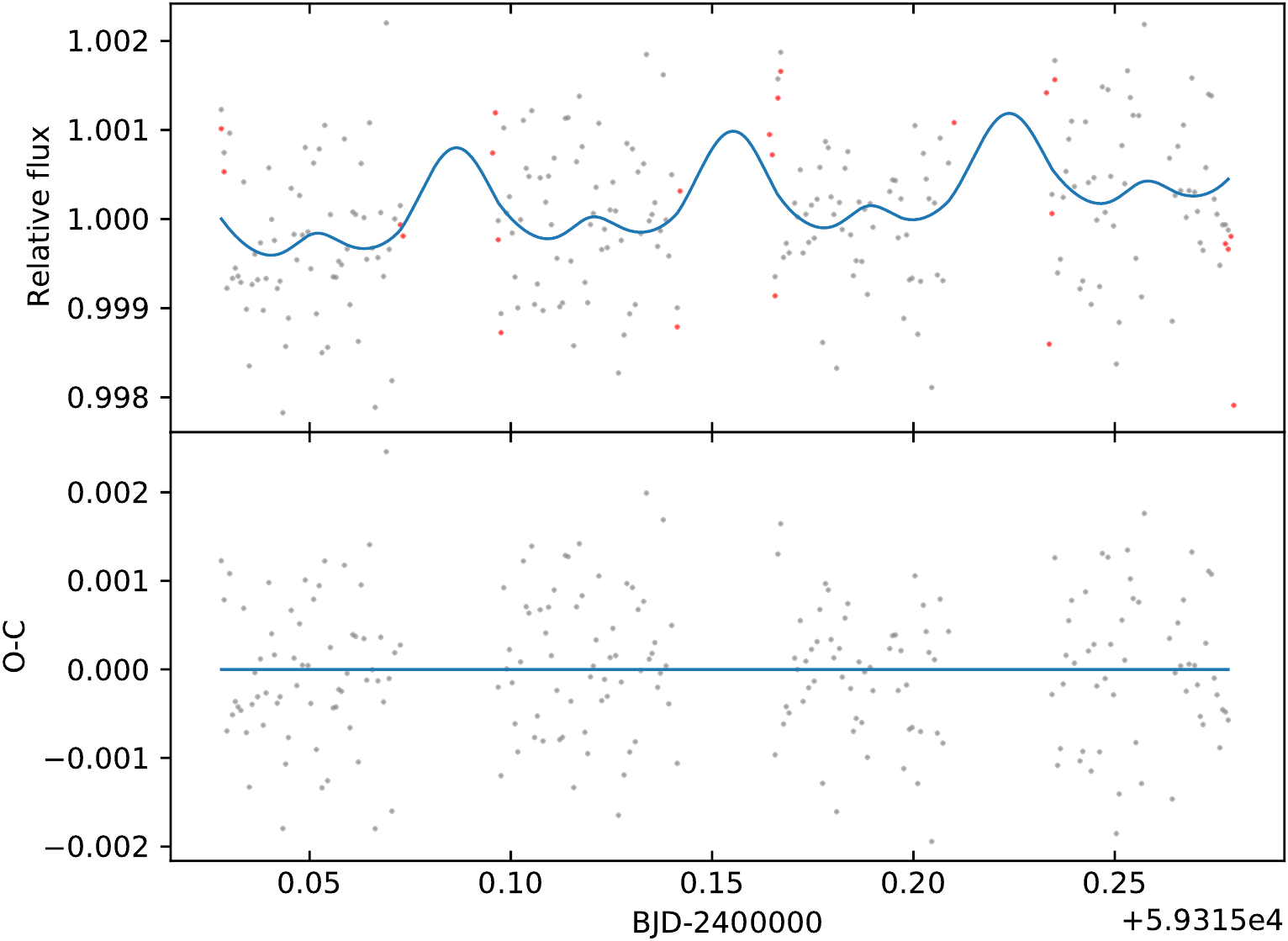}
    \caption{Same as in Fig.~\ref{fig:cheopsVisitFits13} for visits V7 to V9 from top to bottom respectively.}\label{fig:cheopsVisitFits79}
\end{figure}

As for the fit of the \tess\ data, we searched the best-fit parameters by maximizing the likelihood expressed as in Eq.~\ref{eq:logl}. We stopped the MonteCarlo fit after 100,000 steps ($\sim$60 times the autocorrelation time of the chains), so to expect that the chains are sufficiently converged. The fitted parameters, priors and posterior distributions are listed in Table~\ref{tab:CHEOPSfit} and shown in Figs.~\ref{fig:CHEOPSfit1}-\ref{fig:CHEOPSfit9}. We remark that, not having any \cheops\ observation of the transits, the parameters $k$, $q_1$ and $q_2$ cannot be fitted. This explains why they are not listed in the table. The data corrected for correlated noise (instrumental and stellar) and phase folded to the orbital period are shown in Fig.~\ref{fig:phaseFoldedCheops} together with the best fit planetary model computed using the MAP parameters. The best fit of the individual \cheops\ visits is shown in Figs.\ref{fig:cheopsVisitFits13}-\ref{fig:cheopsVisitFits79}.

We also tried the same model discussed above with the reflected light component removed. This last model has a lower AIC and should thus be preferred. Nonetheless, the model including the planetary phase curve has a non negligible relative likelihood of $\sim$33\%. Moreover, being a transiting system in a circular orbit, the eclipse signal must be in the data, as confirmed by the analysis of the \tess\ LCs (Sect.~\ref{sec:TESSfit}). We thus prefer the former, more complex model because it at least puts an upper limit to the eclipse depth in the \cheops\ passband.

\subsection{UVIS light curves}\label{sec:UVISfit}


\citet{Fraine2021} analyzed the \uvis\ data using a simple box model and provide an upper limit to the eclipse depth. We re-analyzed the same data using a more realistic model (Sect.~\ref{sec:modeling}) and the most up to date ephemeris (Sect.~\ref{sec:TESSfit}).

The scheduling of the \uvis\ observations was similar to the ones of \cheops, in the sense that the target has been followed up only for a few hours around a secondary eclipse. For what concerns the modeling of the secondary eclipses and the stellar activity, we thus adopted the same approach described in Sect.~\ref{sec:CHEOPSfit}. The main difference is that we fit a linear trend independently for the forward and reverse scan
\begin{equation}
l_s(t)=c_{0,s}+c_{1,s}\cdot(t-t_{m,s}),\label{eq:lineartermUVIS}
\end{equation}
where $s$ indicates the scan direction (\lq\lq F\rq\rq\ for forward and \lq\lq R\rq\rq\ for reverse) and $t_{m,s}$ is the mid-time of the scan. The main purpose is to allow for different normalization coefficients depending on the scan direction.

\begin{figure}
    \centering
    \includegraphics[width=\linewidth]{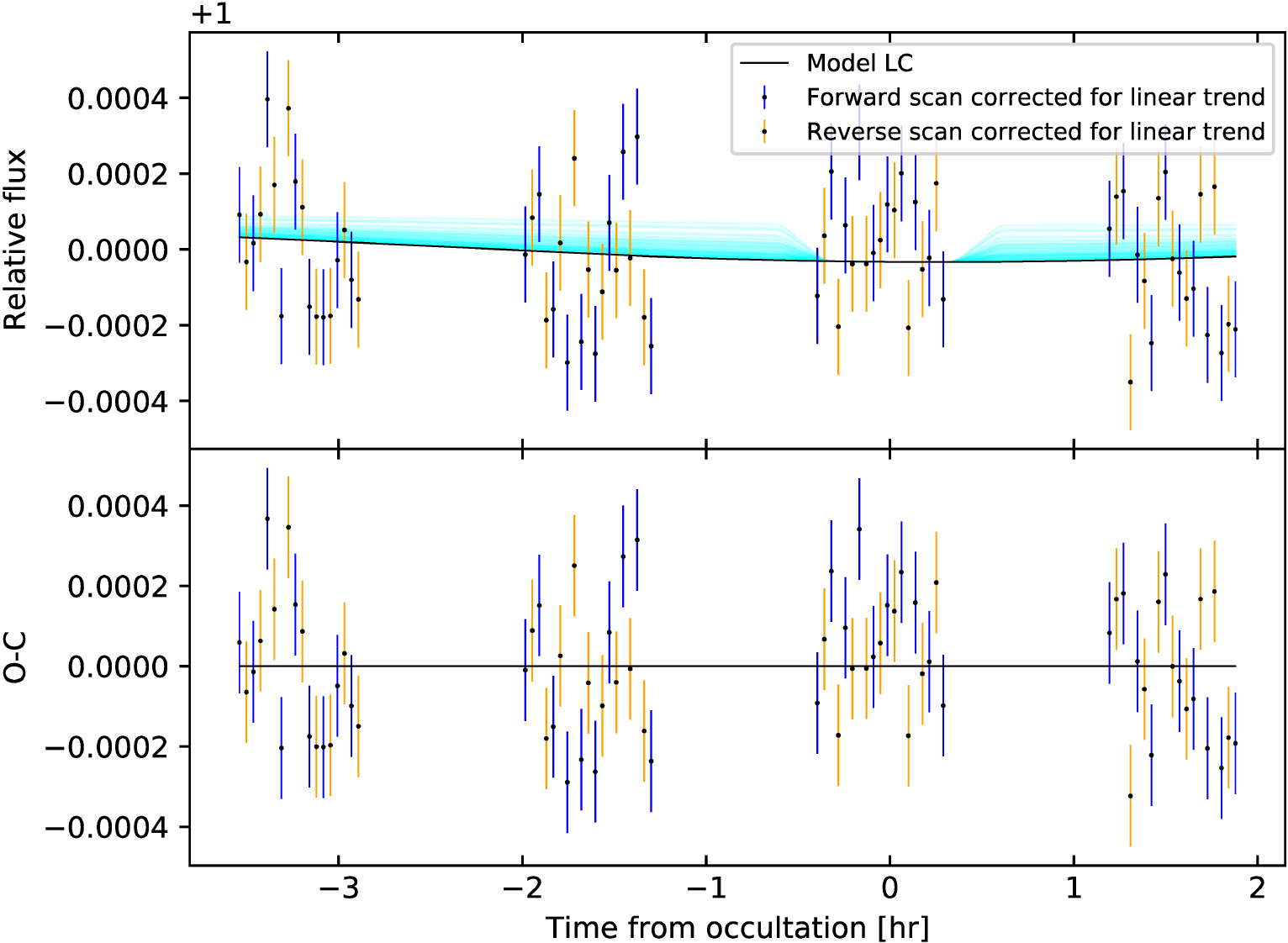}
    \caption{\uvis\ photometry of \wft, corrected for stellar linear trends. The solid line is the best fit model. The corresponding residuals are shown in the bottom panel. { The shaded cyan lines in the top panel show the model computed using 100 random steps in the MCMC chain.}}\label{fig:phaseFoldedUvis}
\end{figure}

Using the same MonteCarlo framework as in Sect.~\ref{sec:TESSfit} and Sect.~\ref{sec:CHEOPSfit}, we ran 10,000 steps to ensure convergence. The outcome of the fit is listed in Table~\ref{tab:UVISfit} and shown in Fig.~\ref{fig:UVISfit}. In Fig.~\ref{fig:phaseFoldedUvis} we plot the UVIS LCs together with the best fit model computed with the MAP values. 

\begin{table*}
\begin{center}
\caption{Model parameters for the fit of the \uvis\ data.}\label{tab:UVISfit}
\begin{tabular}{llllll}
\hline\hline
Jump parameters & Symbol & Units & MAP & C.I.\tablefootmark{a} & Prior\\
\hline
Phase curve amplitude & A$_{\rm p}$ & ppm & 0 & $20^{+30}_{-10}$ & $U$(0,500) \\
Jitter for forward scan & j$_{\rm F}$ & ppm & 140 & 150(30) & $U$(0,2000) \\
Normalization of forward scan & c$_{\rm 0,F}$ & --- & 0.00001 & -0.00001(3) & $U$(-0.002,0.002)\\
Linear trend of forward scan & c$_{\rm 1,F}$ & days$^{-1}$ & 0.0004 & 0.0004(4) & $U$(-0.03,0.03)\\
Jitter for reverse scan & j$_{\rm R}$ & ppm & 80 & 90(40) & $U$(0,2000)\\
Normalization of reverse scan & c$_{\rm 0,R}$ & --- & 0.00001 & -0.00000(3) & $U$(-0.002,0.002)\\
Linear trend of reverse scan & c$_{\rm 1,R}$ & days$^{-1}$ & 0 & 0.0000(3) & $U$(-0.03,0.03)\\
\hline
\\
\hline
\hline
Fixed parameters & Symbol & Units & Value & & Notes \\
\hline
Transit time & $T_0$ & BJD$_{\rm TDB}$-2400000 & 58912.10729 &  & see Table~\ref{tab:TESSfit}\\
Orbital frequency & $\nu_b$ & days$^{-1}$ & 1.2292955 &  & see Table~\ref{tab:TESSfit}\\
Stellar density & $\rho_\star$ & $\rho_\sun$ & 2.20 & & see Table~\ref{tab:TESSfit}\\
Stellar mass & M$_\star$ & M$_{\rm\sun}$ & 0.71 & & see Table~\ref{tab:parameters} \\
RV semi-amplitude & K$_{\rm RV}$ & m/s & 551.7 & & see Table~\ref{tab:parameters} \\
Linear LD coef. & u$_{\rm LLD}$ & --- & 0.652 & & computed with \texttt{LDTk} \\
GD coef. & y$_{\rm GD}$ & --- &  0.521 & & private communication \\
Phase offset & $\Delta\phi$ & --- & 0 &  & \\
Tidal lag & $\Theta$ & rad & 0 &  & \\
\hline
\\
\hline
\hline
Derived parameters & Symbol & Units & MAP & C.I. & Notes\\
\hline
Eclipse depth & $\delta_{\rm ecl}$ & ppm & 0 & $20^{+30}_{-10}$ & 130 (99.9\% upper limit)\\
\hline
\end{tabular}
\tablefoot{
        \tablefoottext{a}{Uncertainties expressed in parentheses refer to the last digit(s).}
}
\end{center}
\end{table*}

\section{Discussion}\label{sec:discussion}

\subsection{Stellar activity}

In the model used to fit the \tess\ LCs we included a GP to detrend the data against the red noise, regardless of its origin. In the case that it is due to stellar activity, the amplitude of the GP ($\sim$1.5~mmag) is about 4 times lower than \citet{Hellier2011}, thus suggesting a smaller photometric signal due to active regions.  This can be due, first of all, to the fact that the \tess\ passband peaks at longer wavelengths, for which the contrast of active regions is lower. Another explanation is that \tess\ has observed \wft\ in a less active state, or that the spot configuration during \tess\ observations was more uniform than during the WASP-South campaign. In any case, everything points to a scenario of a poorly variable star.

In the hypothesis that the red noise is periodic, as suggested by the periodograms in Fig.~\ref{fig:periodogram}, we also tried the SHO kernel for the GP modeling, as it has been reported to be appropriate for periodic signals \citep{Foreman2017}. Nonetheless, a posteriori we found that the SHO kernel is not able to catch the periodic-like correlated noise in the data. According to \citet{Serrano2018}, the time span covered by the \tess\ sectors would be long enough to detect the ~15~day rotation period, but two factors act against a clear detection. The first limitation comes from the small amplitude of the photometric variability. Secondly, the reconstruction of the periodic signal is hampered if the damping timescale of the kernel is shorter than the period of the GP, i.e. the activity signal evolves on timescales shorter than the rotation period. This partially washes out the periodicity of the rotation signal, which can thus no longer be accurately caught by the time-series analysis.

\begin{figure}
    \centering
    \includegraphics[width=\linewidth]{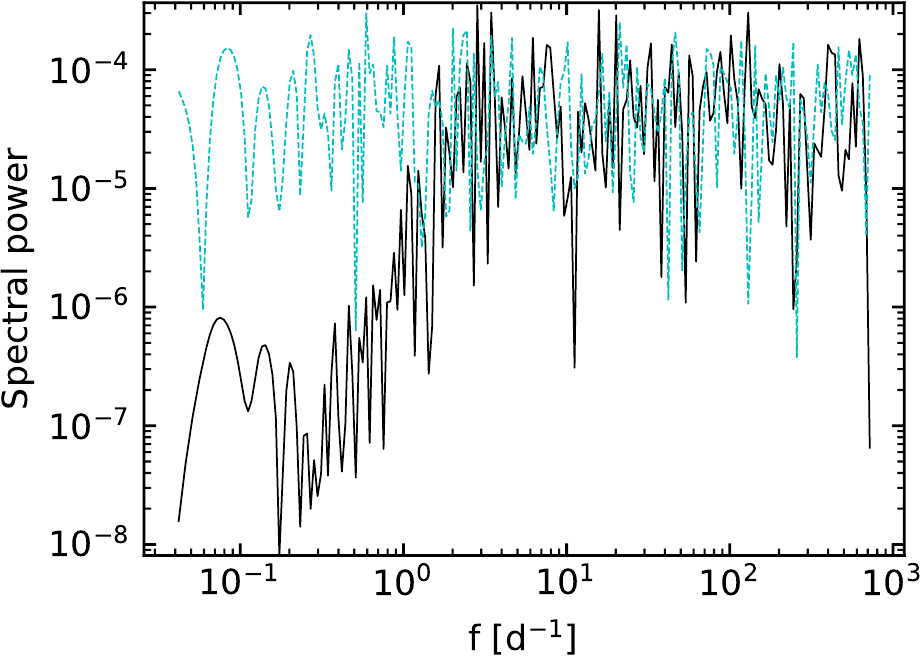}
    \includegraphics[width=\linewidth]{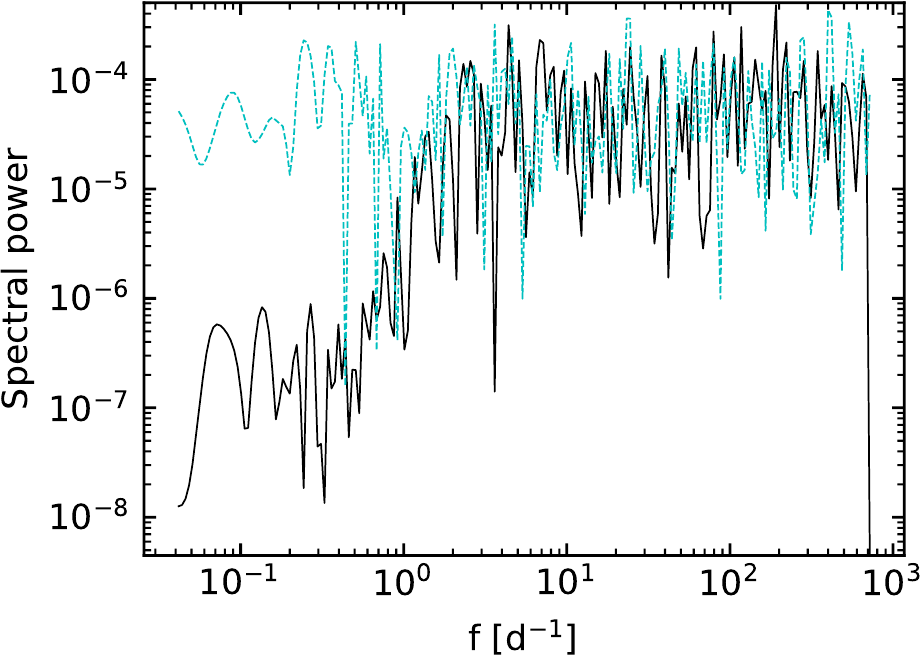}
    \caption{ Periodogram of the residuals of the best fit of sector 9 (top panel) and sector 35 (bottom panel). In each panel, the cyan dashed line shows the periodogram of a randomly generated flat LC with the same time sampling and noise corresponding to the variance of the residuals of the fit.}
    \label{fig:resPeriodogram}
\end{figure}

Given the definition in Eq.~\ref{eq:kernelTESS}, the covariance of the Mat\'ern~3/2 kernel is expected to decay by a factor of 10 after 1~d. This means that any signal with timescales shorter than 1~d can hardly be absorbed by the GP. As a matter of fact, the periodogram of the residuals in Fig.\ref{fig:resPeriodogram} shows that at frequencies higher than $\sim1~[d^{-1}]$ (periods shorter than 1~d) the power spectrum is consistent with white noise, while at shorter frequencies (longer periods) the periodogram drops by 3 orders of magnitude, the missing power being absorbed by the GP. This evidence has a twofold implication. First of all, since the GP absorbs power at period longer than 1~d, it is unlikely that it interferes with the phase curve, whose periodicity is $P_{\rm orb}=0.81347406~d$ (Table~\ref{tab:TESSfit}). Secondly, since the periodogram of the residuals does not show any significant deviation from white noise, we also conclude that there is no evidence of remaining correlated noise in the data, and that the posteriors of the MCMC fit are unbiased.

\begin{figure*}
    \centering
    \includegraphics[width=.45\linewidth]{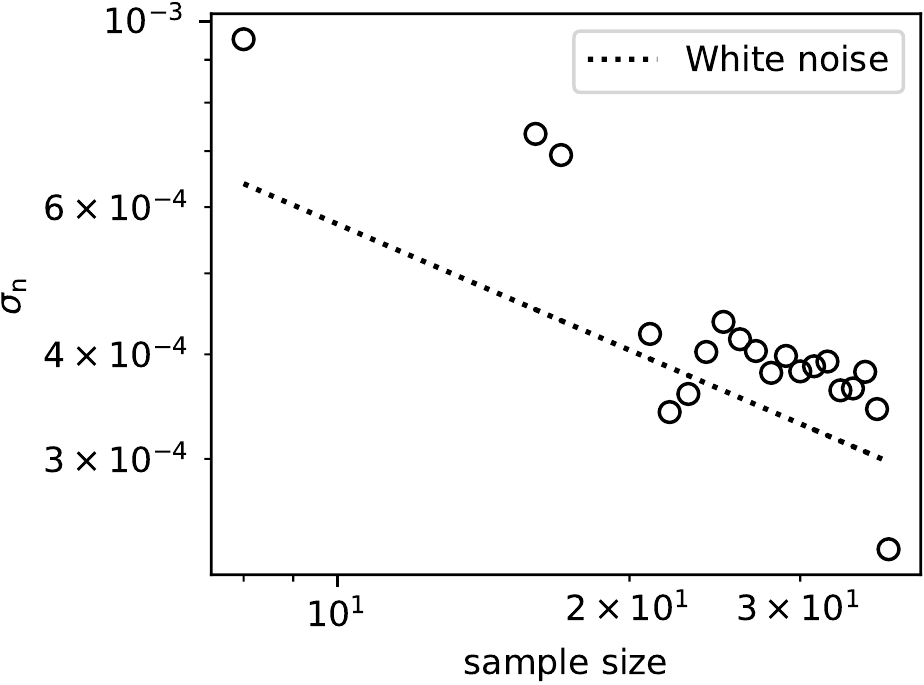}
    \includegraphics[width=.45\linewidth]{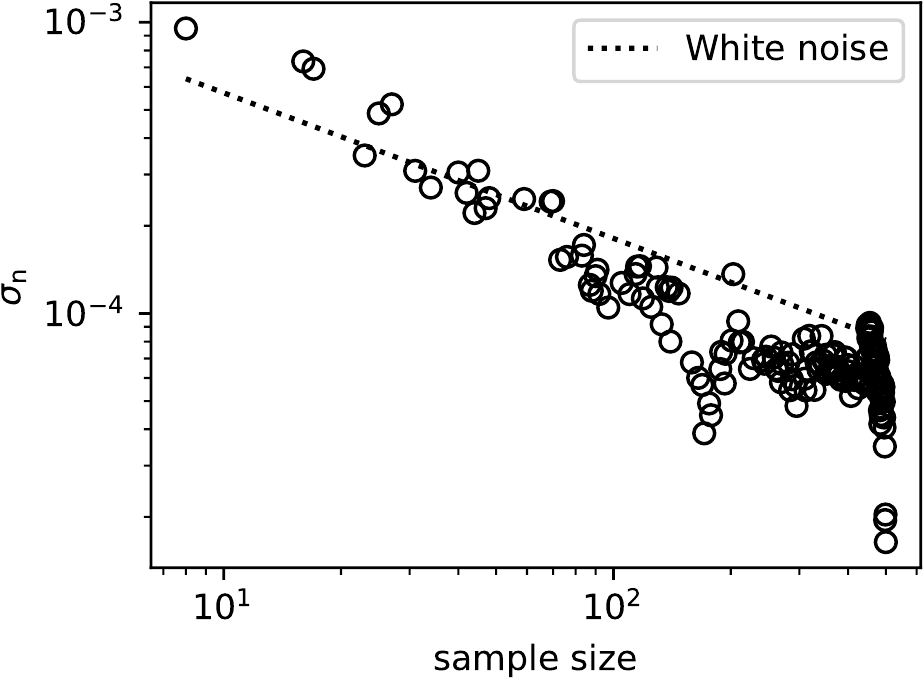}
    \caption{Scaling of the standard error of the average $\sigma_{\rm n}$ as a function of the sample size $n$. Left and right panels show respectively the empirical scaling relation for a time bin corresponding to the transit duration $T_{\rm 14}$=1.245~hr and the orbital period $P_{\rm orb}$=0.81347406~d (see Table~\ref{tab:TESSfit}). In each panel the dotted line shows the expected trend in the assumption of purely white noise. }\label{fig:pont}
\end{figure*}

To tackle the problem of correlated noise we also used the approach proposed by \citet{Pont2006}: we analyzed how the standard error on the average $\sigma_{\rm n}$ scales with the sample size $n$ of residual data points in a time interval corresponding to the transit duration (Fig.~\ref{fig:pont}, left panel) and to the orbital period (Fig.~\ref{fig:pont}, right panel). In both cases, we found that $\sigma_{\rm n}$ scales down as $n^{1/2}$, which is expected in absence of red noise in the residual time series. We thus conclude that the model used to fit the data absorbs both the astrophysical signals and the correlated noise (astrophysical and instrumental), returning unbiased estimate of the planetary parameters. This is consistent with the fact that the jitter terms $j_{\rm o}$ in Eq.~\ref{eq:kernelTESS} are all consistent with 0. The only exception is \tess\ orbit 78, for which we postulate instrumental issue intervening between orbits 77 and 78.

Finally, searching for additional clues on stellar activity, we fit the two \tess\ sectors independently and we found that the corresponding $R_{\rm p}/R_\star$ ratios are consistent within uncertainties. This result indicates that, if active regions are present on the stellar surface, their overall configurations during the two sectors are similar.

\subsection{Orbital parameters}

The analysis of the \tess\ LCs allowed us to update the ephemeris and transit parameters of \wftb. Our determination of the orbital parameters agrees within 3$\sigma$ with previous ground-based studies \citep{Hellier2011,Esposito2017,Wong2020b,Garai2021}.

In particular, we do not find any significant discrepancy among our estimate of $R_p/R_\star$ and the ones reported in literature at previous epochs. This has a twofold implication. At the current photometric precision, the evolution of the activity signal is not significant enough to affect the planetary radius measurement, as it is the case, for example, for more active stars like CoRoT-2 \citep{Czesla2009,SilvaValio2010}. This is consistent with the low activity scenario for \wft, as discussed previously. Moreover, comparing the measurements using different bandpasses, no significant trend with wavelength is detected. This implies that neither stellar activity nor the planetary atmosphere significantly affect the transmission spectrum in the optical domain.

\subsection{Phase curve}

The atmospheric phase curve of \wftb\ has been modeled with two free parameters: the amplitude $A_{\rm p}$ and the phase offset $\Delta\phi$. { Using \tess\ data we obtained a $\sim3\sigma$ detection for the amplitude $A_{\rm p}$ (160$\pm$60~ppm), while other data sets led to a marginal 2$\sigma$ detection (\cheops, $80^{+60}_{-50}$~ppm) or to an upper limit (\uvis, $A_{\rm p}<130$~ppm).

The extended phase coverage of \tess\ data also allowed to investigate the presence of an offset in the atmospheric phase curve, leading to a marginal 2$\sigma$ detection of $\Delta\phi=0.13^{+0.06}_{-0.08}$, that corresponds to an eastward angular offset of $\left(50^{+30}_{-20}\right)^\circ$ with respect to the substellar point. This estimate, despite marginal, is consistent with the eastward offset of $21.1^\circ\pm1.8^\circ$ detected by \citet{Stevenson2017} using Spitzer data.

\citet{Wong2020b} and \citet{Blazek2022} recently published independent extractions of the planetary phase curve from \tess\ LCs using  different flavors of data detrending and modeling. Their results are consistent with ours and confirm the difficulty to reach a clear detection of the planetary emission signal using only \tess\ data, due to the faintness of the reflected light.

}

\subsection{Geometric albedo}\label{sec:geometricAlbedo}

The eclipse depth $\delta_{\rm ecl}$ is a direct measurement of the flux contrast $F_p/F_\star$ between the planet and its parent star when the planetary disk is fully illuminated ($\phi=0.5$), and in principle incorporates the contribution from both the reflection of the stellar spectrum and the thermal emission from the planetary surface.

\begin{figure}
    \centering
    \includegraphics[width=\linewidth]{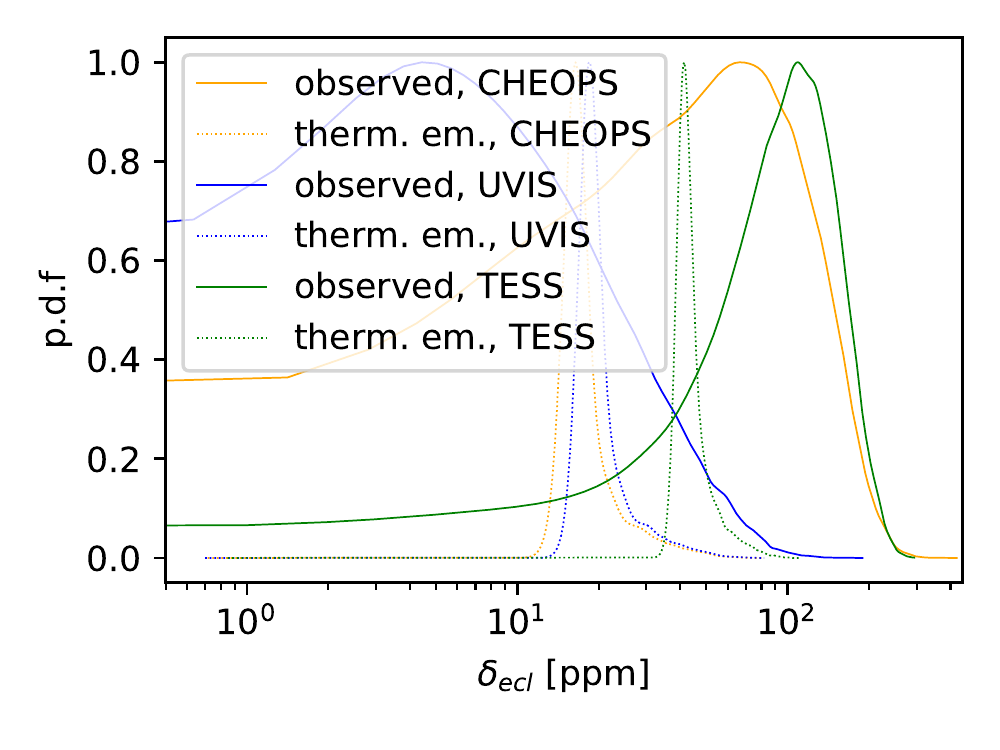}\\
    \includegraphics[width=\linewidth]{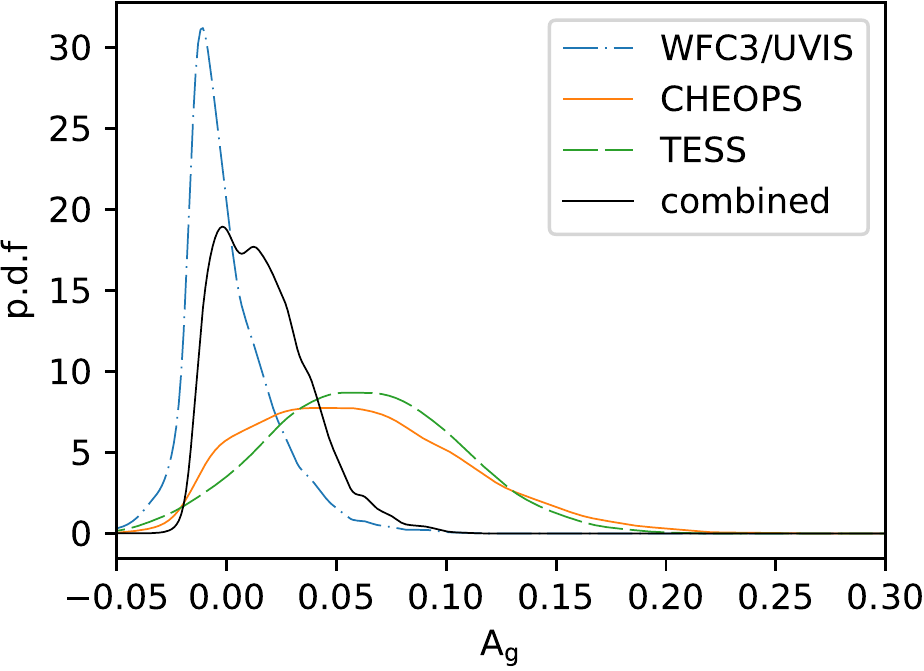}
    \caption{\textit{Top panel - }Posterior distribution functions (PDF) of the thermal contamination (dotted lines) and measured eclipse depths $\delta_{\rm ecl}$ (solid lines) derived from the fit of the \uvis, \cheops\ and \tess\ LCs. For clarity, each function has been rescaled so to peak at 1. \textit{Bottom panel - }Normalized PDFs of the geometric albedo $A_g$ in the \uvis, \cheops\ and \tess\ passbands. The combined PDF is shown as a black solid line.}\label{fig:doccultations}
    \label{fig:my_label}
\end{figure}

\begin{figure*}
    \centering
    \includegraphics[width=\linewidth]{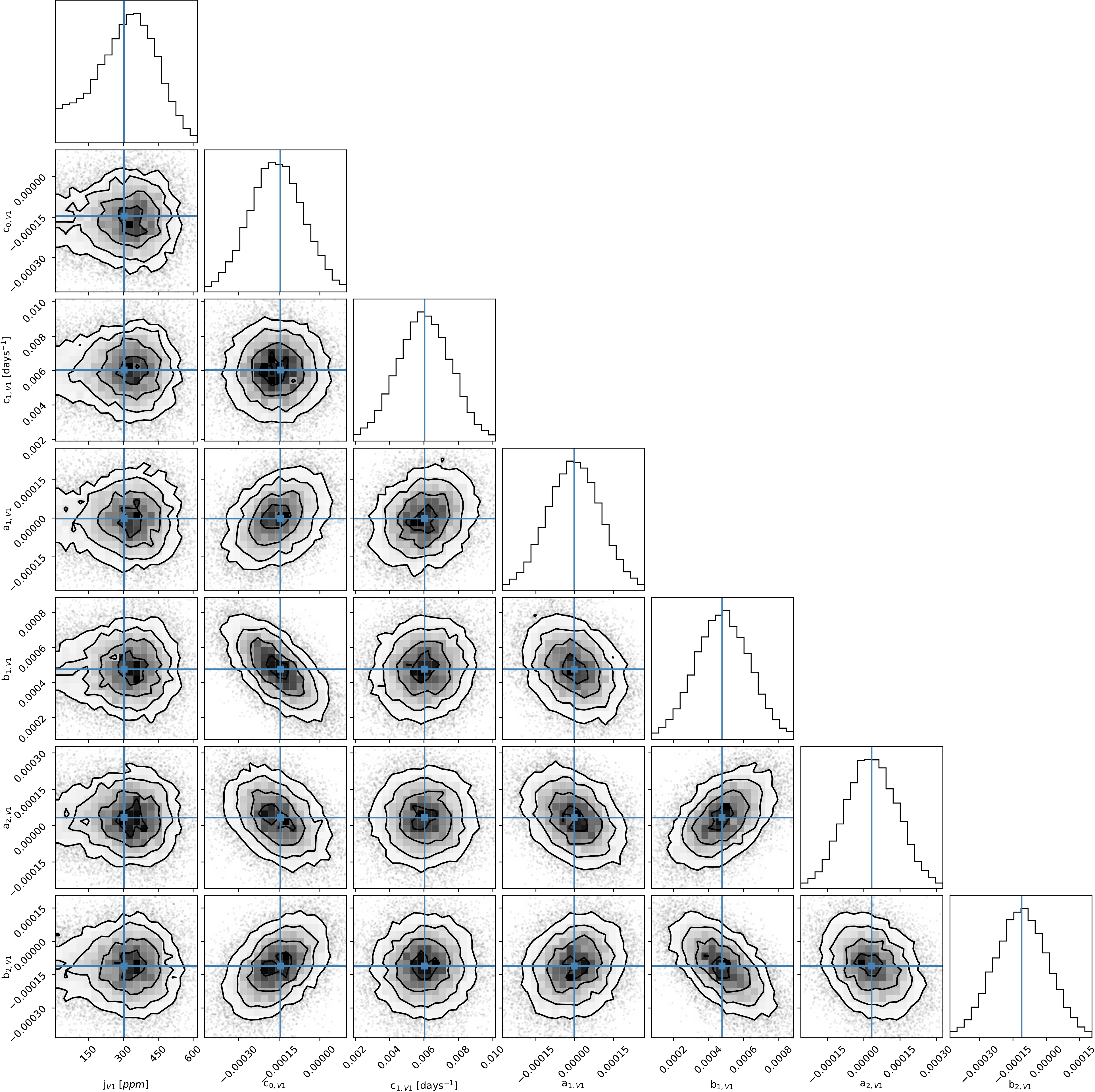}
    \caption{Corner plot of the MCMC chains of the jitter parameter, the linear trend coefficients and the instrumental decorrelation coefficients from the fit of the \cheops\ LCs (see Sect.~\ref{sec:CHEOPSfit}). In each panel, the solid lines mark the MAP values. For plotting purposes, only the coefficients related to visit V1 are shown. The corner plots for the other visits are shown in the next figures. The posterior distribution of $A_p$ is not shown here as it corresponds to what shown in Fig.~\ref{fig:doccultations}. }\label{fig:CHEOPSfit1}
\end{figure*}

\begin{figure*}
    \centering
    \includegraphics[width=\linewidth]{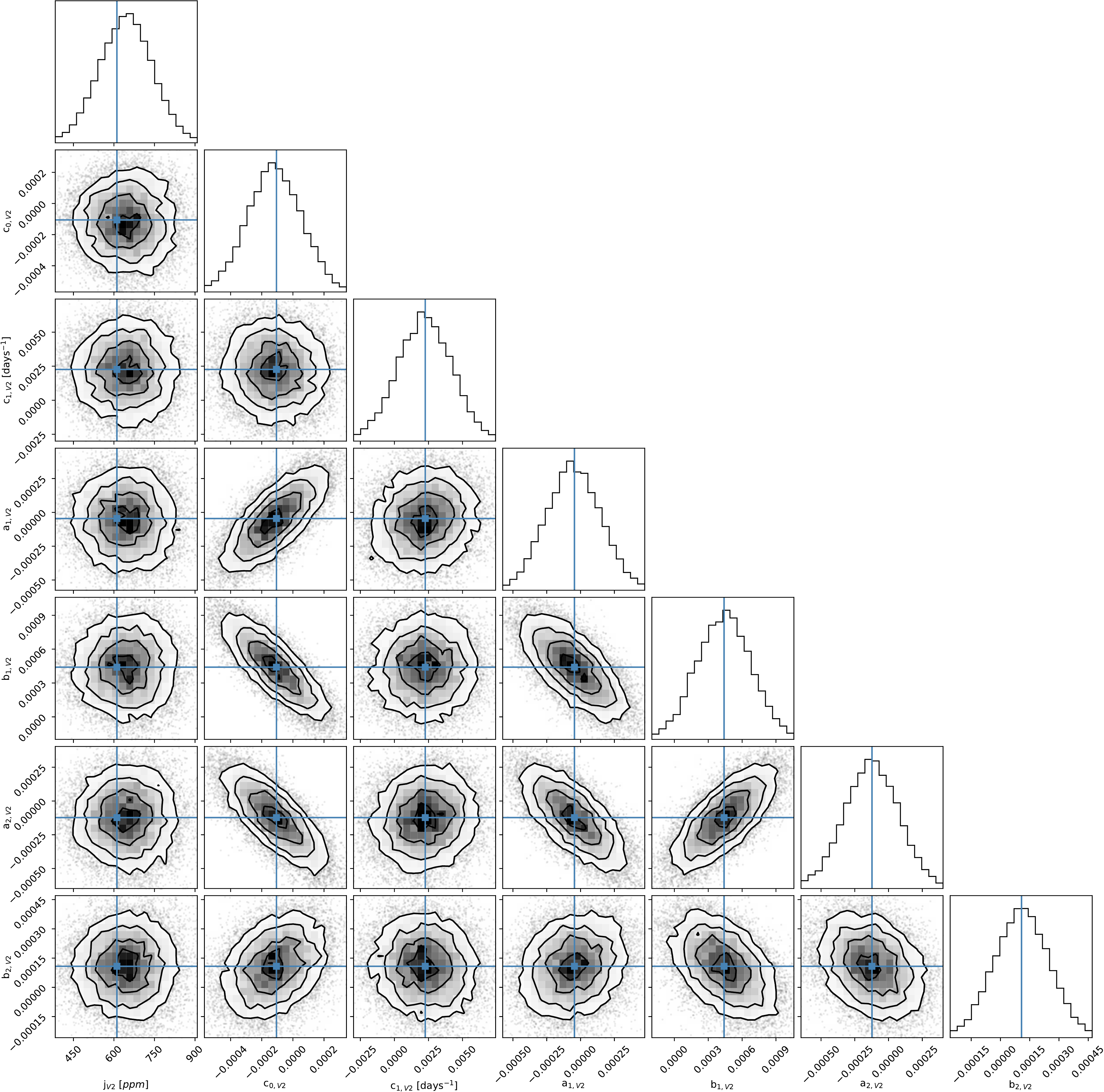}
    \caption{Same as in Fig.~\ref{fig:CHEOPSfit1} for \cheops\ visit V2.}\label{fig:CHEOPSfit2}
\end{figure*}

\begin{figure*}
    \centering
    \includegraphics[width=\linewidth]{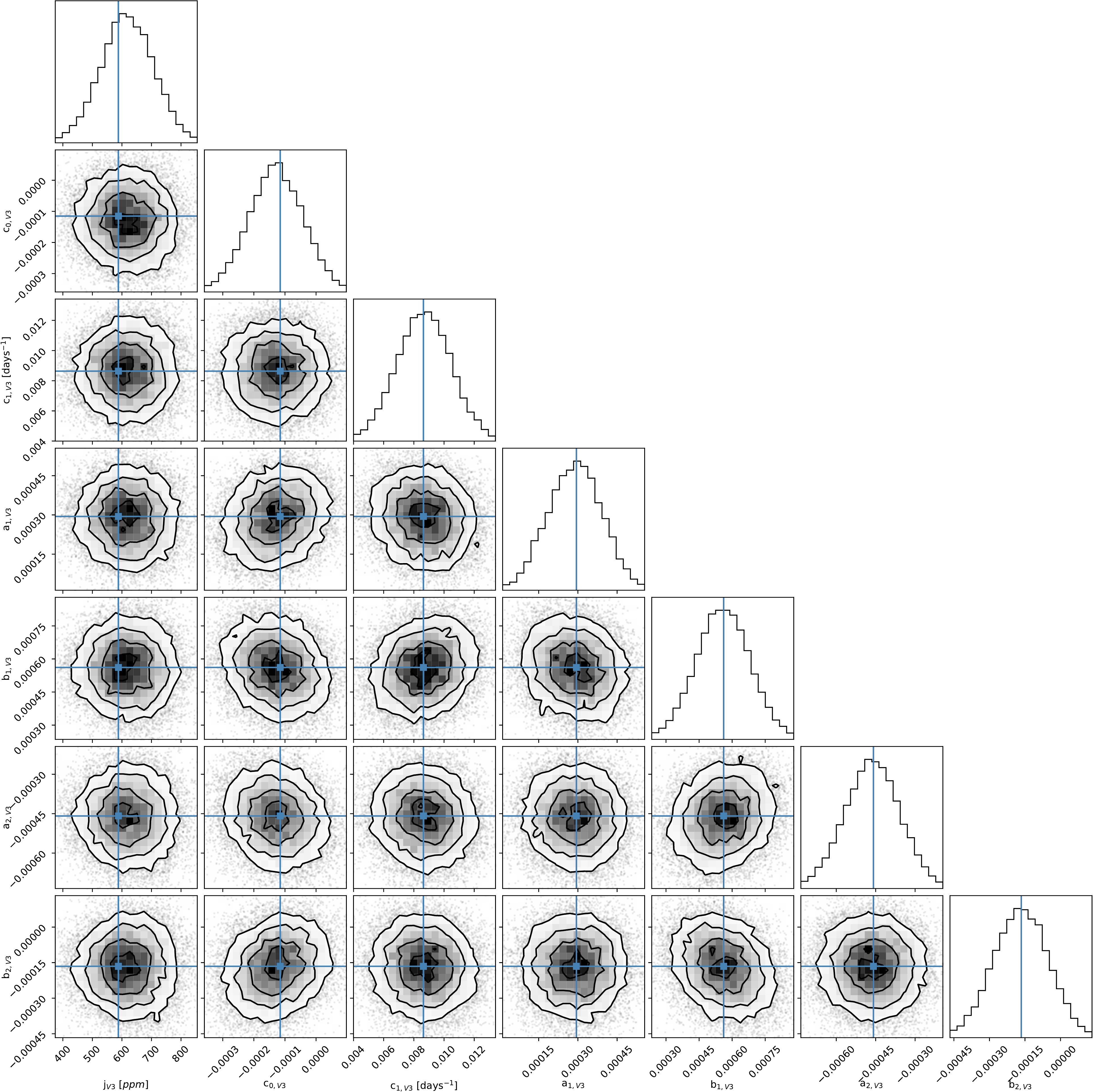}
    \caption{Same as in Fig.~\ref{fig:CHEOPSfit1} for \cheops\ visit V3.}\label{fig:CHEOPSfit3}
\end{figure*}

\begin{figure*}
    \centering
    \includegraphics[width=\linewidth]{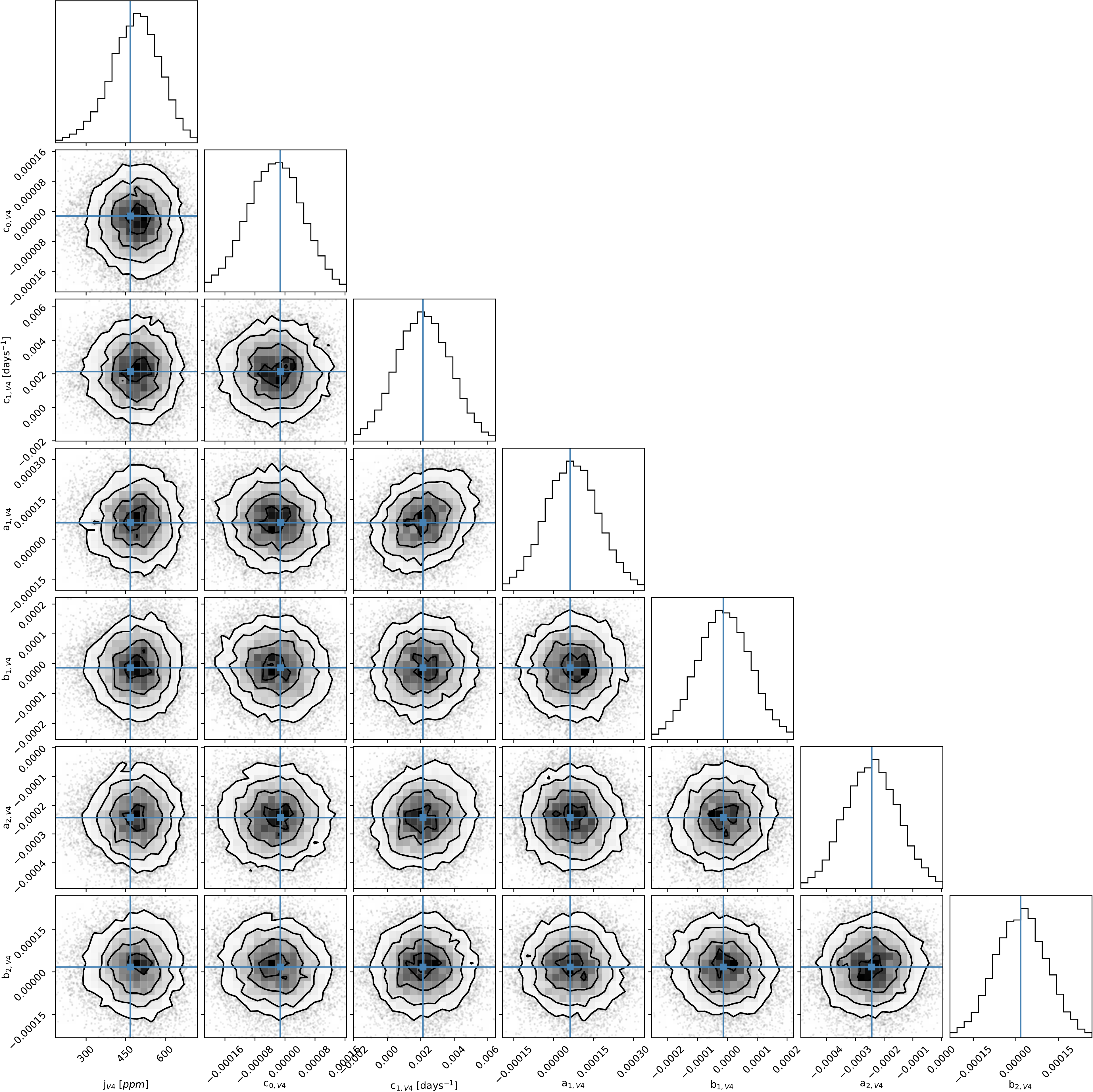}
    \caption{Same as in Fig.~\ref{fig:CHEOPSfit1} for \cheops\ visit V4.}\label{fig:CHEOPSfit4}
\end{figure*}

\begin{figure*}
    \centering
    \includegraphics[width=\linewidth]{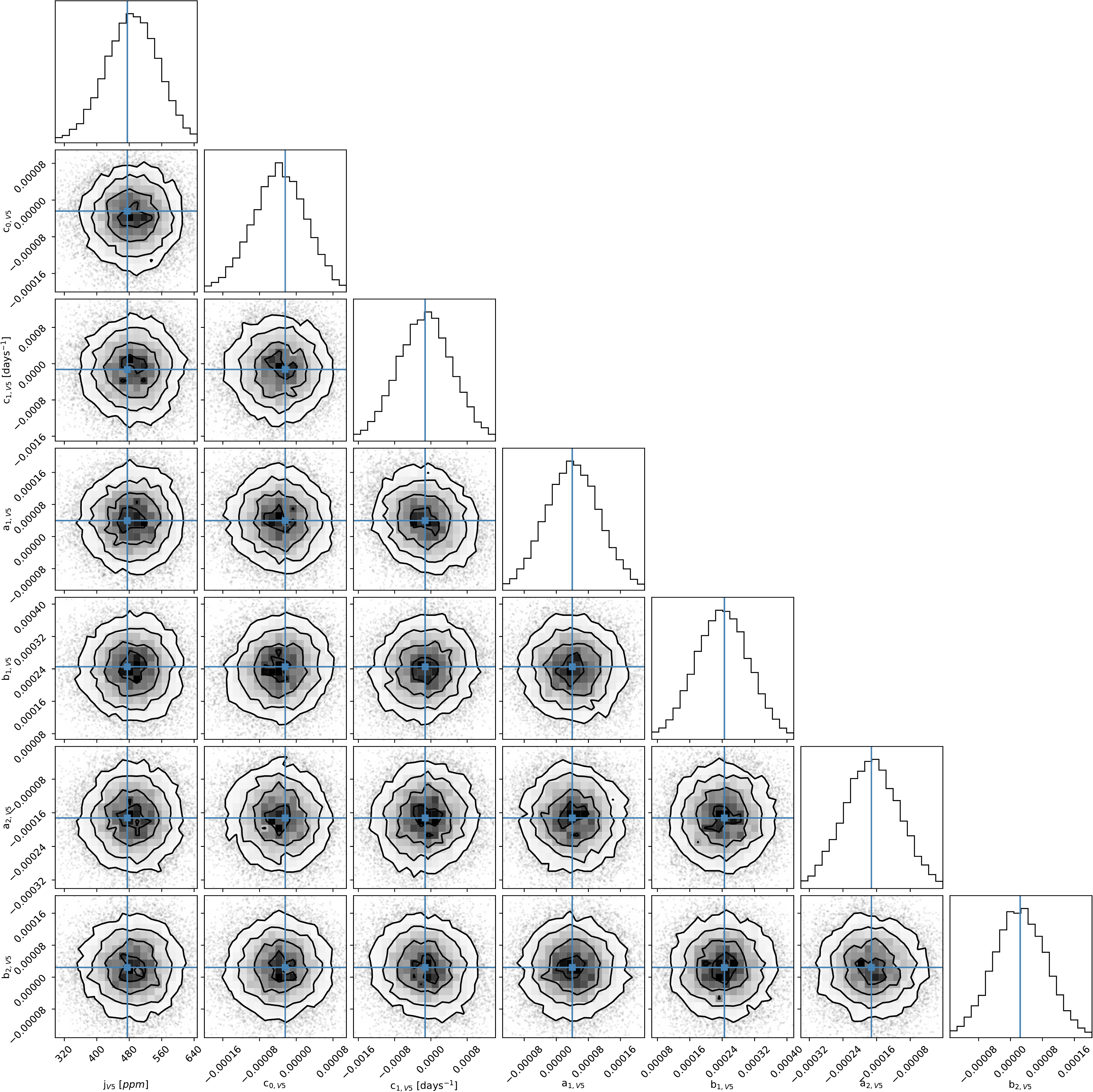}
    \caption{Same as in Fig.~\ref{fig:CHEOPSfit1} for \cheops\ visit V5.}\label{fig:CHEOPSfit5}
\end{figure*}

\begin{figure*}
    \centering
    \includegraphics[width=\linewidth]{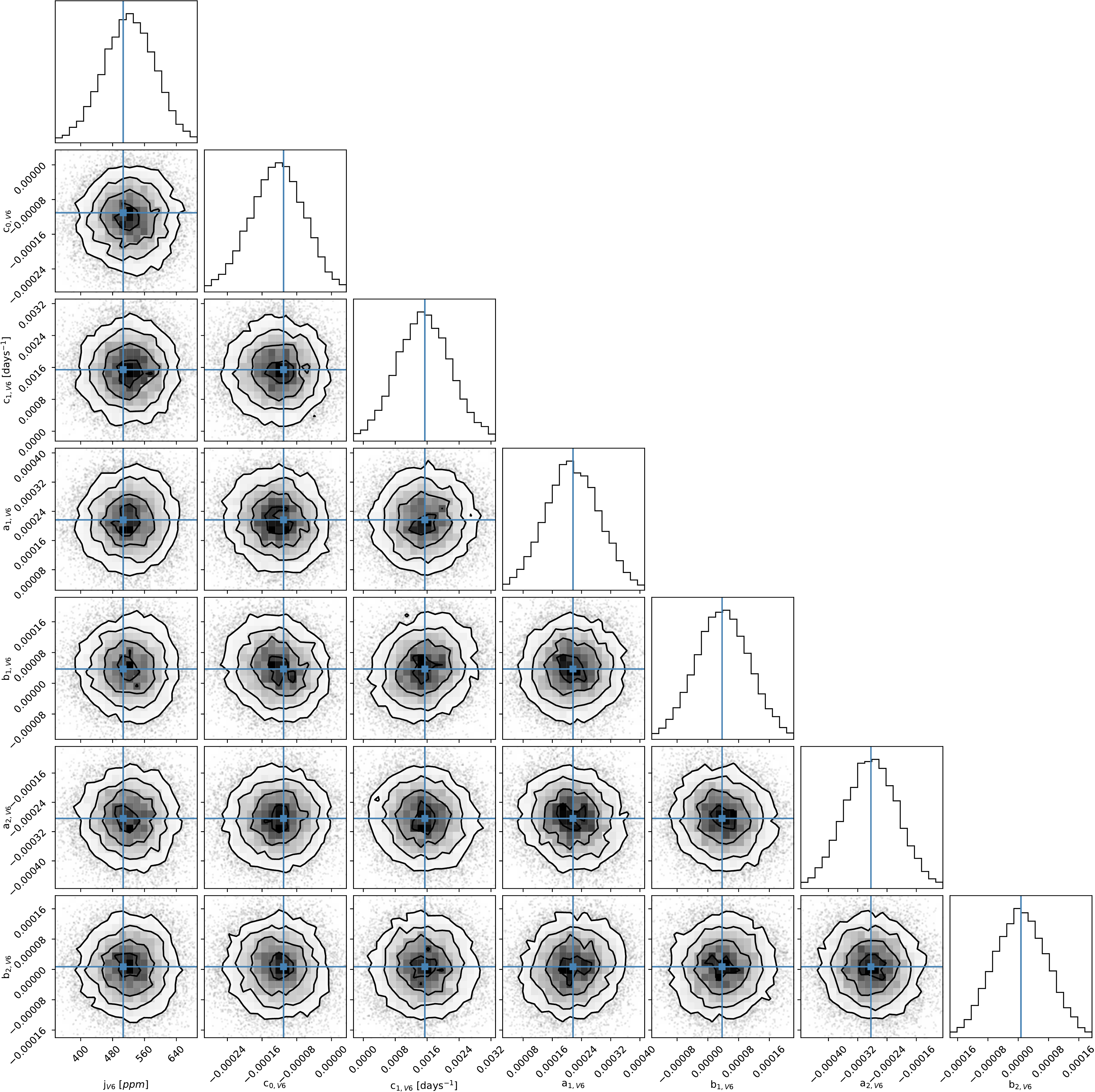}
    \caption{Same as in Fig.~\ref{fig:CHEOPSfit1} for \cheops\ visit V6.}\label{fig:CHEOPSfit6}
\end{figure*}

\begin{figure*}
    \centering
    \includegraphics[width=\linewidth]{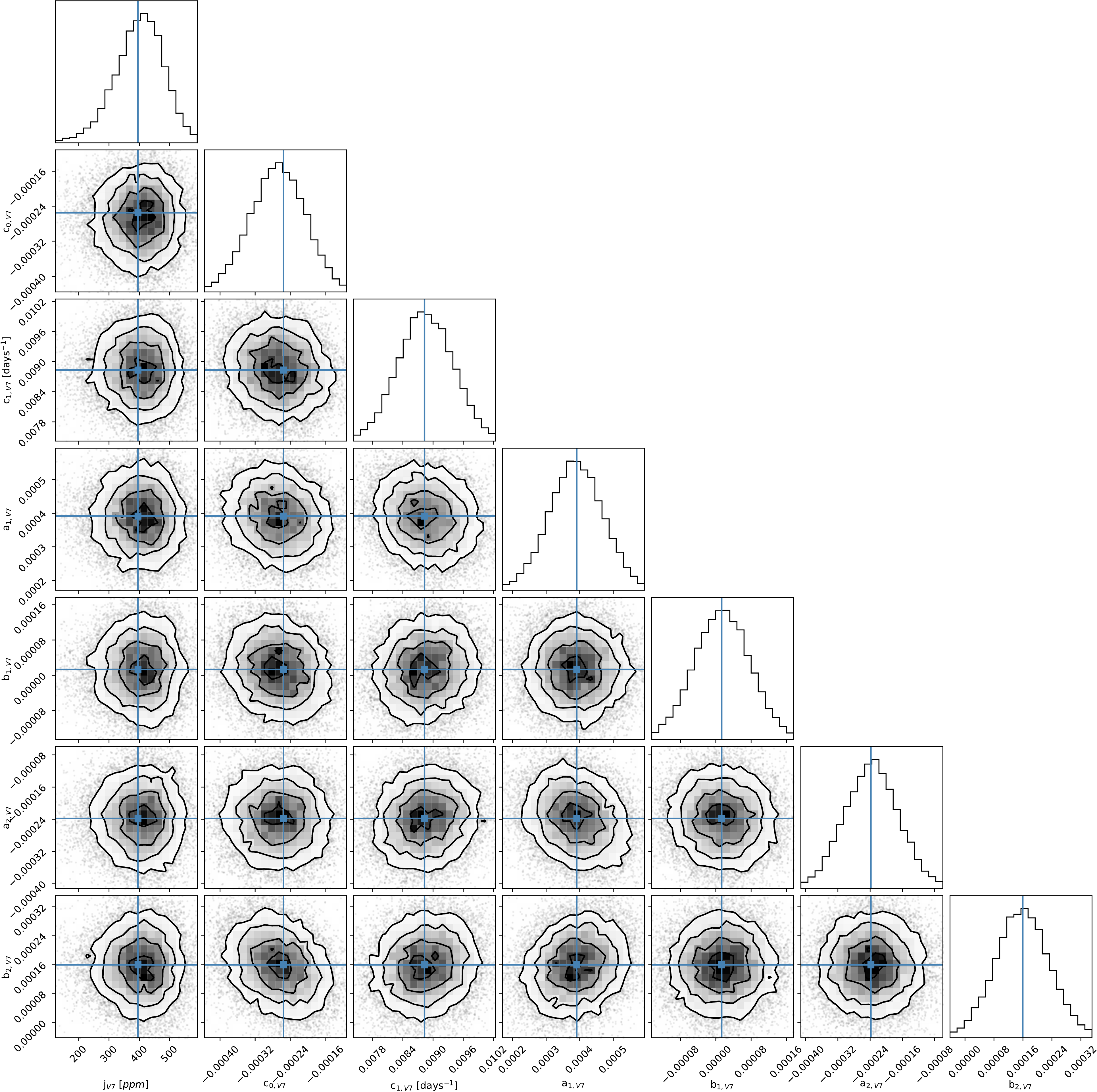}
    \caption{Same as in Fig.~\ref{fig:CHEOPSfit1} for \cheops\ visit V7.}\label{fig:CHEOPSfit7}
\end{figure*}

\begin{figure*}
    \centering
    \includegraphics[width=\linewidth]{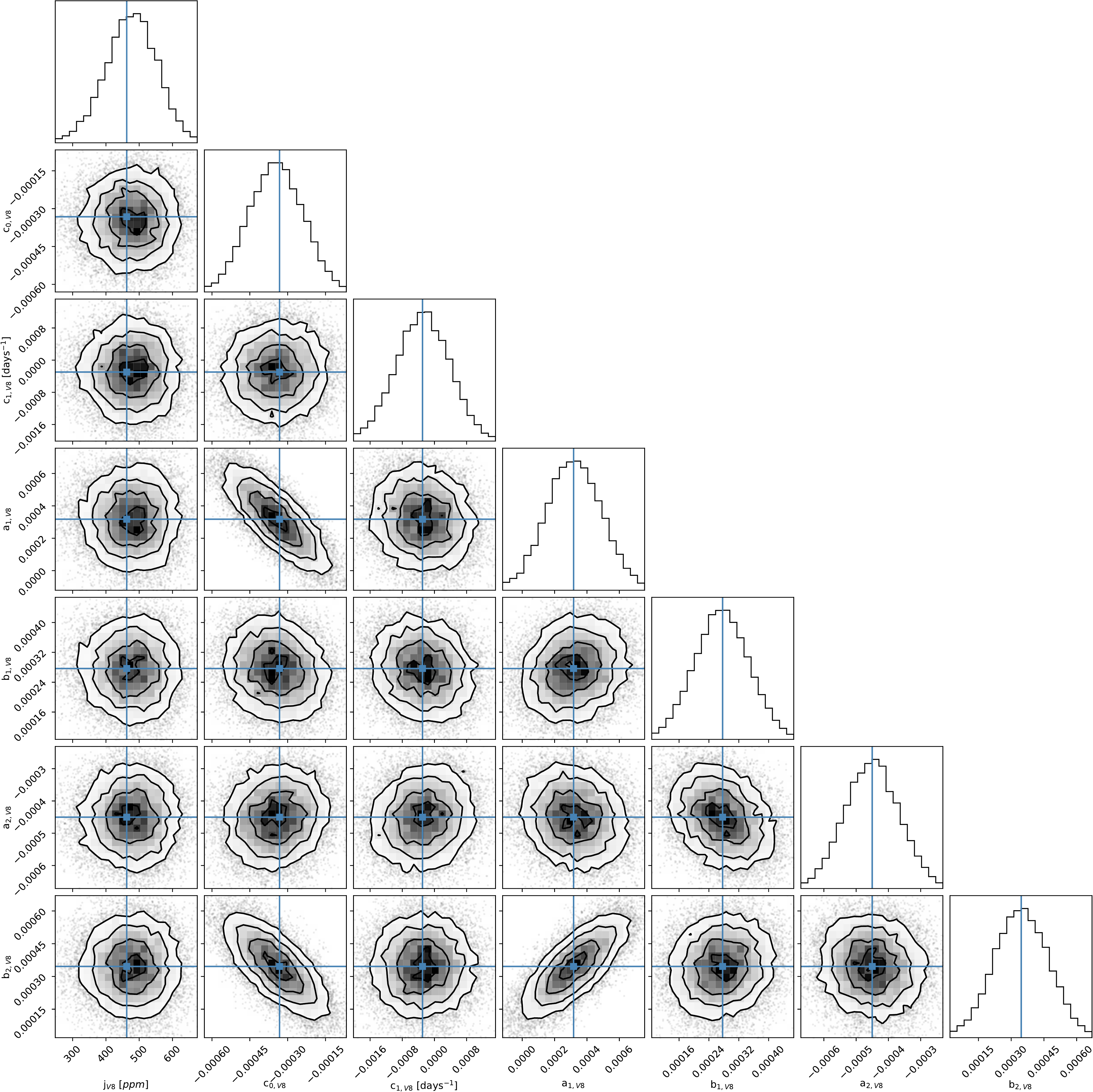}
    \caption{Same as in Fig.~\ref{fig:CHEOPSfit1} for \cheops\ visit V8.}\label{fig:CHEOPSfit8}
\end{figure*}

\begin{figure*}
    \centering
    \includegraphics[width=\linewidth]{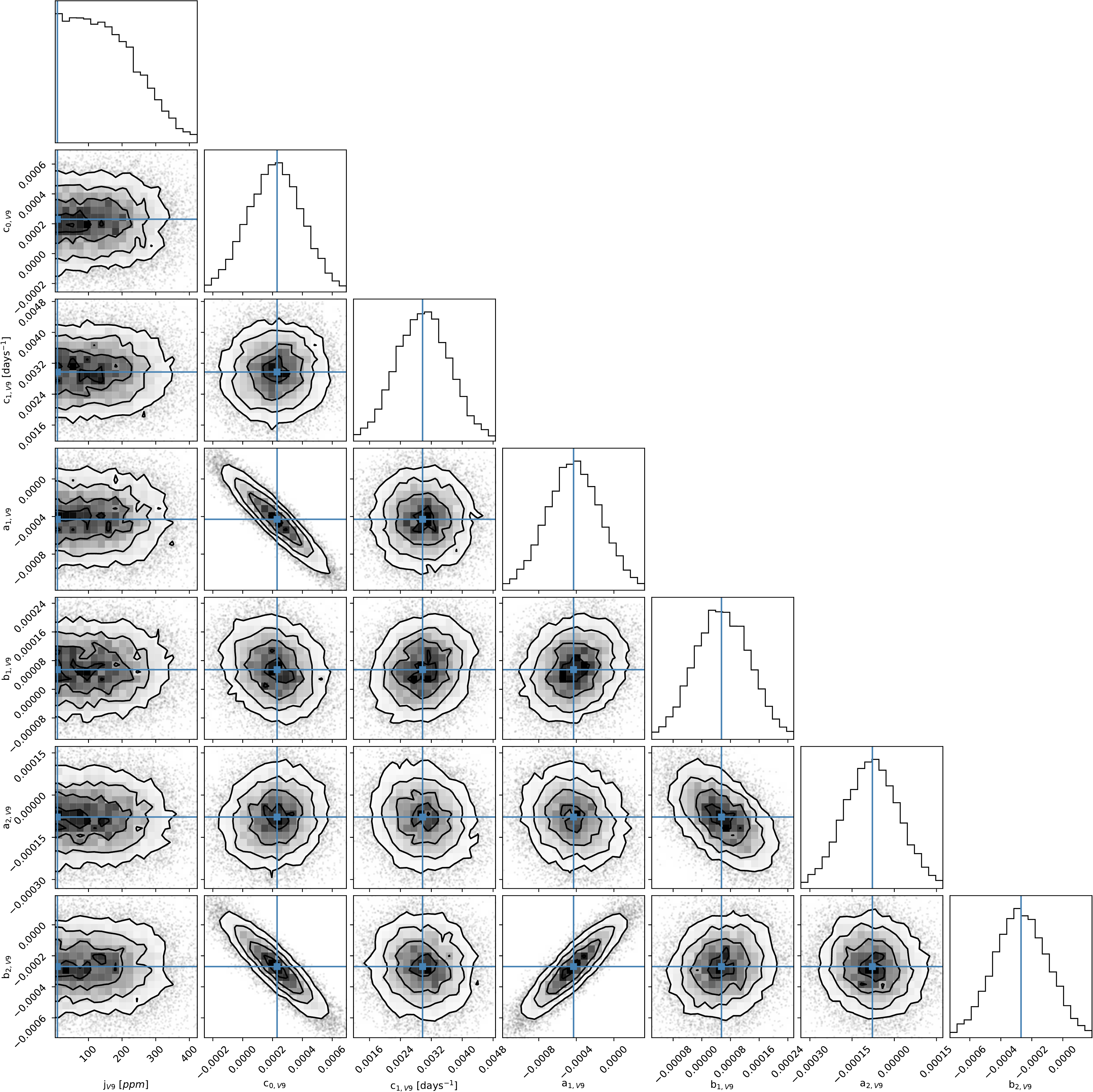}
    \caption{Same as in Fig.~\ref{fig:CHEOPSfit1} for \cheops\ visit V9.}\label{fig:CHEOPSfit9}
\end{figure*}

\begin{figure*}
    \centering
    \includegraphics[width=\linewidth]{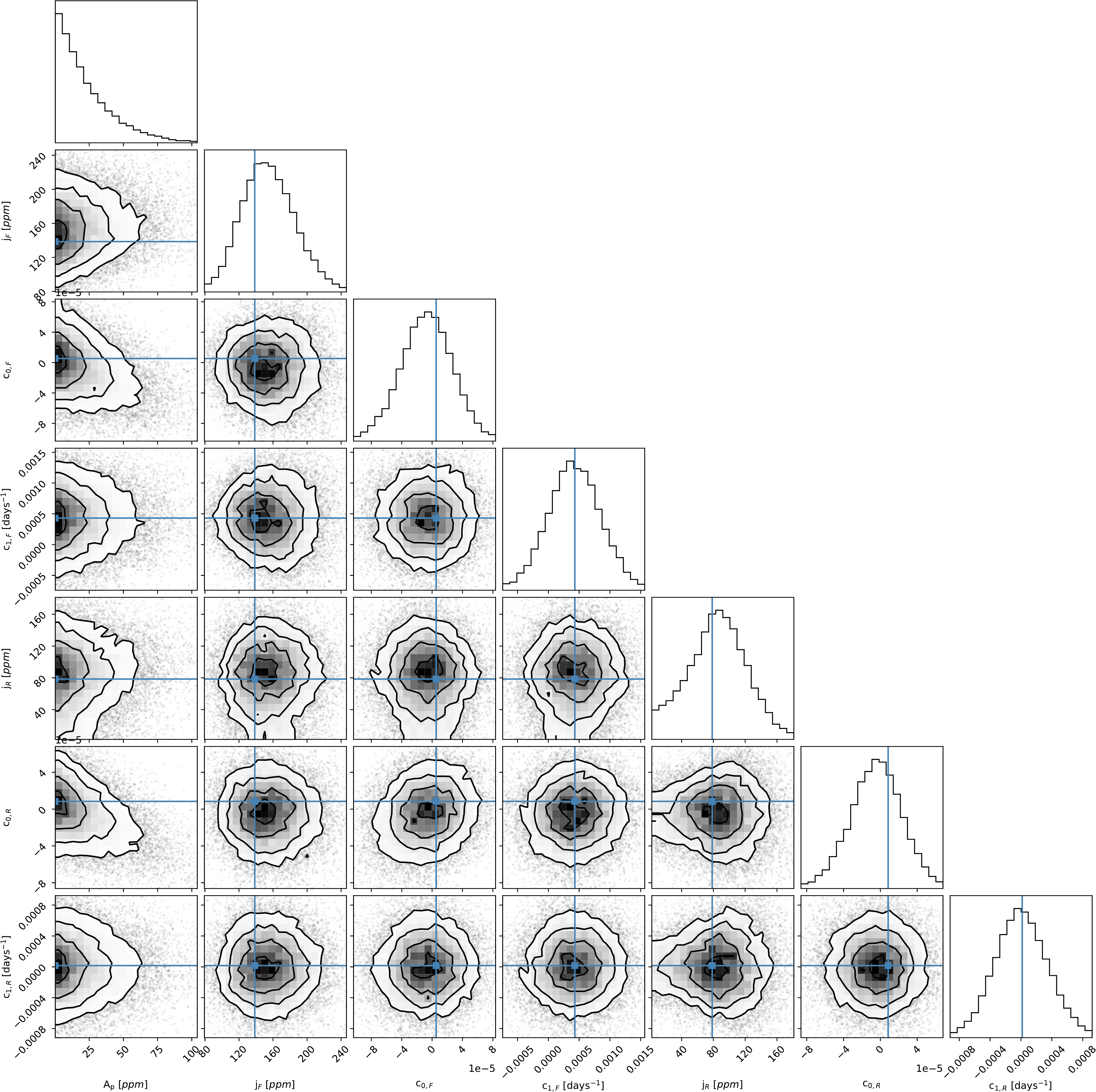}
    \caption{Corner plot of the MCMC chains from the fit of the \uvis\ LCs (see Sect.~\ref{sec:UVISfit}). In each panel, the solid lines mark the MAP values. For plotting purposes, only the coefficients related to visit V1 are shown. The corner plots for the other visits are shown in the next figures. The posterior distribution of $A_p$ is not shown here as it corresponds to what shown in Fig.~\ref{fig:doccultations}. }\label{fig:UVISfit}
\end{figure*}

In our framework (Sect.~\ref{sec:modeling}) $\delta_{\rm ecl}$ is not a free parameter but a quantity derived from the phase curve model. This is particularly important for the fit of the \tess\ LCs, where we allowed for a free phase offset $\Delta\phi$. As a matter of fact, a non zero phase offset makes the eclipse depth smaller than the overall amplitude $A_{\rm p}$ of the phase curve. Aiming at a fully Bayesian analysis of the planetary albedo, for each couple ($A_{\rm p}$, $\Delta\phi$) sampled in the MCMC fit of the LCs we derived the corresponding $\delta_{\rm ecl}$. For the case of \uvis\ and \cheops\ LCs, as we explain in Sect.~\ref{sec:CHEOPSfit}, we artificially fixed the phase offset to zero, which led to exact correspondence with the amplitude of the phase curve $A_{\rm p}$. In Fig.~\ref{fig:doccultations} we show the distribution of the eclipse depths for the \uvis, \cheops\ and \tess\ LCs respectively.

In the case of reflection, the eclipse depth can be expressed as:
\begin{equation}
\delta_{\rm ecl}=\frac{F_p}{F_\star}=A_g\left(\frac{R_p}{a}\right)^2,\label{eq:albedo}
\end{equation}
where \ag\ is the planetary geometric albedo \citep{Seager2010}. In this perspective, the eclipse depth is the observable directly linked the geometric albedo of the planet, but only if the thermal emission in the passband of interest can be either corrected or neglected.

To estimate the thermal emission in the \uvis, \cheops\ and \tess\ passbands we employed the \textsc{Helios-r2} Bayesian retrieval framework. This code was first introduced in \citet{Kitzmann2020} for studying emission spectra of brown dwarfs. It was upgraded in \citet{Wong2021} to perform retrievals of secondary eclipse observations as well. The forward model of \textsc{Helios-r2} calculates the planet's emission spectrum based on a number of free parameters and then converts the result into a secondary eclipse depth using a spectrum for the host star. Since some of the available observations are measured over wide bandpasses, \textsc{Helios-r2} also has the capability to use filter transmission functions to simulate observations in specific filters.

For the retrieval analysis, we used available infrared eclipse depth data summarized in Table~\ref{tab:depths}. Since the number of data points is limited and many observations are only available over wide bandpasses, we choose a rather idealized forward model that describes the atmosphere. As for the retrieval of brown dwarf emission spectra done in \citet{Kitzmann2020}, we described the temperature profile as a piece-wise polynomial. Here we used six first-order elements to parameterize the temperature profile as a function of pressure. {The atmosphere itself is parametrised with 70 computational layers, evenly distributed in logarithmic pressure space.} 
In contrast to the brown dwarf retrievals, however, the temperature profile is allowed to have inversions. The ability of \textsc{Helios-r2} to retrieve inverted profiles has already been demonstrated in \citet{Bourrier2020}.

For higher-quality data, the abundances of chemical species can usually be retrieved freely. For the measurements available for \wftb, a free chemistry approach, however, would be difficult and very degenerate since, with the exception of the HST WFC3/G141 data \citep{Kreidberg2014, Stevenson2014, Stevenson2017}, all observations have been performed over rather wider filter bandpasses. We, therefore, assumed that the atmosphere is in chemical equilibrium for simplification. This allowed us to describe the chemistry by a single free parameter - the overall metallicity [M/H] - instead of retrieving separate mixing ratios for all considered chemical species. To calculate the chemical composition during the evaluation of the forward model, we employed the ultra-fast equilibrium chemistry model \textsc{FastChem}\footnote{\url{https://github.com/exoclime/FastChem}} \citep{Stock2018} in its 2.1 version.

As opacity sources we included the following species: \ch{H2O}, CO, TiO, VO, SH, K, Na, \ch{H2S}, FeH, \ch{CH4}, \ch{CO2}, HCN, MgH, TiH, CrH, and CaH. Together with collision induced absorption of \ch{H2}-\ch{H2} and \ch{H2}-He pairs, this covers all major opacity sources in the wavelength region of the available measurements. We did not use the \ch{H-} continuum or other ions and atoms since their abundances can be considered to be small for the lower atmospheric temperatures of \wftb\ ($\approx$ 2000 K) compared to other exoplanets, like the ultra-hot jupiter KELT-9 b with an equilibrium temperature above 4000 K. As shown in \citet{Kitzmann2018}, the abundance of atoms and ions only strongly increases for temperatures above about 2500 K.

For the planet-to-star radius ratio we used the median posterior obtained fitting the \tess\ LCs (Table~\ref{tab:TESSfit}), while for the planet's surface gravity we used the empirically-derived best fit value combining the fitted orbital parameters (Table~\ref{tab:TESSfit}) as described in \citet{Southworth2007}. In total our retrieval has eight free parameters (seven parameters to describe the temperature profile and one for the chemical composition) which are summarized in Table \ref{tab:retrieval_parameters}. 

\begin{table}
\caption{List of free parameters and prior distributions used in the retrieval calculations.}             
\label{tab:retrieval_parameters}      
\centering          
\begin{tabular}{l l l}     
\hline\hline       
Parameter & \multicolumn{2}{c}{Prior} \\                     
          & Type   & Values\\
\hline
\textit{Temperature profile}   & \\
T$_1$                          & uniform       & 1000 -- 5000 K\\
b$_{i=1,\dots,6}$              & uniform       & 0.1 -- 2.0 \\
\hline
\textit{Equilibrium chemistry} & \\
$[\mathrm M/ \mathrm H]$       & uniform       & 0.1 -- 3\\
\hline                  
\end{tabular}
\end{table}

Our retrieval analysis only used the infrared measurements, which are unlikely to be affected by contributions of a geometric albedo, unless large cloud particles are present in the atmosphere.

\begin{figure}
    \centering
    \includegraphics[width=0.47\linewidth]{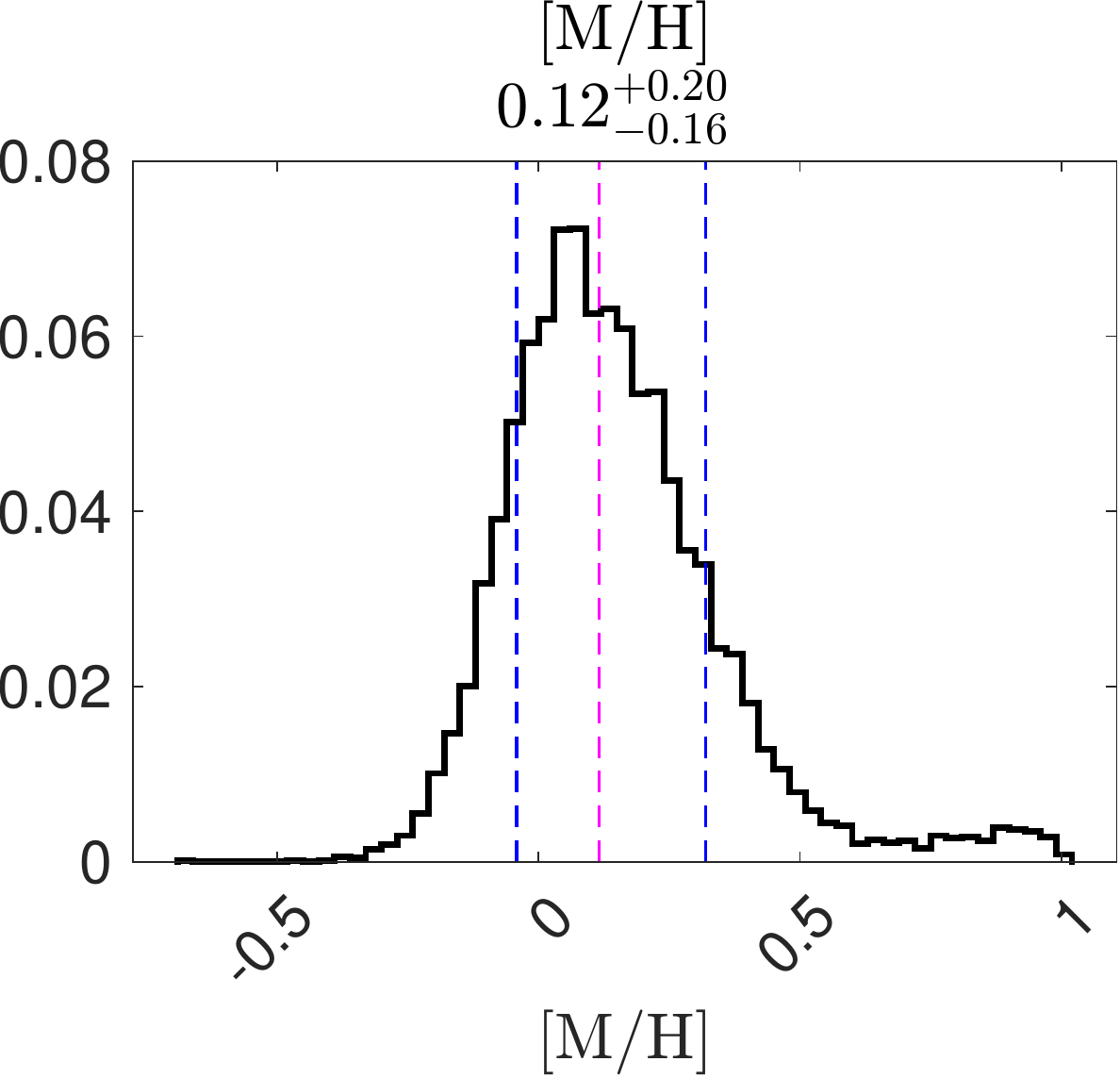}
    \hspace{0.01\linewidth}
    \includegraphics[width=0.47\linewidth]{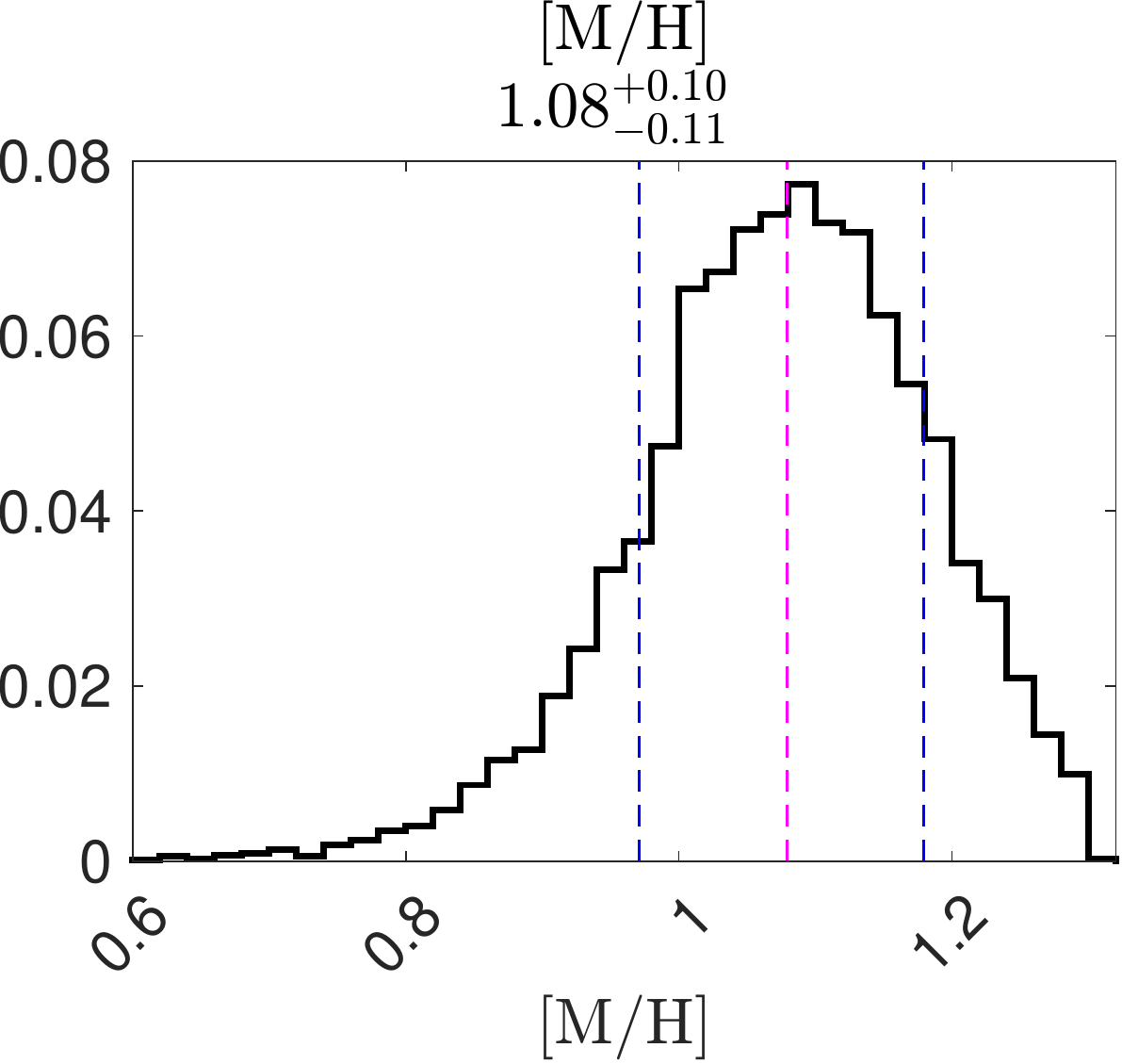}
    \includegraphics[width=0.49\linewidth]{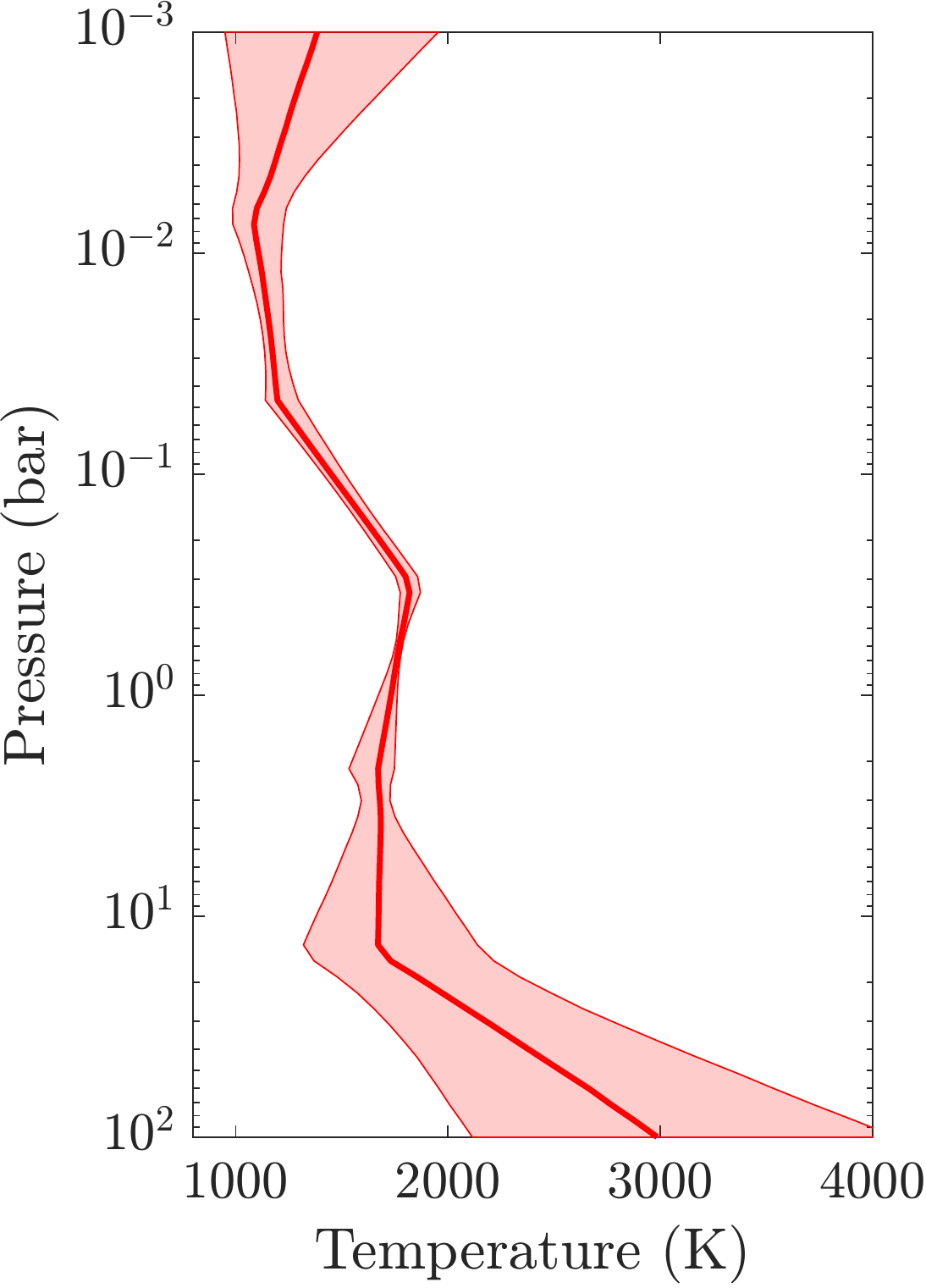}\includegraphics[width=0.49\linewidth]{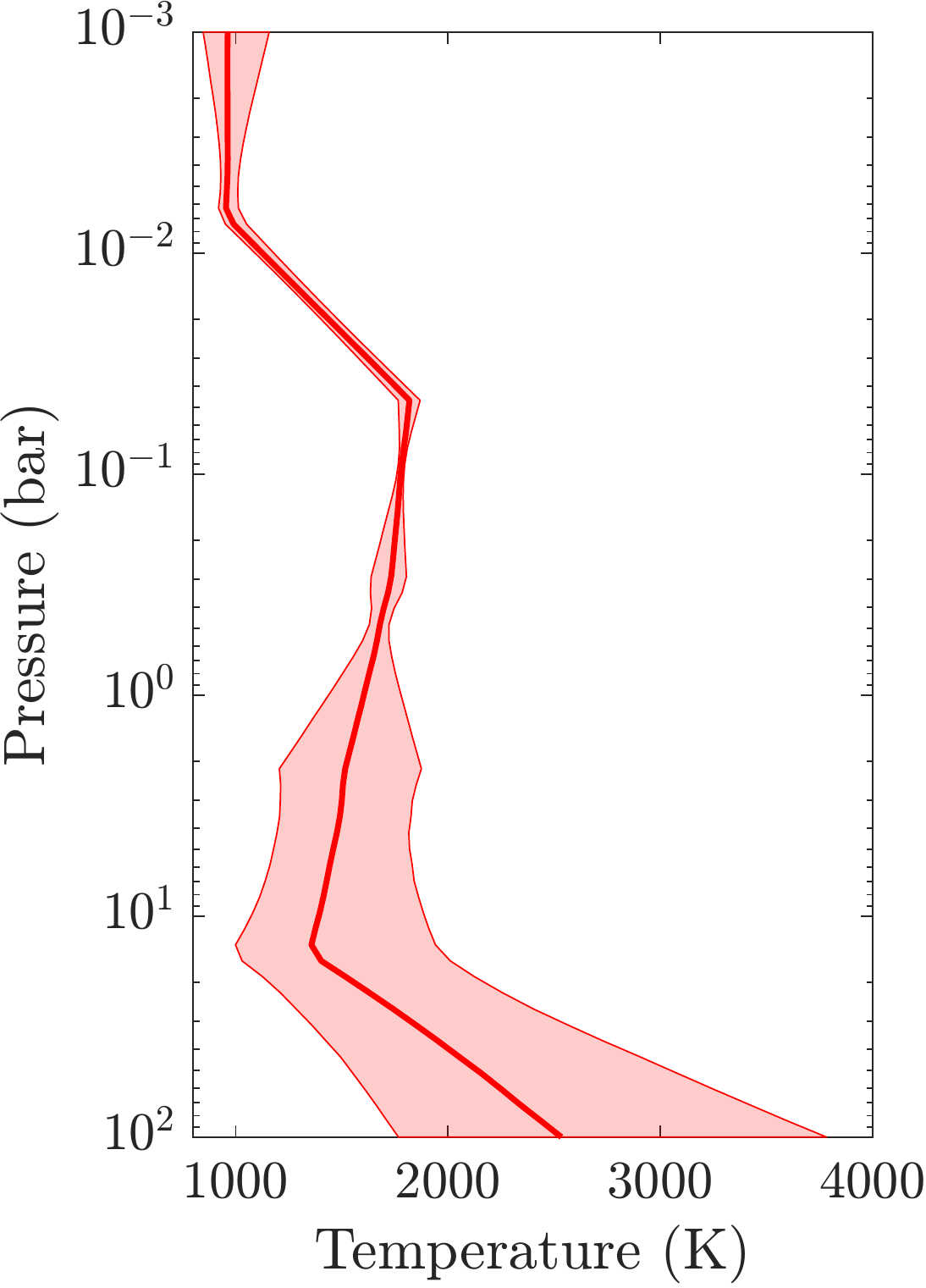}
    \caption{\textit{Left column - } Posterior distributions for the secondary-eclipse retrieval of \wftb. The upper panel shows the posterior for the metallicity $[M/H]$, where the vertical, magenta line indicates the median of the distribution and the blue lines the corresponding 1-$\sigma$ intervals. The lower panel depicts the retrieved temperature-pressure profile. The thick, red line is the median temperature profile, while the shaded, red area is the 1-$\sigma$ interval of the temperature distribution. \textit{Right column - }Same as in the left column for the high metallicity mode of the posterior distribution of [M/H].}
    \label{fig:retrieval_posteriors}
\end{figure}

{The resulting temperature structure and the posterior distribution of metallicity [M/H] are shown in Fig.~\ref{fig:retrieval_posteriors}. The results suggest a metallicity of $0.1\pm0.2$ that is consistent with {slightly enhanced} solar element abundances, in agreement with \citet{Stevenson2017} (0.3–1.7$\times$solar). This result is also consistent with the derived metallicity of the host star listed in Table \ref{tab:parameters}. We note, however, that the data also supports a bimodal posterior distribution of the metallicity, with an additional higher-metallicity solution of [M/H]$\approx 1.08$ or roughly more than ten times solar metallicity. The corresponding posterior distributions are shown in the left column of Fig.~\ref{fig:retrieval_posteriors}. Since the host star is close to solar metallicity, we decided to isolate the low-metallicity peak from the posterior as our preferred solution by choosing an appropriate prior.} 

The temperature-pressure profile depicted in the lower panel of Fig.~\ref{fig:retrieval_posteriors} is almost isothermal near pressures of 0.1 bar, without signs of a strong temperature inversion. This suggests that very strong shortwave absorbers like Fe, \ch{Fe+}, or TiO are not a dominant opacity source in this atmosphere. The tightest constraints on the temperature profile are obtained for pressures between 1 bar and 0.1 bar. In the lower and upper atmosphere, on the other hand, the temperature-pressure profile is essentially prior-dominated.

\begin{figure}
    \centering
    \includegraphics[width=\linewidth]{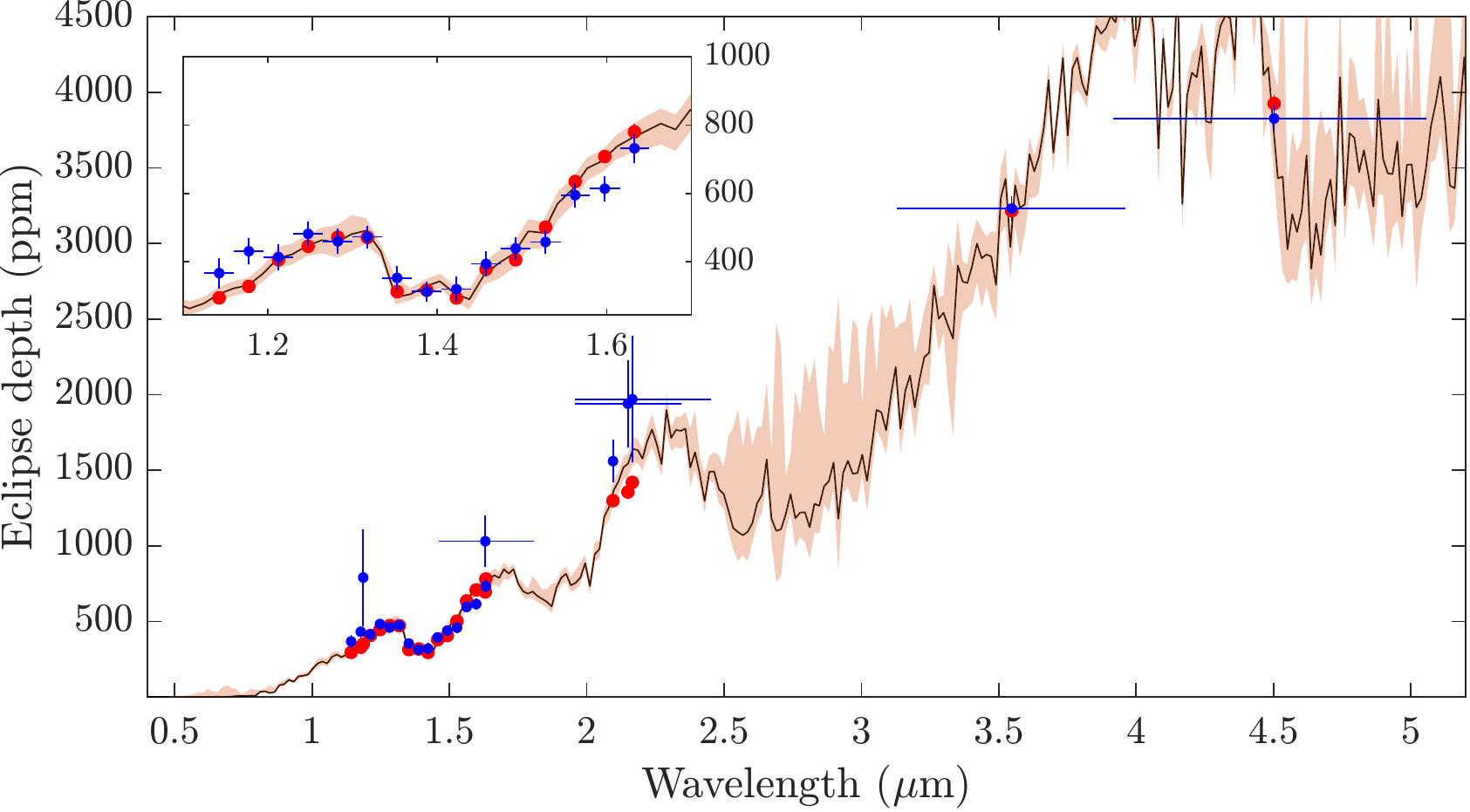}
    \includegraphics[width=\linewidth]{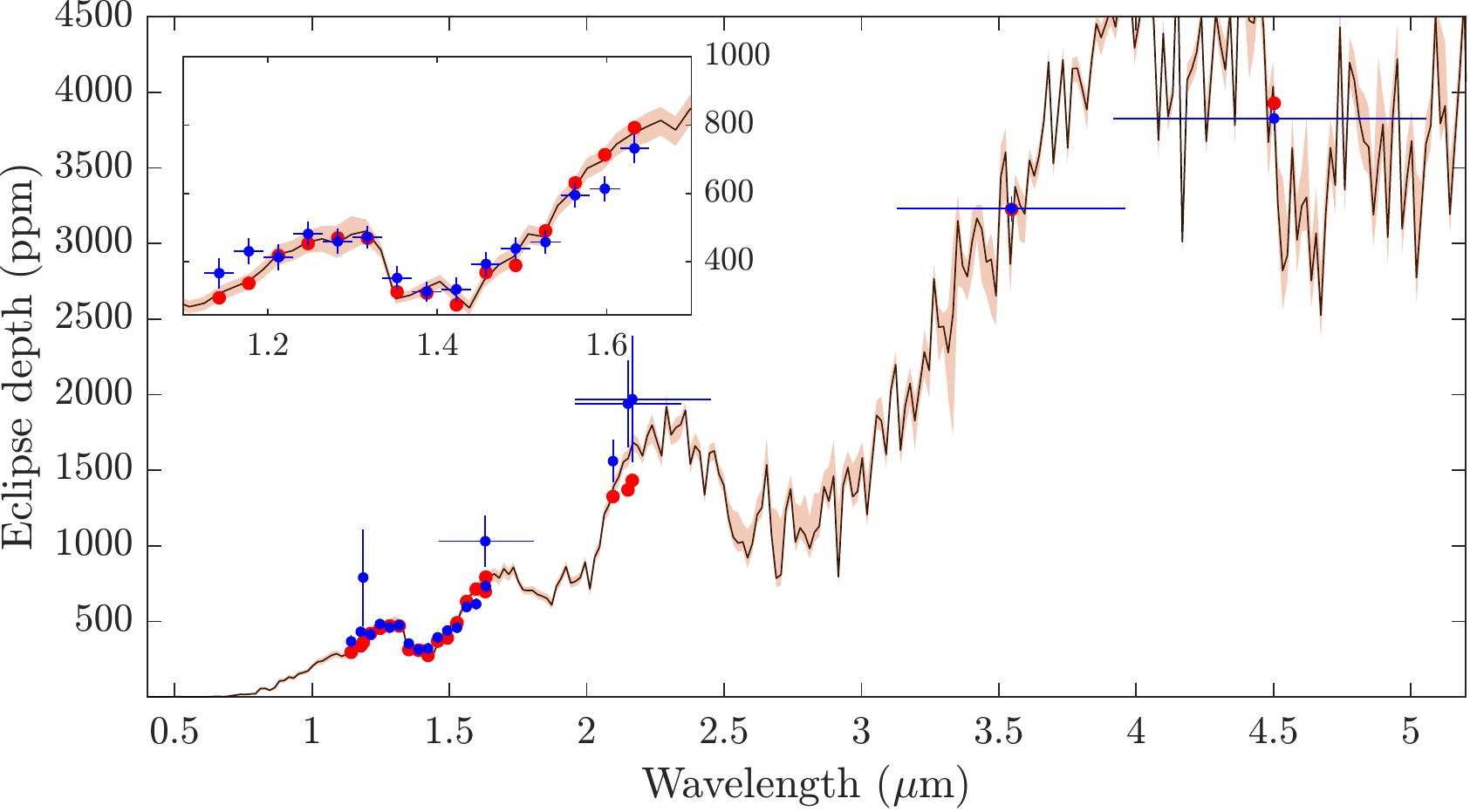}
    \caption{\textit{Top panel - }Posterior distributions of the spectra and bandpass-integrated eclipse depths in units of ppm of the retrieval. The black line indicates the median spectrum, while the red, shaded area is the 1-$\sigma$ interval. Blue points denote the measured eclipse depths with their corresponding error bars while red points are the median, bandpass-integrated values from the posterior spectra. Their red error bars again indicate their corresponding 3-$\sigma$ intervals. The inset plot shows a magnification of the HST WFC3/G141 data. \textit{Bottom panel - } Same as in the top panel for the high metallicity mode returned by our atmospheric retrieval.}
    \label{fig:retrieval_post_spectra}
\end{figure}

Figure \ref{fig:retrieval_post_spectra} shows the posterior eclipse depths for all bandpasses that have been included in the retrieval. The inset plot is a magnification of the HST WFC3/G141 data. Due to the lower error bars of the space-based data (HST WFC3/G141 and Spitzer), the retrieval is mostly driven by their reported eclipse depth. The ground-based data, on the other hand, seems to have a consistent shift towards higher eclipse depths compared to those obtained from the space telescopes. Due to their much larger errors, however, their impact on the resulting posteriors is only very minor. 

In Fig.~\ref{fig:retrieval_post_spectra} we also plot, for comparison, the retrieved atmospheric model corresponding to the high metallicity mode in the $[M/H]$ posterior distribution. The two models, which are almost indistinguishable, fit all the data equally well, explaining the bimodality in the posterior distribution. At the current level of data precision we cannot a priori distinguish between the two solutions. We prefer the low metallicity mode based on the expectation that the metallicity of the planet should not significantly differ from that of its host star.

\begin{table*}
\begin{center}
\caption{Eclipse depth measurements of \wftb.}\label{tab:depths}
\begin{tabular}{lll}
\hline\hline
Passband or wavelength ($\mu$m) & Depth (ppm) & Ref.\\
\hline
UVIS/F350LP & <140 & this work\\
CHEOPS & 80$^{+60}_{-50}$ & this work\\
TESS & 70$^{+50}_{-40}$ & this work\\
1.186 & 790$\pm$320 & \citet{Gillon2012}\\
2.095 & 1560$\pm$140 & \citet{Gillon2012}\\
H & 1030$\pm$170 & \citet{Wang2013}\\
K$_S$ & 1940$\pm$290 & \citet{Wang2013}\\
GROND K & 1970$\pm$420 & \citet{Chen2014}\\
1.1425 & 367$\pm$45 & \citet{Kreidberg2014,Stevenson2014,Stevenson2017}\\
1.1775 & 431$\pm$39 & \citet{Kreidberg2014,Stevenson2014,Stevenson2017}\\
1.2125 & 414$\pm$38 & \citet{Kreidberg2014,Stevenson2014,Stevenson2017}\\
1.2475 & 482$\pm$36 & \citet{Kreidberg2014,Stevenson2014,Stevenson2017}\\
1.2825 & 460$\pm$37 & \citet{Kreidberg2014,Stevenson2014,Stevenson2017}\\
1.3175 & 473$\pm$33 & \citet{Kreidberg2014,Stevenson2014,Stevenson2017}\\
1.3525 & 353$\pm$34 & \citet{Kreidberg2014,Stevenson2014,Stevenson2017}\\
1.3875 & 313$\pm$30 & \citet{Kreidberg2014,Stevenson2014,Stevenson2017}\\
1.4225 & 320$\pm$36 & \citet{Kreidberg2014,Stevenson2014,Stevenson2017}\\
1.4575 & 394$\pm$36 & \citet{Kreidberg2014,Stevenson2014,Stevenson2017}\\
1.4925 & 439$\pm$33 & \citet{Kreidberg2014,Stevenson2014,Stevenson2017}\\
1.5275 & 458$\pm$35 & \citet{Kreidberg2014,Stevenson2014,Stevenson2017}\\
1.5625 & 595$\pm$36 & \citet{Kreidberg2014,Stevenson2014,Stevenson2017}\\
1.5975 & 614$\pm$37 & \citet{Kreidberg2014,Stevenson2014,Stevenson2017}\\
1.6325 & 732$\pm$42 & \citet{Kreidberg2014,Stevenson2014,Stevenson2017}\\
Spitzer/IRAC3.6 & 3231$\pm$60 & \citet{Stevenson2017}\\
Spitzer/IRAC4.5 & 3827$\pm$84 & \citet{Stevenson2017}\\
\hline
\end{tabular}
\end{center}
\end{table*}

In a post-process procedure of the retrieval, we estimated the predicted thermal emission in all short-wave bandpasses to decontaminate the measured eclipse depths. In particular, we computed the contribution to the eclipse depth in the \uvis, \cheops\ and \tess\ LCs from the thermal emission by integrating the emission spectrum in the corresponding passbands. We did this computation for each step sampled by the retrieval analysis, thus obtaining for each instrument a population of contamination estimates that follows the posterior distribution of the expected thermal emission in the corresponding passband (Fig.~\ref{fig:doccultations}). Afterwards, we randomly extracted an equally long subset of samples from the posterior distributions of the observed eclipse depth in the three passbands. Finally, for each instrument we decontaminated the observed eclipse depths by subtracting, sample by sample, the corresponding contamination from thermal emission. We thus obtained the reflection-only depths and the corresponding \ag\ by means of Eq.~\ref{eq:albedo}. The posterior distributions of the optical geometric albedos \ag\ in the three passbands are shown in Fig.~\ref{fig:doccultations}: they are consistent among themselves and indicate that \wftb\ has a low geometric albedo. To get the most precise upper limit on \ag\, we assumed no wavelength-dependence for \ag\ and multiply the three posterior distributions obtaining the combined posterior, shown in black in Fig.~\ref{fig:doccultations} after normalization. Following this last distribution function we can state that the albedo of the planet is lower than 0.087 at 99.9\% confidence. { This result is broadly consistent with the theoretical predictions of \citet{Sudarsky2000} and with ensembles of optical geometric albedos measured by Kepler \citep{Heng2013} and \tess\ \citep{Wong2021}.}

\section{Conclusions}\label{sec:conclusion}

In this work we provide a thorough analysis of the \wft\ system. Using publicly available spectra we carried out a detailed spectroscopic characterization of the host star, confirming it as a late-K main sequence dwarf with solar metallicity and slow rotation, the latter supported also by spectroscopic and photometric activity indicators.

For a detailed characterization of the exoplanet, we analyzed the publicly available \uvis\ and \tess\ photometry, together with dedicated \cheops\ observations of eleven individual occultations. We retrieved the transit, occultation and phase curve shapes while jointly modeling systematics (stellar and instrumental), ellipsoidal variations, gravity darkening and Doppler boosting.

We obtained a tentative detection of $A_p=100\pm50$ ppm (Table~\ref{tab:TESSfit}) of the phase curve in the \tess\ data, and we found marginal evidence for a $\left(50^{+30}_{-20}\right)^\circ$ eastward phase offset (that is a peak before occultation). Previous infra-red observations also detected similar eastward peaks \citep{Stevenson2014,Stevenson2017,Morello2019}. The eastward offsets seen in the infra-red are in line with atmospheric circulation models of hot Jupiters that predict advection of hot material by means of equatorial jet streams \citep{Showman2002}. In this paradigm however, reflection-dominated phase curves are expected to peak after occultation, as reflective clouds are advected torwards the day side via the morning terminator \citep{Esteves2015}. { \wftb\ however might deviate from this paradigm as the object is predicted to form clouds at all longitudes and at optical depths which cannot be probed by the data analyzed in this work \citep{Helling2020,Venot2020}}. A local peak in reflectivity could be explained if local thermal and chemical conditions at the evening terminator are conducive to cloud condensation as well as cloud retention at observable altitudes.

Using all available eclipse depths in the infra-red, we perform a comprehensive modeling of the dayside atmosphere of \wftb. Our retrieval indicates that the metallicity of the atmosphere [M/H]=$0.1\pm$0.2 is consistent with the stellar counterpart, and compatible with previous estimates by \citet{Stevenson2017}. 
The retrieval also suggests a non-inverted pressure-temperature profile at the sub-stellar point, which is common for mildly-irradiated hot Jupiters \citep[e.g.][]{DiamondLowe2014}.

The model inferred from the infrared eclipses allowed us to extrapolate the thermal emission spectrum to optical wavelengths. We thus estimated the thermal emission contamination in the \uvis, \cheops\ and \tess\ passbands and decontaminated the observed eclipse depths. This allowed us to put an upper limit to the geometric albedo \ag\ of the planet of 0.087 with a 99.9\% confidence level, in agreement with \citet{Blazek2022}. \wftb\ is thus quite similar in this regard to other \lq\lq dark mirrors\rq\rq\ such as, for example, TrES-2~b \citep[$A_{\rm g}=0.0136^{+0.0022}_{-0.0033}$,][]{Barclay2012}, WASP-104~b \citep[$A_{\rm g}<0.03$,][]{Mocnic2018} and 51~Peg~b \citep[$A_{\rm g}<0.20$,][]{Scandariato2021, Spring2022}. As discussed in, for example, \citet{Marley2013cctp.book..367M}, the potential formation of high-temperature condensates for a solar-composition atmosphere starts around 2000~K, depending on pressure. Our inferred, high temperatures close to 2000~K at pressures between 1 bar and 0.1 bar on the dayside of \wftb, therefore, suggests that clouds are potentially either absent or form at pressures higher than those probed by our measurements. The low dayside albedos also indicate the absence of hazes that might form at higher altitudes.

\begin{acknowledgements}

CHEOPS is an ESA mission in partnership with Switzerland with important contributions to the payload and the ground segment from Austria, Belgium, France, Germany, Hungary, Italy, Portugal, Spain, Sweden, and the United Kingdom. The CHEOPS Consortium would like to gratefully acknowledge the support received by all the agencies, offices, universities, and industries involved. Their flexibility and willingness to explore new approaches were essential to the success of this mission. 
LBo, GBr, VNa, IPa, GPi, RRa, GSc, VSi, and TZi acknowledge support from CHEOPS ASI-INAF agreement n. 2019-29-HH.0. 
ML acknowledges support of the Swiss National Science Foundation under grant number PCEFP2\_194576. 
ABr was supported by the SNSA. 
PM acknowledges support from STFC research grant number ST/M001040/1. 
MF and CMP gratefully acknowledge the support of the Swedish National Space Agency (DNR 65/19, 174/18). 
SH gratefully acknowledges CNES funding through the grant 837319. 
V.V.G. is an F.R.S-FNRS Research Associate. 
ACC and TW acknowledge support from STFC consolidated grant numbers ST/R000824/1 and ST/V000861/1, and UKSA grant number ST/R003203/1. 
YA and MJH acknowledge the support of the Swiss National Fund under grant 200020\_172746. 
We acknowledge support from the Spanish Ministry of Science and Innovation and the European Regional Development Fund through grants ESP2016-80435-C2-1-R, ESP2016-80435-C2-2-R, PGC2018-098153-B-C33, PGC2018-098153-B-C31, ESP2017-87676-C5-1-R, MDM-2017-0737 Unidad de Excelencia Maria de Maeztu-Centro de Astrobiologí­a (INTA-CSIC), as well as the support of the Generalitat de Catalunya/CERCA programme. The MOC activities have been supported by the ESA contract No. 4000124370. 
S.C.C.B. acknowledges support from FCT through FCT contracts nr. IF/01312/2014/CP1215/CT0004. 
XB, SC, DG, MF and JL acknowledge their role as ESA-appointed CHEOPS science team members. 
ACC acknowledges support from STFC consolidated grant numbers ST/R000824/1 and ST/V000861/1, and UKSA grant number ST/R003203/1. 
This project was supported by the CNES. 
The Belgian participation to CHEOPS has been supported by the Belgian Federal Science Policy Office (BELSPO) in the framework of the PRODEX Program, and by the University of Liège through an ARC grant for Concerted Research Actions financed by the Wallonia-Brussels Federation. 
L.D. is an F.R.S.-FNRS Postdoctoral Researcher. 
This work was supported by FCT - Fundação para a Ciência e a Tecnologia through national funds and by FEDER through COMPETE2020 - Programa Operacional Competitividade e Internacionalizacão by these grants: UID/FIS/04434/2019, UIDB/04434/2020, UIDP/04434/2020, PTDC/FIS-AST/32113/2017 \& POCI-01-0145-FEDER- 032113, PTDC/FIS-AST/28953/2017 \& POCI-01-0145-FEDER-028953, PTDC/FIS-AST/28987/2017 \& POCI-01-0145-FEDER-028987, O.D.S.D. is supported in the form of work contract (DL 57/2016/CP1364/CT0004) funded by national funds through FCT. 
B.-O.D. acknowledges support from the Swiss National Science Foundation (PP00P2-190080). 
DG gratefully acknowledges financial support from the CRT foundation under Grant No. 2018.2323 ``Gaseousor rocky? Unveiling the nature of small worlds''. 
M.G. is an F.R.S.-FNRS Senior Research Associate. 
KGI is the ESA CHEOPS Project Scientist and is responsible for the ESA CHEOPS Guest Observers Programme. She does not participate in, or contribute to, the definition of the Guaranteed Time Programme of the CHEOPS mission through which observations described in this paper have been taken, nor to any aspect of target selection for the programme. 
This work was granted access to the HPC resources of MesoPSL financed by the Region Ile de France and the project Equip@Meso (reference ANR-10-EQPX-29-01) of the programme Investissements d'Avenir supervised by the Agence Nationale pour la Recherche. 
This work was also partially supported by a grant from the Simons Foundation (PI Queloz, grant number 327127). 
Acknowledges support from the Spanish Ministry of Science and Innovation and the European Regional Development Fund through grant PGC2018-098153-B- C33, as well as the support of the Generalitat de Catalunya/CERCA programme. 
LMS gratefully acknowledges financial support from the CRT foundation under Grant No. 2018.2323 ‘Gaseous or rocky? Unveiling the nature of small worlds’. 
S.G.S. acknowledge support from FCT through FCT contract nr. CEECIND/00826/2018 and POPH/FSE (EC). 
GyMSz acknowledges the support of the Hungarian National Research, Development and Innovation Office (NKFIH) grant K-125015, a PRODEX Institute Agreement between the ELTE E\"otv\"os Lor\'and University and the European Space Agency (ESA-D/SCI-LE-2021-0025), the Lend\"ulet LP2018-7/2021 grant of the Hungarian Academy of Science and the support of the city of Szombathely. 
S.S. has received funding from the European Research Council (ERC) under the European Union’s Horizon 2020 research and innovation programme (grant agreement No 833925, project STAREX).

This research has made use of the SVO Filter Profile Service (http://svo2.cab.inta-csic.es/theory/fps/) supported from the Spanish MINECO through grant AYA2017-84089.

G.S. acknowledges P.M. because, hey!, writing a paper is fun, but writing a paper while making a cake for his 6-month old nephew is more fun.
\end{acknowledgements}



\section{Posterior distributions of the model parameters and best fit of \tess\ light curves}

\FloatBarrier

\section{Posterior distributions of the model parameters and best fit of \cheops\ light curves}

\longtab{
\begin{longtable}{lllllll}
\caption{Model parameters for the fit of the \cheops\ data.}\label{tab:CHEOPSfit}\\
\hline\hline
Jump parameters & Symbol & Units & Prior & MAP & C.I.\tablefootmark{a}\\
\hline
\endfirsthead
\caption{continued.}\\
\hline\hline
Jump parameters & Symbol & Units & Prior & MAP & C.I.\tablefootmark{a}\\
\hline
\endhead
\hline
\endfoot
Phase curve amplitude & A$_{\rm p}$ & ppm & $U$(0,500) & 70 & $80^{+60}_{-50}$\\
Jitter for V1 & j$_{\rm V1}$ & ppm & $U$(0,2000) & 300 & 320$^{+120}_{-160}$\\
Normalization of V1 & c$_{\rm 0,V1}$ & -- & $U$(-0.002,0.002) & -0.0001 & -0.0002(1)\\
Linear term for V1 & c$_{\rm 1,V1}$ & days$^{-1}$ & $U$(-0.03,0.03) & 0.006 & 0.006(2)\\
Roll-angle\dots & a$_{\rm 1,V1}$ & -- & $U$(-0.02,0.02) & 0 & 0.0000(1)\\
\dots detrending\dots & b$_{\rm 1,V1}$ & -- & $U$(-0.02,0.02) & 0.0005 & 0.0005(2)\\
\dots coefficients\dots & a$_{\rm 2,V1}$ & -- & $U$(-0.02,0.02) & 0 & 0.0000(1)\\
\dots for V1 & b$_{\rm 2,V1}$ & -- & $U$(-0.02,0.02) & -0.0001 & -0.0001(1)\\
Jitter for V2 & j$_{\rm V2}$ & ppm & $U$(0,2000) & 600 & 600(100)\\
Normalization of V2 & c$_{\rm 0,V2}$ & -- & $U$(-0.002,0.002) & -0,0001 & -0.0001(2)\\
Linear term for V2 & c$_{\rm 1,V2}$ & days$^{-1}$ & $U$(-0.03,0.03) & 0.002 & 0.002(2)\\
Roll-angle\dots & a$_{\rm 1,V2}$ & -- & $U$(-0.02,0.02) & 0 & 0.0000(2)\\
\dots detrending\dots & b$_{\rm 1,V2}$ & -- & $U$(-0.02,0.02) & 0.0004 & 0.0004(2)\\
\dots coefficients\dots & a$_{\rm 2,V2}$ & -- & $U$(-0.02,0.02) & -0.0001 & -0.0001(2)\\
\dots for V2 & b$_{\rm 2,V2}$ & -- & $U$(-0.02,0.02) & 0.0001 & 0.0001(1)\\
Jitter for V3 & j$_{\rm V3}$ & ppm & $U$(0,2000) & 590 & 620(90)\\
Normalization of V3 & c$_{\rm 0,V3}$ & -- & $U$(-0.002,0.002) & -0.00011 & -0.00013(9)\\
Linear term for V3 & c$_{\rm 1,V3}$ & days$^{-1}$ & $U$(-0.03,0.03) & 0.009 & 0.009(3)\\
Roll-angle\dots & a$_{\rm 1,V3}$ & -- & $U$(-0.02,0.02) & 0.0003 & 0.0003(1)\\
\dots detrending\dots & b$_{\rm 1,V3}$ & -- & $U$(-0.02,0.02) & 0.0003 & 0.0006(1)\\
\dots coefficients\dots & a$_{\rm 2,V3}$ & -- & $U$(-0.02,0.02) & -0.0005 & -0.0005(1)\\
\dots for V3 & b$_{\rm 2,V3}$ & -- & $U$(-0.02,0.02) & -0.0002 & -0.0002(1)\\
Jitter for V4 & j$_{\rm V4}$ & ppm & $U$(0,2000) & 500 & 500(100)\\
Normalization of V4 & c$_{\rm 0,V4}$ & -- & $U$(-0.002,0.002) & -0.00001 & -0.00002(7)\\
Linear term for V4 & c$_{\rm 1,V4}$ & days$^{-1}$ & $U$(-0.03,0.03) & 0.002 & 0.002(2)\\
Roll-angle\dots & a$_{\rm 1,V4}$ & -- & $U$(-0.02,0.02) & 0 & 0.0001(1)\\
\dots detrending\dots & b$_{\rm 1,V4}$ & -- & $U$(-0.02,0.02) & -0.00001 & -0.00001(9)\\
\dots coefficients\dots & a$_{\rm 2,V4}$ & -- & $U$(-0.02,0.02) & -0.00024 & -0.00024(9)\\
\dots for V4 & b$_{\rm 2,V4}$ & -- & $U$(-0.02,0.02) & 0.00002 & 0.00002(9)\\
Jitter for V5 & j$_{\rm V5}$ & ppm & $U$(0,2000) & 480 & 490(70)\\
Normalization of V5 & c$_{\rm 0,V5}$ & -- & $U$(-0.002,0.002) & -0.00002 & -0.00003(6)\\
Linear term for V5 & c$_{\rm 1,V5}$ & days$^{-1}$ & $U$(-0.03,0.03) & -0.0001 & -0.0001(6)\\
Roll-angle\dots & a$_{\rm 1,V5}$ & -- & $U$(-0.02,0.02) & 0.00004 & 0.00004(7)\\
\dots detrending\dots & b$_{\rm 1,V5}$ & -- & $U$(-0.02,0.02) & 0.00025 & 0.00024(7)\\
\dots coefficients\dots & a$_{\rm 2,V5}$ & -- & $U$(-0.02,0.02) & -0.00017 & -0.00017(7)\\
\dots for V5 & b$_{\rm 2,V5}$ & -- & $U$(-0.02,0.02) & 0.00002 & 0.00002(7)\\
Jitter for V6 & j$_{\rm V6}$ & ppm & $U$(0,2000) & 510 & 520(70)\\
Normalization of V6 & c$_{\rm 0,V6}$ & -- & $U$(-0.002,0.002) & -0.00011 & -0.00012(6)\\
Linear term for V6 & c$_{\rm 1,V6}$ & days$^{-1}$ & $U$(-0.03,0.03) & 0.0015 & 0.0015(7)\\
Roll-angle\dots & a$_{\rm 1,V6}$ & -- & $U$(-0.02,0.02) & 0.00022 & 0.00022(7)\\
\dots detrending\dots & b$_{\rm 1,V6}$ & -- & $U$(-0.02,0.02) & 0.00004 & 0.00004(7)\\
\dots coefficients\dots & a$_{\rm 2,V6}$ & -- & $U$(-0.02,0.02) & -0.00028 & -0.00028(7)\\
\dots for V6 & b$_{\rm 2,V6}$ & -- & $U$(-0.02,0.02) & 0.00001 & 0.00001(7)\\
Jitter for V7 & j$_{\rm V7}$ & ppm & $U$(0,2000) & 400 & 410(80)\\
Normalization of V7 & c$_{\rm 0,V7}$ & -- & $U$(-0.002,0.002) & -0.00025 & -0.00026(6)\\
Linear term for V7 & c$_{\rm 1,V7}$ & days$^{-1}$ & $U$(-0.03,0.03) & 0.0088 & 0.0088(4)\\
Roll-angle\dots & a$_{\rm 1,V7}$ & -- & $U$(-0.02,0.02) & 0.00039 & 0.00039(5)\\
\dots detrending\dots & b$_{\rm 1,V7}$ & -- & $U$(-0.02,0.02) & 0.00001 & 0.00001(6)\\
\dots coefficients\dots & a$_{\rm 2,V7}$ & -- & $U$(-0.02,0.02) & -0.00024 & -0.00024(7)\\
\dots for V7 & b$_{\rm 2,V7}$ & -- & $U$(-0.02,0.02) & 0.00016 & 0.00016(7)\\
Jitter for V8 & j$_{\rm V8}$ & ppm & $U$(0,2000) & 460 & 480(80)\\
Normalization of V8 & c$_{\rm 0,V8}$ & -- & $U$(-0.002,0.002) & -0.0003 & -0.0003(1)\\
Linear term for V8 & c$_{\rm 1,V8}$ & days$^{-1}$ & $U$(-0.03,0.03) & -0.0003 & -0.0003(7)\\
Roll-angle\dots & a$_{\rm 1,V8}$ & -- & $U$(-0.02,0.02) & 0.0003 & 0.0003(2)\\
\dots detrending\dots & b$_{\rm 1,V8}$ & -- & $U$(-0.02,0.02) & 0.00028 & 0.00028(7)\\
\dots coefficients\dots & a$_{\rm 2,V8}$ & -- & $U$(-0.02,0.02) & -0.00045 & -0.00045(8)\\
\dots for V8 & b$_{\rm 2,V8}$ & -- & $U$(-0.02,0.02) & 0.0003 & 0.0003(1)\\
Jitter for V9 & j$_{\rm V9}$ & ppm & $U$(0,2000) & 0 & 140$^{+120}_{-90}$\\
Normalization of V9 & c$_{\rm 0,V9}$ & -- & $U$(-0.002,0.002) & 0.0002 & 0.0002(2)\\
Linear term for V9 & c$_{\rm 1,V9}$ & days$^{-1}$ & $U$(-0.03,0.03) & 0.0030 & 0.0030(7)\\
Roll-angle\dots & a$_{\rm 1,V9}$ & -- & $U$(-0.02,0.02) & -0.0004 & -0.0004(3)\\
\dots detrending\dots & b$_{\rm 1,V9}$ & -- & $U$(-0.02,0.02) & 0.00006 & 0.00006(8)\\
\dots coefficients\dots & a$_{\rm 2,V9}$ & -- & $U$(-0.02,0.02) & -0.0001 & -0.0001(1)\\
\dots for V9 & b$_{\rm 2,V9}$ & -- & $U$(-0.02,0.02) & -0.0003 & -0.0003(4)\\
\hline
Fixed parameters & Symbol & Units & Value  &  & Notes \\
\hline
Transit time & $T_0$ & BJD$_{\rm TDB}$-2400000 & 58912.10730 &  & see Table~\ref{tab:TESSfit}\\
Orbital frequency & $\nu_b$ & days$^{-1}$ & 1.2292955 &  & see Table~\ref{tab:TESSfit}\\
Stellar density & $\rho_\star$ & $\rho_\sun$ & 2.20 & & see Table~\ref{tab:TESSfit}\\
Stellar mass & M$_\star$ & M$_{\rm\sun}$ & 0.71 & & see Table~\ref{tab:parameters} \\
RV semi-amplitude& K$_{\rm RV}$ & m/s & 551.7 & & see Table~\ref{tab:parameters} \\
Linear LD coef. & u$_{\rm LLD}$ & --- & 0.648 & & computed with \texttt{LDTk} \\
GD coef. & y$_{\rm GD}$ & --- &  0.537 & & from \citet{Claret2021} \\
Phase offset & $\Delta\phi$ & --- & 0 &  & \\
Tidal lag & $\Theta$ & rad & 0 &  & \\
\hline
Derived parameters & Symbol & Units & MAP & C.I. & Notes \\
\hline
Eclipse depth & $\delta_{\rm ecl}$ & ppm & 70 & $80^{+60}_{-50}$ & 280 (99.9\% upper limit)\\
\hline
\end{longtable}
\tablefoot{
        \tablefoottext{a}{Uncertainties expressed in parentheses refer to the last digit(s).}
}
}

\FloatBarrier

\FloatBarrier

\section{Posterior distributions of the model parameters to fit \uvis\ light curves}

\end{document}